\documentclass[usenatbib,usegraphicx,usedcolumn]{mn2e}
\usepackage{bm}
\usepackage{amssymb,amsmath}
\usepackage{graphicx}
\usepackage{natbib}
\usepackage{program}
\usepackage{color}
\usepackage{float}

\usepackage[backref,breaklinks,colorlinks,citecolor=blue]{hyperref}  

\def\simlt{\lower.5ex\hbox{$\; \buildrel < \over \sim \;$}}
\def\simgt{\lower.5ex\hbox{$\; \buildrel > \over \sim \;$}}

\graphicspath{{./}{figures/}}

\title[First Galaxies in BECDM Cosmology] {Galaxy Formation with BECDM - II. Cosmic Filaments and First Galaxies}
\author[Mocz et. al.]{Philip Mocz$^{1}$\thanks{E-mail: pmocz@astro.princeton.edu (PM)}\thanks{Einstein Fellow}, 
Anastasia Fialkov$^{2}$,
Mark Vogelsberger$^{3}$,
Fernando Becerra$^{4}$, \newauthor 
Xuejian Shen$^{5}$,
Victor H. Robles$^{6}$, 
Mustafa A. Amin$^{7}$, 
Jes\'us Zavala$^{8}$, \newauthor 
Michael Boylan-Kolchin$^{9}$, 
Sownak Bose$^{4}$, 
Federico Marinacci$^{10}$, \newauthor 
Pierre-Henri Chavanis$^{11}$, 
Lachlan Lancaster$^{1}$, 
and Lars Hernquist$^{4}$ \\
$^{1}$Department of Astrophysical Sciences, Princeton University, 4 Ivy Lane, Princeton, NJ, 08544, USA \\
$^{2}$Institute of Astronomy, University of Cambridge, Madingley Road, Cambridge CB3 0HA, UK\\
$^{3}$Department of Physics, Kavli Institute for Astrophysics and Space Research, M.I.T., Cambridge, MA 02139, USA\\
$^{4}$Harvard-Smithsonian Center for Astrophysics, 60 Garden Street, Cambridge, MA 02138, USA \\
$^{5}$TAPIR, California Institute of Technology, Pasadena, CA 91125, USA \\
$^{6}$Department of Physics and Astronomy, University of California, Irvine, CA 92697,USA\\
$^{7}$Physics \& Astronomy Department, Rice University, Houston, Texas 77005-1827, USA\\
$^{8}$Center for Astrophysics and Cosmology, Science Institute, University of Iceland, Dunhagi 5, 107 Reykjavik, Iceland \\
$^{9}$Department of Astronomy, The University of Texas at Austin, 2515 Speedway, Stop C1400, Austin, TX 78712-1205, USA \\
$^{10}$Department of Physics \& Astronomy, University of Bologna, via Gobetti 93/2, 40129 Bologna, Italy\\
$^{11}$Laboratoire de Physique Th\'eorique, CNRS, Universit\'e Paul Sabatier, 118 route de Narbonne 31062 Toulouse, France\\
}

\begin{document}
\pagerange{\pageref{firstpage}--\pageref{lastpage}} \pubyear{2019}
\maketitle

\label{firstpage}

\begin{abstract} 
Bose-Einstein Condensate Dark Matter (BECDM; also known as Fuzzy Dark Matter) is motivated by fundamental physics and has recently received significant attention as a serious alternative to the established Cold Dark Matter (CDM) model. We perform cosmological simulations of BECDM gravitationally coupled to baryons and investigate structure formation at high redshifts ($z \ga 5$) for a boson mass $m=2.5\cdot 10^{-22}~{\rm eV}$, exploring the dynamical effects of its wavelike nature on the cosmic web and the formation of first galaxies. Our BECDM simulations are directly compared to CDM as well as to simulations where the dynamical quantum potential is ignored and only the initial suppression of the power spectrum is considered -- a Warm Dark Matter-like (``WDM'') model often used as a proxy for BECDM. Our simulations confirm that ``WDM'' is a good approximation to BECDM on large cosmological scales even in the presence of the baryonic feedback. Similarities also exist on small scales, with primordial star formation happening both in isolated haloes and continuously along cosmic filaments; the latter effect is not present in CDM. Global star formation and metal enrichment in these first galaxies are delayed in BECDM/``WDM'' compared to the CDM case: in BECDM/``WDM'' first stars form at $z\sim 13$/$13.5$ while in CDM star formation starts at $z\sim 35$. The signature of BECDM interference, not present in ``WDM'', is seen in the evolved dark matter power spectrum: although the small scale structure is initially suppressed, power on kpc scales is added at lower redshifts. Our simulations lay the groundwork for realistic simulations of galaxy formation in BECDM.
\end{abstract}



\section{Introduction}
\label{Sec:Intro}

The Lambda Cold Dark Matter ($\Lambda$CDM) theory has proven to be quite successful in describing the observable Universe:
it explains both the homogeneity of the Universe on the largest cosmological scales and the structure of the cosmic web \citep{PlanckCP:2016}. Numerical simulations based on $\Lambda$CDM are able to correctly predict statistical properties of the observed different populations of galaxies, including Milky Way-like disk galaxies \citep[e.g.,][]{2014Natur.509..177V,2014MNRAS.444.1518V,Pillepich:2017,2018MNRAS.480..800H,2017MNRAS.467..179G,2018MNRAS.475..676S,2019arXiv190907976V}. However, on small-scales, certain challenges have been claimed to afflict the CDM model \citep{2015PNAS..11212249W,2017ARA&A..55..343B} such as the ``cusp-core problem'' \citep[$\Lambda$CDM predicts cuspy haloes instead of the observed cored haloes;][]{moore1994,flores1994,2004MNRAS.351..903G,2009MNRAS.397.1169D,2010AdAst2010E...5D};
the ``missing satellites problem'' \citep{klypin1999, moore1999} and the related problem with the abundance of
isolated dwarfs \citep{2009ApJ...700.1779Z,2011ApJ...739...38P,2015MNRAS.454.1798K};
and the ``too-big-to-fail problem'' \citep[in $\Lambda$CDM large dark matter subhaloes are too dense towards their centers when compared to the brightest observed dwarf satellite galaxies][]{boylan-kolchin2011,2012MNRAS.422.1203B}. 
Many of these do, however, rely on comparison of observational data with collisionless simulations. Indeed, the inclusion of baryon physics has been demonstrated to reduce or reconcile many of these issues \citep[e.g.,][]{2018PhRvL.121u1302K, 2019arXiv190410471O, 2019MNRAS.486..655D}. Still, $\Lambda$CDM theory is not universally accepted partially because the nature of its Cold Dark Matter (CDM) component remains a mystery. Multiple ground-based searches for a Weakly Interactive Massive Particle (WIMP) of mass $10-1000$~GeV, the most natural CDM candidate, have found no convincing evidence of dark matter, placing limits on its mass and the interaction strength with baryons \citep{2018RPPh...81f6201R}. 

The lack of discovery of a particle associated with the $\Lambda$CDM model resulted in a large variety of alternative dark matter scenarios, many of which are immune to small-scale problems of $\Lambda$CDM (for a recent review on different DM models and their impact on structure formation see \citealt{2019Galax...7...81Z}). One theoretical particle candidate for dark matter could be axions (axion dark matter), such as the standard quantum chromodynamics (QCD) axions of mass $10^{-3}$--$10^{-6}~{\rm eV}$ which resolve the strong CP problem \citep{1977PhRvL..38.1440P}. Axion-like particles are also predicted by string theory, which suggests the existence of a plethora of particles with masses over a broad range $ 10^{-33}$--$10^{-10}~{\rm eV}$ \citep[e.g.,][]{2010PhRvD..81l3530A,2019PhRvD..99f3517V}. Another popular dark matter candidate is sterile neutrino (see recent review by \citealt{2019PrPNP.104....1B}) or other types of Warm Dark Matter (WDM) -- fermionic particles of typical masses of a few keV. In some of the above-mentioned scenarios, dark matter affects cosmic structure formation on observable scales (e.g., axions of mass $\sim 10^{-22}~{\rm eV}$ suppress formation of galaxies of masses below $\sim 10^9$ M$_{\odot}$), thus making it possible to constrain or rule out such dark matter candidates using cosmology. 


In this paper we focus on Bose-Einstein Condensate Dark Matter \citep[BECDM, also known as Fuzzy Dark Matter, FDM;][]{Hu:2000} which, although being similar to CDM on large cosmological scales \citep{2018PhRvD..97h3519M}, has recently received close attention since it can provide a solution to the small-scale problems of $\Lambda$CDM. BECDM are ultra-light (axion-like) boson particles of masses around $10^{-22}~{\rm eV}$ which, due to the uncertainty principle, experience a quantum potential that prevents gravitational collapse on the de Broglie wavelength scale of few kpc (given cosmological velocities of $10$--$100~{\rm km}~{\rm s}^{-1}$). As a result, such dark matter is expected to form solitonic cores (rather than cusps) in galaxy centers and render fewer dwarf galaxies than what is predicted for $\Lambda$CDM
\citep{Hu:2000,2000NewA....5..103G,2000CQGra..17L...9G,2001PhRvD..63f3506M,2017PhRvD..95d3541H}.
Furthermore, filaments, and not necessarily haloes as in $\Lambda$CDM, are the sites of primordial star formation in BECDM \citep{Hirano:2017, moczPRL}, in a similar way to how the same process occurs in WDM models \citep{2003ApJ...591L...1Y,Gao:2007}. Finally, due to the quantum nature of BECDM, the cosmic web exhibits interference patterns \citep{2014NatPh..10..496S,moczPRL}. The bosons may also experience self-interactions, which may further affect cosmic structure formation \citep[][not considered in the current work]{2018PhRvD..98b3009C,2018PhRvD..97b3529D}. 

Owing to the kpc size of its de Broglie wavelength, BECDM can be tested with current and upcoming observations. At present, the majority of observational constraints place the boson mass at $m\gtrsim 10^{-22}~{\rm eV}$ \citep{2018MNRAS.476.3063H,2018arXiv180800464A,2019MNRAS.485.2861C,2019arXiv190906381L}, with Lyman-$\alpha$ constraints being potentially the most strict $m\gtrsim 10^{-21}~{\rm eV}$ \citep{Irsic:2017,2019MNRAS.482.3227N}. For a summary of various recent astrophysical constraints on the BECDM particle, see \cite{2019arXiv190804790Y}. However, the majority of these constraints in the literature are derived using approximate modeling of BECDM. Until recently, the study of structure formation in BECDM has been done using analytic methods as well as BECDM-only simulations \citep[although see][for the first approximate hydrodynamical treatment of BECDM minus wave effects]{Hirano:2017}. Using analytic tools, \citet{Hu:2000} showed that perturbations grow linearly on scales much larger than the Jeans scale at equality, $k_{{\rm Jeq}} = 9 \left(m/10^{-22}\rm{eV}\right)^{1/2}$ Mpc$^{-1}$, but oscillate on smaller scales leading to a suppression of clustering power and the subsequent deficit of dwarf galaxies. BECDM-only numerical simulations confirmed this picture: modest-resolution simulations verified the suppression of low-mass haloes \citep{Woo:2009}, while higher resolution simulations characterized the formation of solitonic cores \citep[e.g.,][]{2014NatPh..10..496S, 2016PhRvD..94d3513S,2017MNRAS.471.4559M}; although, we remark that these simulations did not use self-consistent initial conditions predicted by the BECDM model \citep{axionCAMB}, but artificial ones instead to highlight/exaggerate growth of structure. Simulations with WDM, although ignoring the quantum effects of BECDM, also serve as a guideline for understanding structure formation in BECDM. For instance, the fact that first star forming objects in WDM are filamentary \citep{2003ApJ...591L...1Y,Gao:2007} suggests that also in BECDM first stars will form in dense filaments rather than spherical haloes \citep{Hirano:2017}. In this paper we explore structure and galaxy formation in BECDM using the first fully self-consistent simulations of BECDM coupled to baryonic physics, which were recently presented by \cite{moczPRL}. Such simulations are indispensable for future validation or rejection of BECDM.

In order to explore the observable effects of BECDM, we have developed a spectral BECDM solver and used it to investigate properties of idealized, virialized BECDM haloes \citep[][Paper I]{2017MNRAS.471.4559M}. We have also integrated the solver into the hydrodynamics code {\sc Arepo}\footnote{{\sc Arepo} has been used to carry out state-of-the-art large-scale cosmological simulations, including the \textit{Illustris-TNG} project \citep{Pillepich:2017,2018MNRAS.475..676S}.} \citep{2010MNRAS.401..791S,2016MNRAS.455.1134P}, 
and performed first-of-their-kind hydrodynamical simulations with BECDM and explored star-forming filamentary structures at $z\sim 5.5$ \citep{moczPRL}. In this work we look deeper into the simulated universe and explore structure and galaxy formation across cosmic time. For reference, we also compare our BECDM simulations to $\Lambda$CDM and WDM-like\footnote{Our WDM-like particles are collisionless (like CDM) but feature a truncated initial power spectrum of BECDM.} (``WDM'') simulations with the same initial seed for the perturbations.

This paper is organized as follows. 
In Section~\ref{sec:num} we set the stage describing the mathematical formalism behind BECDM and summarize the numerical method that we employ.
In Section~\ref{sec:sim} we discuss the setup of our cosmological simulations, and present large-scale results in Section~\ref{sec:lss}.
In Sections~\ref{sec:dm}, \ref{sec:gas}, and \ref{sec:stars}  we explore the formation of dark matter structure,  evolution of structure in gas and formation of first star-forming objects, respectively. 
Observational prospects with telescopes such as the James Webb Space Telescope are discussed in Section~\ref{sec:JWST}. 
Main differences between BECDM and CDM, as well as between BECDM and ``WDM'', are summarized in Section~\ref{sec:sum}.
We offer our concluding remarks in Section~\ref{sec:conc}. 


\section{Numerical Methodology}
\label{sec:num}

In this section we outline  the mathematical framework that governs BECDM, the numerical methods, and the physics included in our cosmological simulations. We also offer comments on the computational challenges in simulating BECDM compared to CDM.

\subsection{BECDM cosmology}

BECDM is governed by the Schr\"odinger-Poisson (SP) equations. 
In an expanding universe, these are:
\begin{equation}
i\hbar \left(\frac{\partial \psi}{\partial t} +\frac{3}{2}H \right) = -\frac{\hbar^2}{2m}\nabla^2\psi + m V\psi
\end{equation}
\begin{equation}
\nabla^2 V = 4\pi G(\rho-\overline{\rho})
\end{equation}
where $\psi$ is the wave function describing the scalar-field boson in the non-relativistic limit, 
$\rho\equiv\lvert\psi\rvert^2$ is the dark matter density field, $\overline{\rho}$ is the volume-average density, $m$ is the boson (axion) mass, and $H\equiv \dot{a}/a$ is the rate of Hubble expansion where at redshift $z$ the scale factor is $a\equiv 1/(1+z)$.

In terms of comoving coordinate $\mathbf{x}$ (with the physical distance $\mathbf{r}\equiv a\mathbf{x}$) the SP equations become:
\begin{equation}
i\hbar \frac{\partial \psi_{\rm c}}{\partial t} = -a^{-2}\frac{\hbar^2}{2m}\nabla_{\rm c}^2\psi_{\rm c} + a^{-1} m V_{\rm c}\psi_{\rm c}
\label{eqn:coSP1}
\end{equation}
\begin{equation}
\nabla_{\rm c}^2 V_{\rm c} = 4\pi G(\rho_{\rm c}-\overline{\rho}_{\rm c})
\label{eqn:coSP2}
\end{equation}
where we have defined comoving quantities, relating to physical quantities as:
\begin{equation}
\rho_{\rm c} \equiv a^3\rho,
\,\,\,
\psi_{\rm c} \equiv a^{3/2}\psi,
\,\,\,
\nabla_{\rm c} \equiv a\nabla,
\,\,\,
V_{\rm c} \equiv aV.
\end{equation}

The SP equations, which describe a self-gravitating superfluid, can be recast into a fluid formulation per the \cite{1927ZPhy...40..322M} transformation, which can aid with physical intuition.
Decomposing the wavefunction as by its amplitude $\mathcal{R}$ and phase $S/\hbar$:
\begin{equation}
\psi_{\rm c} = \mathcal{R} {\rm e}^{iS/\hbar}
\end{equation}
and defining a velocity ($\mathbf{v}_{\rm M}$; Madelung velocity) as the gradient of the phase:
\begin{equation}
\mathbf{v}_{\rm M} \equiv \nabla S / m
\label{eqn:vmadel}
\end{equation}
the Schr\"odinger equation can then be written as:
\begin{equation}
\frac{\partial \rho_{\rm c}}{ \partial t}
+ \nabla_{\rm c} \cdot \left(
\rho_{\rm c} \mathbf{v}_{\rm M}
\right) = 0,
\end{equation}
\begin{equation}
\frac{\partial \mathbf{v}_{\rm M}}{\partial t}
+ a^{-2} \mathbf{v}_{\rm M} \cdot\nabla_{\rm c} \mathbf{v}_{\rm M} 
= -a^{-1} \nabla_{\rm c}V_{\rm c}
+ a^{-2} \frac{\hbar^2}{2m^2}\nabla_{\rm c}
\left(\frac{\nabla^2\mathcal{R}}{\mathcal{R}}\right).
\end{equation}
Aside from the quantum potential term on the r.h.s. (defined as $V_{\rm Q}\equiv -\frac{\hbar^2}{2m}\nabla^2\mathcal{R}/\mathcal{R}$), the evolution equations
look like those of classical evolution of individual particles under self-gravity, in the spirit of a Bohmian interpretation of quantum mechanics.
The quantum potential is responsible for quantum wave effects, including dispersion and interference, which can provide support against localized collapse due to self-gravity.

In contrast, CDM, a collisionless fluid as opposed to a superfluid, is governed by the Vlasov-Poisson (VP) equations.
In terms of canonical coordinates these are:
\begin{equation}
\frac{\partial f}{\partial t}
+ a^{-2} \mathbf{p}\cdot \frac{\partial f}{\partial \mathbf{x}}
- a^{-1}\nabla_{\rm c} V_{\rm c} \cdot \frac{\partial f}{\partial \mathbf{p}} = 0,
\label{eqn:coVP1}
\end{equation}
\begin{equation}
\nabla_{\rm c}^2 V_{\rm c} = 4\pi G(\rho_{\rm c}-\overline{\rho}_{\rm c})
\label{eqn:coVP2}
\end{equation}
where $f=f(x,p,t)$ is the distribution function, 
$\mathbf{p}=a^2\dot{\mathbf{x}}$ is the canonical momentum coordinate
(c.f. peculiar velocity is given by $\mathbf{v}=a\dot{\mathbf{x}}$),
and the density is given by $\rho_{\rm c} = \int f \,d^3p$.

On scales much larger than the local de-Broglie wavelength
\begin{equation}
\lambda_{\rm dB} = \frac{h}{m \mathbf{v}_{\rm M}}
\end{equation}
(which is on the order of a kpc for the axion masses and typical velocities inside low mass haloes that we consider here)
the SP and the VP equations approximate each other, even in the case of multi-stream flows \citep{1993ApJ...416L..71W,2017PhRvD..96l3532K,2018PhRvD..97h3519M}. 
Thus on the largest spatial scales, one expects BECDM to behave just like CDM.
In the SP equations, the parameter $\hbar/m$ controls the
level of macroscopic quantum-ness (``fuzziness'') in the equations.
In the limit $\hbar/m\to 0$ (i.e., heavy particles), the SP equations may be expected to recover the classical VP limit \citep{2017PhRvD..96l3532K, 2018PhRvD..97h3519M}. In particular \citet{2018PhRvD..97h3519M} have shown that the gravitational force-field converges to the classical collisionless limit as $\mathcal{O}(m^{-1})$ despite the density field not converging due to order unity fluctuations. Hence baryonic physics coupled to the dark matter converges to classical CDM behavior as well in the limit $\hbar/m\to 0$.

On smaller scales, the SP system is expected to show quantum wave phenomena, such as soliton cores (stable, ground-state eigenmodes where the uncertainty principal/quantum potential opposes the gravitational collapse; \citealt{2014NatPh..10..496S}),
vortex lines and reconnection \citep{2017MNRAS.471.4559M}, interference patterns, quantum tunneling, and 
non-classical phenomena if there exist jumps in the wavefunction phase (e.g., colliding cores can bounce off each other; \citealt{2016PhRvD..94d3513S}). 

Finally, we point out the scaling symmetry of the SP and VP equations:
\begin{equation}
\{ x,t,\rho,m \} \to \{ \alpha x, 
\beta t, \beta^{-2} \rho, \alpha\beta^{-2} m\},
\label{eqn:scaling}
\end{equation}
where $\alpha$ and $\beta$ are arbitraty scaling parameters.
The VP equations are scale-free, whereas
the SP equations have a single-scale (the de Broglie wavelength) set by the value of $\hbar/m$.

We note the SP equations for BECDM are obtained by taking the non-relativistic limit of a Universe with a scalar field. Higher order corrections may include an attractive self-potential which is of interest for future work because it has been recently shown that even a weak coupling strength can cause an instability in structure formation \citep{2018PhRvD..97b3529D}.

\subsubsection{Spectral Method}

The SP equations in comoving form (Eqns~\ref{eqn:coSP1} and \ref{eqn:coSP2}) are evolved numerically using the spectral method we have developed and used in \citet{2017MNRAS.471.4559M} -- see reference for details.
The method uses a $2^{\rm nd}$-order time-stepping method and gives exponential convergence in space. The time-steps are decomposed into a kick-drift-kick leapfrog-like scheme, where each `kick' and `drift' are unitary operators acting on the wavefunction.
This makes it natural to automatically couple the method to particle-based $N$-body techniques that evolve gas and star particles in the simulation on the same sub-timestep spacings.
In our simulations, the BECDM is fully coupled to baryons by including the baryonic contribution to the potential in the Poisson equation 
(Eqn~\ref{eqn:coSP2}) and using the full gravitational potential in Eqn~\ref{eqn:coSP1}.

The spectral method proves to be useful at capturing interference patterns and
vortex lines, which turn out to be an integral feature of filaments and haloes in these cosmological simulations \citep{2017MNRAS.471.4559M}. Vortex lines are locations where the density $\rho=0$, but are sites of quantized vorticity in the fluid (the rest of the fluid is vortex free, as $\nabla\times\mathbf{v}_{\rm M}=0$ since the velocity is the gradient of a scalar -- the phase).
An alternative approach would be to work in the fluid (Madelung) formulation, and use fluid solver methods such as the smooth-particle-hydrodynamics approach
developed by \cite{2015PhRvE..91e3304M}, but the 
fluid formulation may prove difficult to capture vortex lines accurately, as the density vanishes and the velocity is formally infinite here.
The spectral method is able to capture vortex lines without difficulty, owing to its exponential spatial convergence and the fact that the calculations are done in terms of the phase rather than its gradient (i.e., the velocity).

The spectral method is limited to uniform grids. 
The resolution requirements are strict to be able to resolve soliton cores expected to be in the centers of haloes
\citep{2014NatPh..10..496S}. The soliton comoving radius scales approximately as \citep{2014PhRvL.113z1302S}:
\begin{equation}
x_{\rm c} \sim 2\left(\frac{m}{2.5\cdot10^{-22}~{\rm eV}}\right)^{-1} \left(\frac{1+z}{7}\right)^{1/2} \left(\frac{M_{\rm halo}}{10^9~M_\odot}\right)^{-1/3}~{\rm kpc}
\end{equation}
and is on order of few kpc for the selected axion mass. 
Furthermore, because the velocity is defined as the gradient of the phase, 
the spatial resolution defines a maximum
velocity that can be numerically represented as:
\begin{equation}
v_{\rm M, max}=\frac{\hbar}{m}\frac{\pi}{\Delta x}.
\end{equation}
One needs to resolve the velocity dispersion of the highest mass halo expected to be formed in the cosmological box of a given volume.
These limitations mean that with present computational supercomputing resources, one is limited to simulated box sizes of a few $h^{-1}$ Mpc and is confined to the high-redshift regime for axion masses of interest around $10^{-22}$--$10^{-21}$ eV. 

\subsubsection{Discussion of various numerical approaches for BECDM and their advantages and limitations}
\label{Sec:methods_comp}

BECDM is  more computationally challenging  than CDM, due to the need to resolve kpc-scale interference patterns and large velocities (present throughout the halo) and soliton cores. Here we outline the advantages and disadvantages of different methods (spectral, adaptive grid finite difference, and SPH and mesh-free finite-volume methods for the Madelung formulation).

Spectral methods are an ideal choice for the Schr\"odinger-Poisson equations, owing to the smoothness of the wavefunction, and the unitary nature of the discretization \citep{2017MNRAS.471.4559M}. Spectral accuracy allows for machine-precision control of spatial truncation errors. The only limitation of the method is that the largest wave-number in the solution (corresponding to the smallest scale of $2\pi/k_{\rm max}$) needs to be resolved. To robustly model BECDM, one needs to resolve the de Broglie wavelength, which  is of order a kpc for our choice of the boson mass. Thus, the box size for a cosmological simulation is limited to a few comoving Mpc for a resolution of $1024^3$. This limits applications of spectral methods to systems with low values of  maximal velocities, such as low-mass  first galaxies (in haloes of mass  $<10^{11}~M_\odot$), such as the work presented by \citet{moczPRL} and the one here. Due to the resolution requirements we stop our simulations at $z=5.5$. It is not possible at the current resolution for us to evolve the solution further to lower redshifts. This is because the halo masses would increase, as well as their velocity dispersion, which would not be resolved and the spectral method would break. Moreover, the physical soliton size becomes smaller in these cases; more massive haloes have more compact solitons, and soliton size on the fixed comoving grid shrinks as the inverse of the scale factor $1/a$.

To achieve a slightly larger cosmological box, one may use an adaptive refined mesh as done by \citet{2014NatPh..10..496S}. The wave function can be evolved with a finite difference technique in this case (taking special care that truncation errors of the complex field do not exponentially blow up the solution). This approach makes it possible to capture the evolution of BECDM in  larger volumes utilizing the fact that haloes and filaments show interference patterns which need to be resolved, but voids, which have only a single velocity, are feature-free (and make up the majority of the volume). Still, such simulations are limited to box sizes $<10$~Mpc, whereas $\Lambda$CDM simulations can be much larger \citep[$>100$~Mpc -- Gpc,][]{2005Natur.435..629S,2019arXiv190905273V}.

The SP equation can also be rewritten in fluid (Madelung) form, and solved with an SPH discretization, a concept shown in \cite{2015PhRvE..91e3304M}. The method has a theoretical limitation, which is that velocities can formally diverge where the density is vanishing, and it remains to be shown that such regions (such as vortex lines in 3D) can be accurately resolved. Second, at least $\mathcal{O}(10)$ particles are needed per interference wave-crest in order to resolve it (i.e., the SPH smoothing length needs to be smaller than the local interference scale). An SPH approach has been developed by \cite{2018MNRAS.478.3935N} for the study of large-scale ($>10$~Mpc) cosmic structure. However, their simulations do not have enough resolution to resolve interference crests or the soliton core at the halo centers, which we are interested in. The SPH numerical method is able to capture the Mpc-scale effects of the quantum potential (e.g. suppressed power spectrum) while missing small-scale features. 
The fluid formulation has also been discretized recently for the mesh-free finite-volume method \citep{2019MNRAS.489.2367H}.
However, it remains to be shown that these Madelung methods can accurately capture coherent interference patterns in the simulations. 

A review of some of the numerical methods for BECDM is also given by \cite{2018FrASS...5...48Z}.  An open-source pseudo-spectral solver (\textsc{PyUltraLight}) exists for Python as well \citep{2018JCAP...10..027E}.

\subsection{Baryons}

Dark matter is coupled to baryons (gas, stars) in the universe through gravity.
The baryons themselves experience complex physical processes, which in our simulations are covered by {\sc Arepo} and include sub-grid models for primordial and metal-line cooling, chemical enrichment, stochastic star formation with a density
threshold of $0.13~{\rm cm}^{-3}$, supernova feedback via kinetic winds, and instantaneous uniform reionization at $z\sim 6$ \citep{2013MNRAS.436.3031V,2014MNRAS.438.1985T,Pillepich:2017,2018MNRAS.475..676S}.
Such models have been used in the \textit{Illustris} and \textit{Illustris-TNG} projects which aimed at reproducing 
the observed properties of galaxies in a $\Lambda$CDM Universe.
We refer the reader to the papers cited in this paragraph for the full details of the baryonic physics implementation. For the purposes of this work, we used the fiducial framework of \cite{2013MNRAS.436.3031V}. Feedback from supermassive black holes was not included as it would not be relevant for the low mass first haloes studied here.

The stellar feedback models have been tuned previously to CDM simulations, with free parameters constrained based on the overall star formation efficiency using smaller scale simulations \citep{2013MNRAS.436.3031V}. We use the same model for our BECDM simulations here. Whether the tuned feedback parameters in smaller CDM simulations apply to BECDM simulations requires further study.

Stellar feedback affects low mass galaxies most. Our subgrid model for feedback is meant to describe the effects of Type II supernovae (SNII),
and uses the local star formation rate to set the mass loading of 
stellar winds driven by the energy available through SNII \citep{2013MNRAS.436.3031V,2018MNRAS.473.4077P}. Reduced star formation in BECDM compared to CDM will also lessen the impact of such feedback, unless the mass loading factor is re-tuned.

\section{Simulations}
\label{sec:sim}
The spectral BECDM solver has been implemented in the {\sc Arepo} code \citep{2010MNRAS.401..791S}, a high-performance parallel code for solving gravity and (magneto)hydrodynamics which incorporates  star formation via sub-grid prescription. Here we describe our simulation setups and output.

We carry out three types of cosmological simulations with a full physical treatment of the baryons:
\begin{enumerate}
    \item CDM: the dark matter component is CDM, with proper CDM initial conditions (Gaussian random field conditions evolved to $z=127$ via second-order Lagrangian perturbation theory).
    \item BECDM: all  dark matter is in the form of BECDM. We model both the initial  suppression in the  power spectrum \citep[using {\sc axionCAMB},][]{axionCAMB}
        and posterior evolution which accounts for the quantum potential;
    \item ``WDM'': an intermediate approach where we simulate the dark matter as collisionless (like CDM) but with BECDM initial conditions. In this type of simulation, we explore just the effect of the suppression in the power spectrum on the subsequent structure formation and evolution, and ignore the dynamical effects of the quantum potential. We refer to this simulation as a ``WDM'' simulation, in quotes, because full-physics WDM should include the contribution of thermal velocities which we ignore here. Such an analogy  between BECDM and WDM has been made in the past \citep{2017PhRvD..95d3541H,Hirano:2017}: our boson particle mass of $m=2.5\cdot 10^{-22}~{\rm eV}$ corresponds roughly to a WDM particle mass of $1.4$~keV.  
\end{enumerate}
Finally, to isolate the effect of baryonic feedback,  we run three corresponding dark matter-only simulations.

\subsection{Generating initial conditions for BECDM}

Initial power spectra for the BECDM cosmology are generated 
at $z=127$ using {\sc axionCAMB} and assuming that all dark matter is in the form of axions. The code adds a relativistic axion fluid (with boson mass $m$) component to the universe, and calculates its early evolution using the first order perturbed Klein-Gordon equations. At late times, the treatment uses a WKB approximation that matches the axion fluid to an effective fluid with equation of state zero and a scale-dependent sound-speed.

The initial power spectra match that of CDM on large scales ($>$~Mpc), and the main effect of BECDM is to introduce a cutoff on small scales
due to the length-scale introduced by the de Broglie wavelength of the axion. For an axion mass of $m=2.5\cdot 10^{-22}$, 
the cutoff is approximately at $L_{\rm cutoff}\simeq 1.4h^{-1}$~Mpc. The cutoff scales should approximately scale as $m^{-1/2}$ per the scaling symmetry of the SP equations (Eqn~\ref{eqn:scaling}). Stated more accurately, the power drops by a factor of $2$ relative to CDM at \citep{Hu:2000}:
\begin{equation}
    k_{1/2} \simeq  4.5\left(m/10^{-22}~{\rm eV}\right)^{4/9} {\rm Mpc}^{-1}.
\end{equation}
We note, again that Lyman-$\alpha$ constraints prefer heavier dark matter particle masses (barring uncertainties in the equation of state of intergalactic gas, and the effect of galactic winds on the gas distribution and thermal state). In the case of WDM, the $2$-sigma constraint is $m>3.5~{\rm keV}$ \citep{2017PhRvD..96b3522I}. Our choice of the  axion mass is justified by: (1) we have yet to learn how well the WDM analogy actually approximates BECDM in the nonlinear regime, (2) we wish to explore  a case where quantum effects are significant on the simulated scales -- higher axion mass, fully consistent with the Lyman-$\alpha$ constraints, result in smaller de Broglie wavelengths which is below the resolution limit of our simulations. 
We point out that while the general effects (due to a primordial cutoff) are generally the same in ``WDM'' and in BECDM, the physics driving this is markedly different: free-streaming in one case, versus quantum pressure in the other.

The power spectra shapes we obtain are fed into {\sc NGenIC} \citep{2005Natur.435..629S,2012MNRAS.426.2046A}, which then generates initial conditions in physical space based on the Zeldovich approximation (second-order Lagrangian perturbation theory). {\sc NGenIC} perturbs the positions of equal mass particles on a uniform 3D grid.
We generate new uniform-grid initial conditions for our spectral method via the standard cloud-in-cell (CIC) binning of the output of {\sc NGenIC}. Furthermore, {\sc NGenIC} returns particle velocities, which we use to specify the phase of the wavefunction. Converting velocities to the phase is accomplished in a straightforward manner by taking the divergence of both sides of Eqn~\ref{eqn:vmadel} and solving the resulting Poisson equation using standard Poisson-solver methods.

All our simulations have the same random seed (i.e., same randomly drawn Fourier mode phases and amplitudes up to normalization by the initial power spectral shape of the cosmology). This allows for direct comparison of structures that form under each cosmology we consider.

\subsection{Convergence}

We have carried out various numerical convergence tests on the box size and resolution to ensure that our production run is reliable. Our box  size, resolution, and particle mass were chosen so that we could trust the internal structure of haloes down to $z\sim5$ in our final production run. We have found that the spectral method has excellent exponential convergence properties as long as power does not try to build up beyond the smallest resolved spatial scale (i.e., the smallest soliton has to be resolved). Beyond this limit, the code shows clear numerical artefacts (e.g., `ringing')  so the breakdown of convergence is easy to tell -- this is a nice property of spectral methods. Such ringing is not observed in the production run. 
At half resolution, a $512^3$ run is converged to our production run down to redshift $z\sim 7$.

At half resolution ($512^3$), twice the box size, the power spectrum in the overlapping region of spatial scales have been found to match at $z=9$. We do not have this test at later redshifts because to get converged answers on the box size test would require doing another production quality run at twice the box size. But the findings show that the early growth of these first objects, and their internal structure, is not affected by the box size, which is our main focus. However, it has been well studied that box size can affect the \textit{clustering} of haloes, but will not affect the \textit{internal structure} of haloes \citep{2006MNRAS.370..691P}. 

\subsection{Summary of the Simulation setup}

Following the resolution requirements outlined above, we choose a box size of $1.7h^{-1}$~Mpc, a $1024^3$ resolution spectral grid used for dark matter. 
We also include $512^3$ baryon gas particles, giving us a mass resolution of $\sim 10^3 M_\odot$. The simulations were run from $z=127$ to $z=5.5$. 

The simulations use cosmological parameters of 
$\Omega_{\rm m}= 0.3089$, $\Omega_{\rm \Lambda}= 0.6911$, $\Omega_{\rm b}= 0.0486$, $h=0.6774$ consistent with the Planck observations of temperature and polarization anisotropies of the CMB \citep{PlanckCP:2016}. Since we are simulating a small cosmological volume, we have decided to enhance clustering by boosting $\sigma_8$ from $ 0.8$ to $1.4$ \citep[e.g.,][]{Naoz2012}.
This is done because we are interested in capturing the first structures that form (which arise from higher-$\sigma$ rarer fluctuations) rather than focusing on the statistics of halo masses. 

We have selected an ``interesting'' initial seed for structure formation out of 20 randomly generated initial conditions to contain a nearly-spherical halo and a well defined  filament. This was done with an aim to explore the  variety of structures that could form in a BECDM universe. The drawback of  our selection process is that the simulated volume is not representative.

We have carried out our simulations on the Stampede 2 supercomputing cluster, part of the Texas Advanced Computing Center (TACC) at the University of Texas, and the Odyssey cluster supported by the FAS Division of Science, Research Computing Group at Harvard University. The computational cost for the BECDM simulation is $\sim 3$ million CPU core hours. 
The other simulations are cheap (by more than a factor of $20$) because they use $8$ times fewer particles and have a less stringent timestep criterion \citep{2017MNRAS.471.4559M}.

\section{Large scale structure}
\label{sec:lss}

\begin{figure*}
\begin{center}
\setlength\tabcolsep{1.5pt}
\begin{tabular}{cccccc}
& $z=63$ & $z=31$ & $z=15$ & $z=7$  & $z=5.5$ \\ 
\rotatebox{90}{\,CDM - dm} & 
\includegraphics[width=0.19\textwidth]{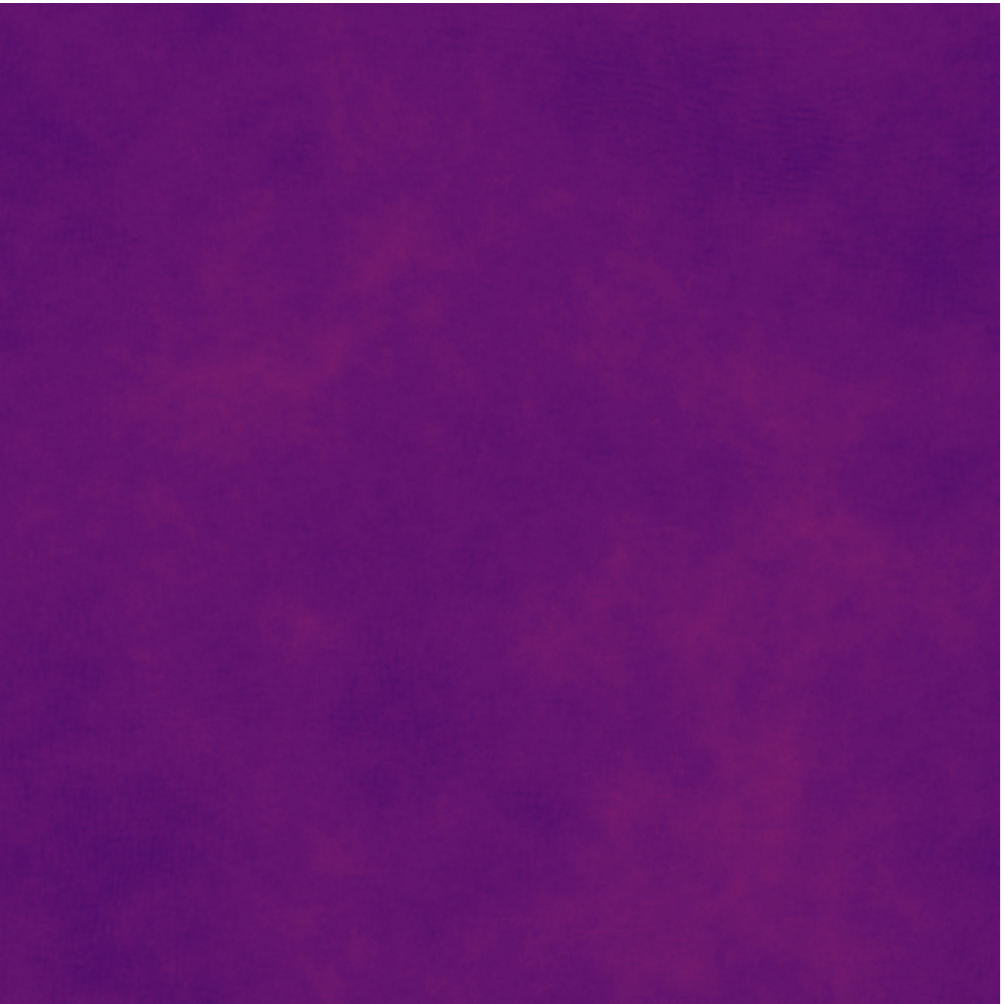}  &
\includegraphics[width=0.19\textwidth]{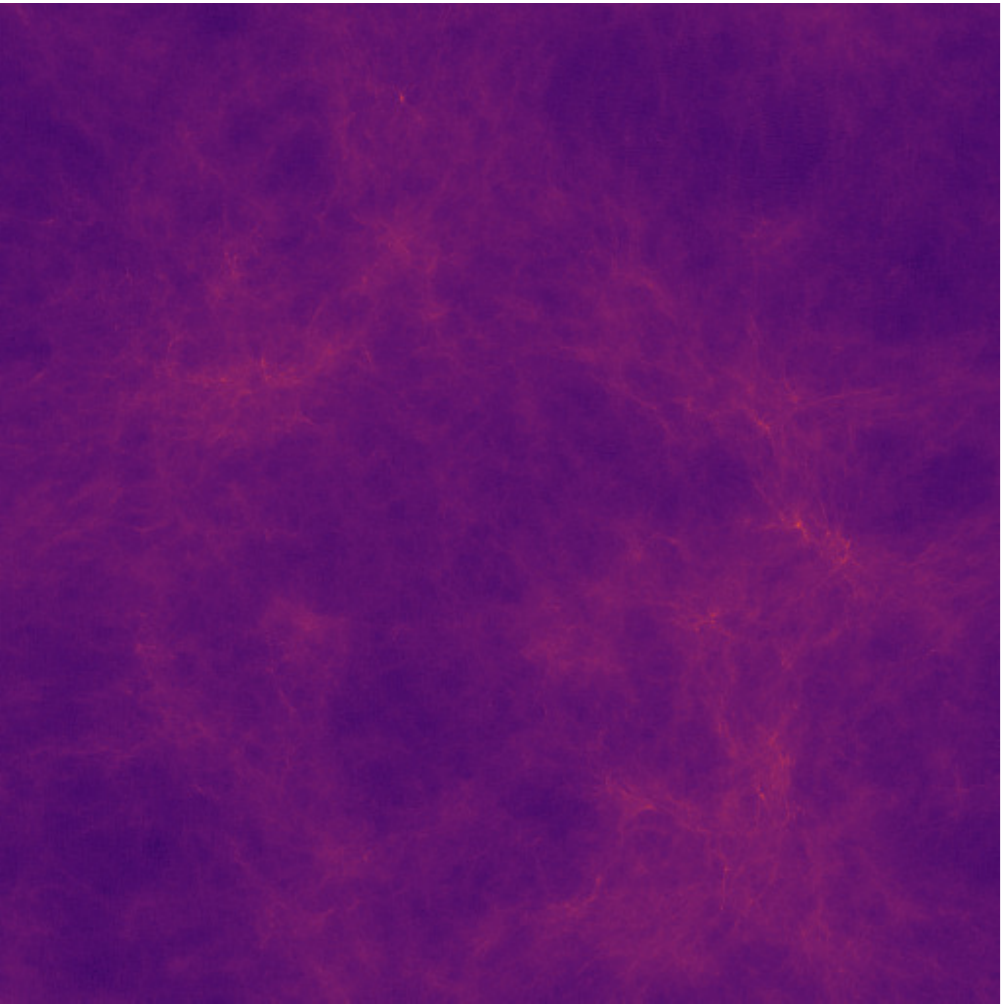}  &
\includegraphics[width=0.19\textwidth]{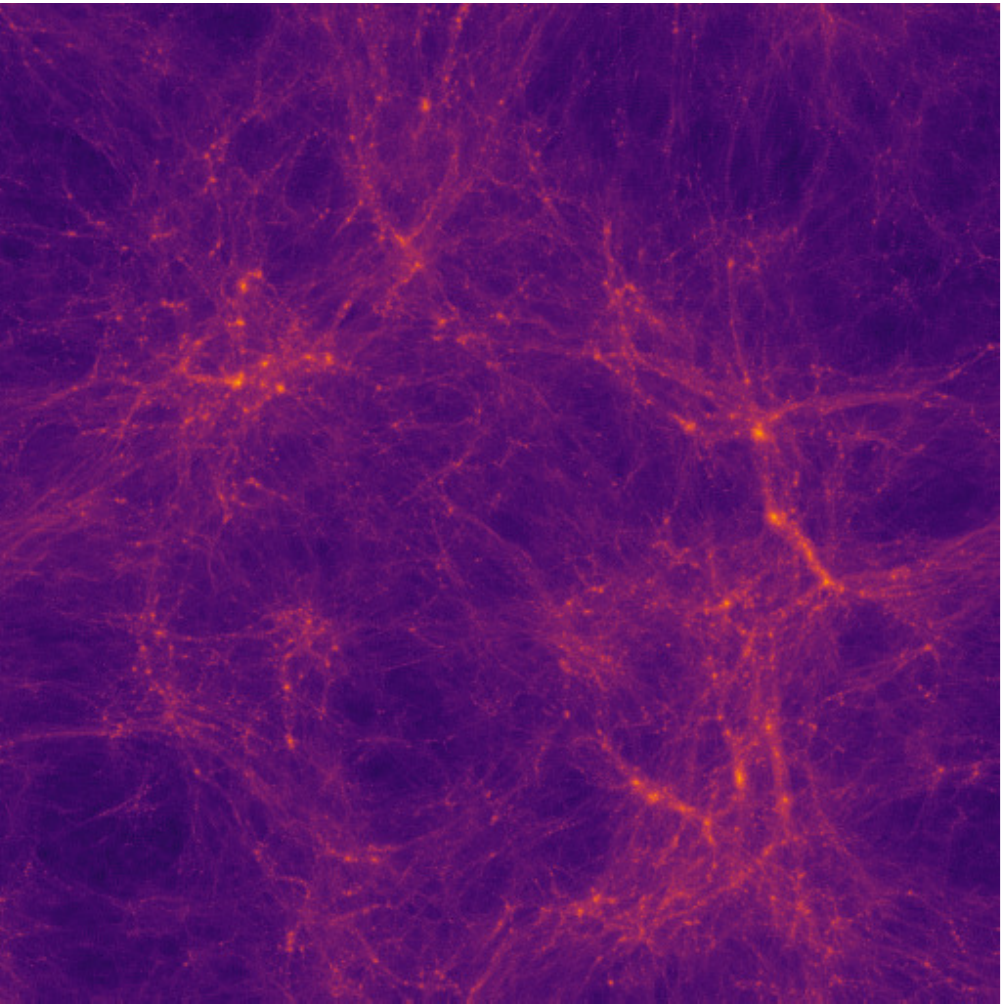}  &
\includegraphics[width=0.19\textwidth]{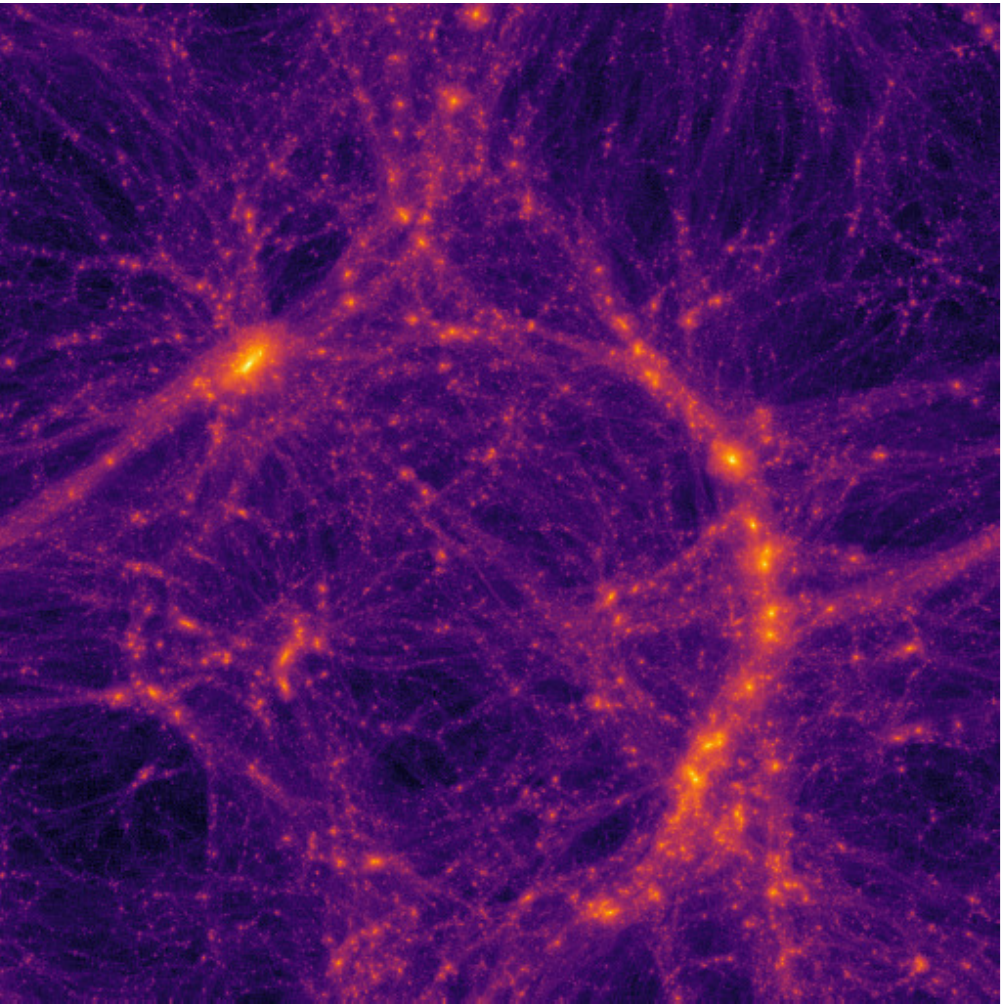}  &
\includegraphics[width=0.19\textwidth]{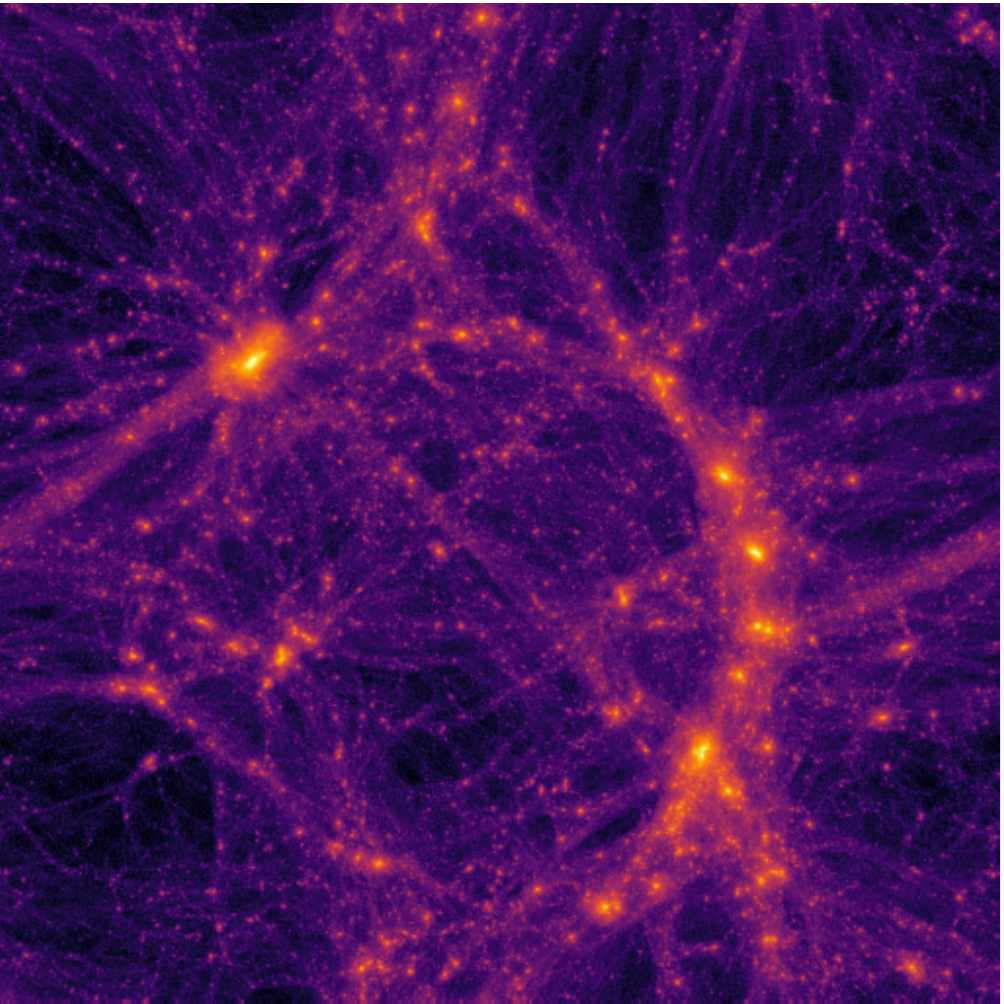}\\
\rotatebox{90}{\,``WDM'' - dm} & 
\includegraphics[width=0.19\textwidth]{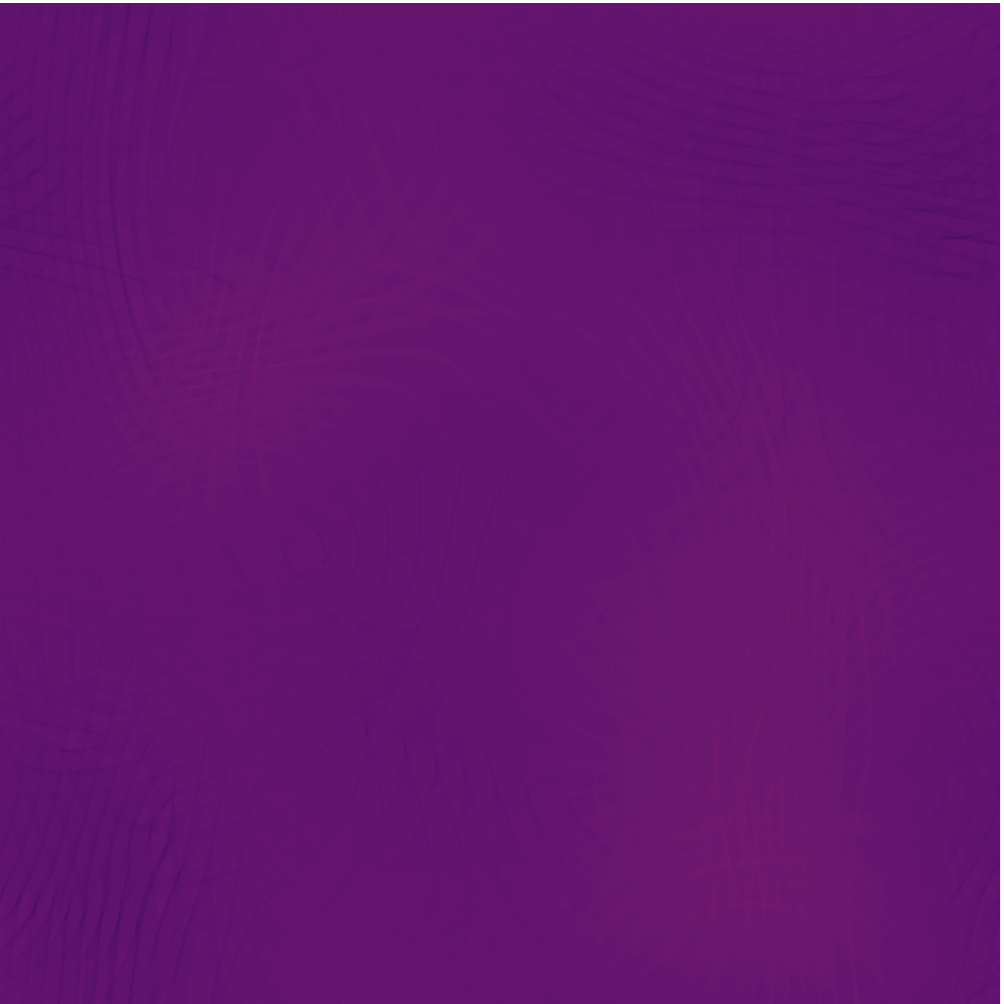}  &
\includegraphics[width=0.19\textwidth]{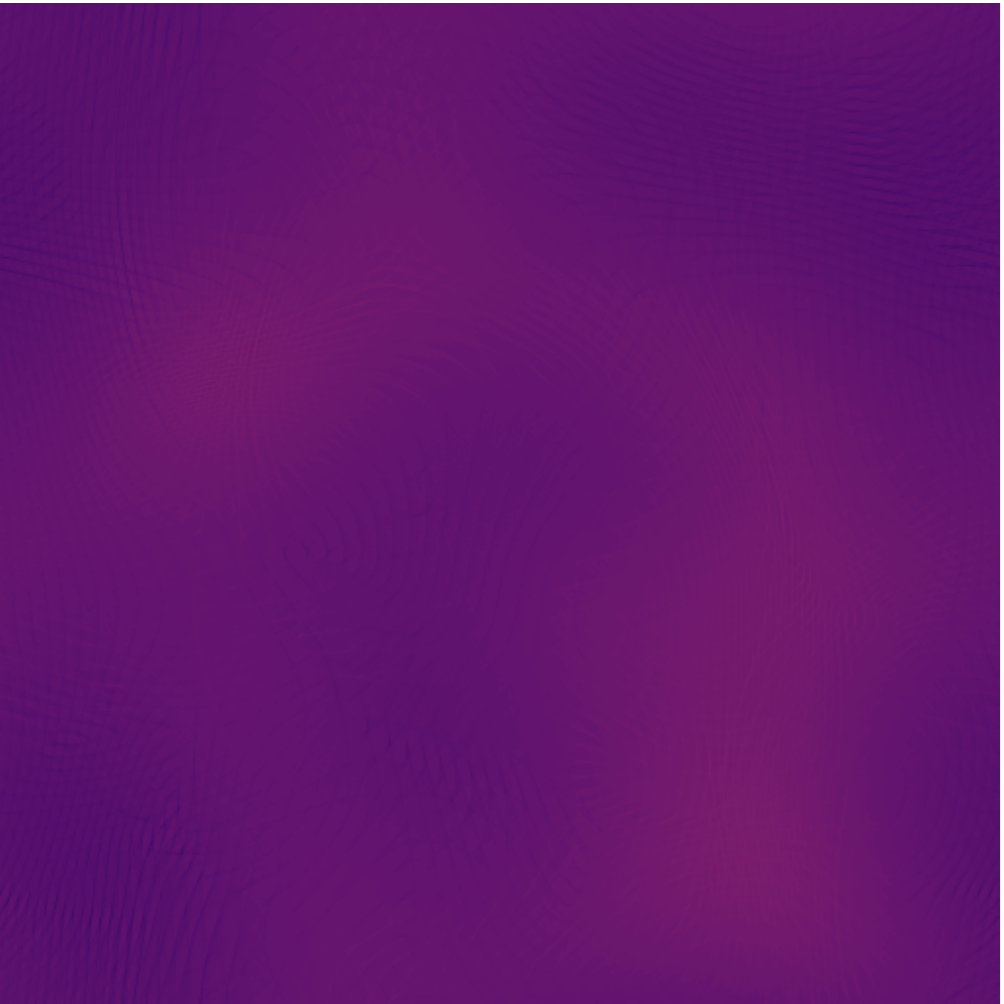}  &
\includegraphics[width=0.19\textwidth]{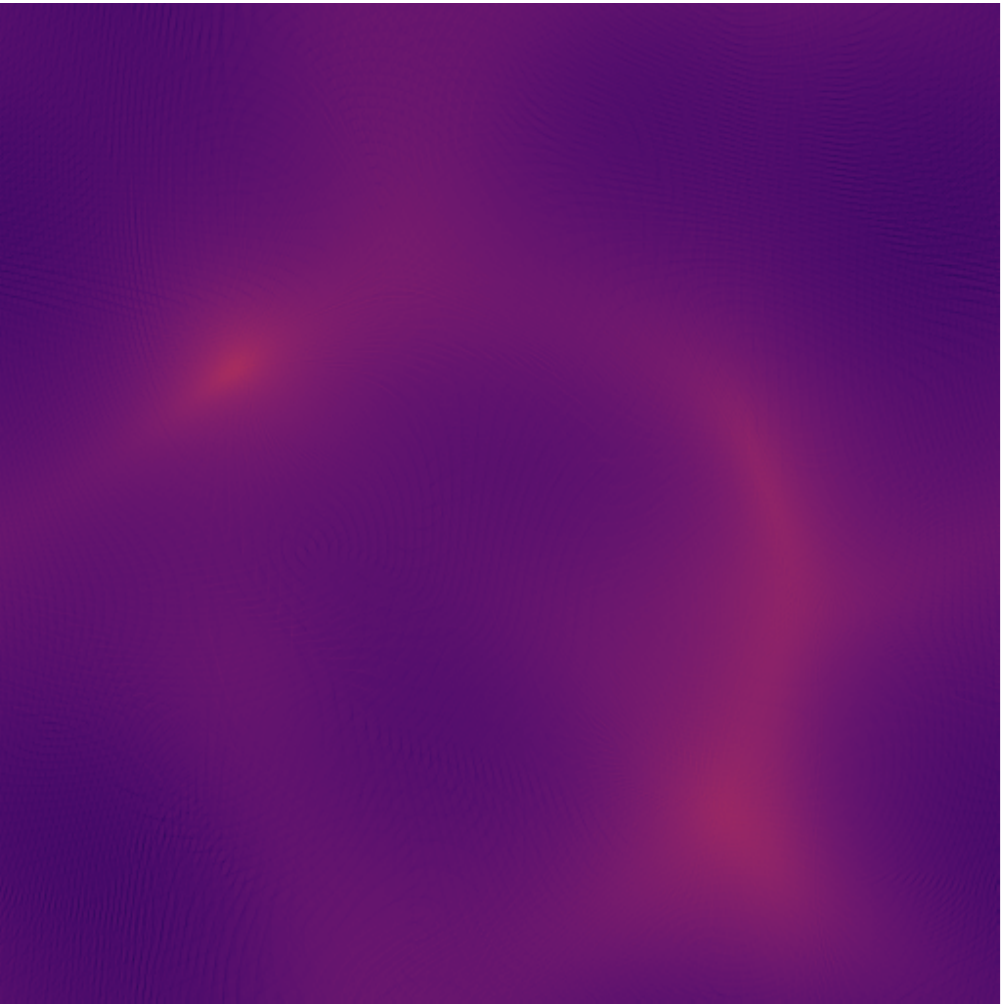}  &
\includegraphics[width=0.19\textwidth]{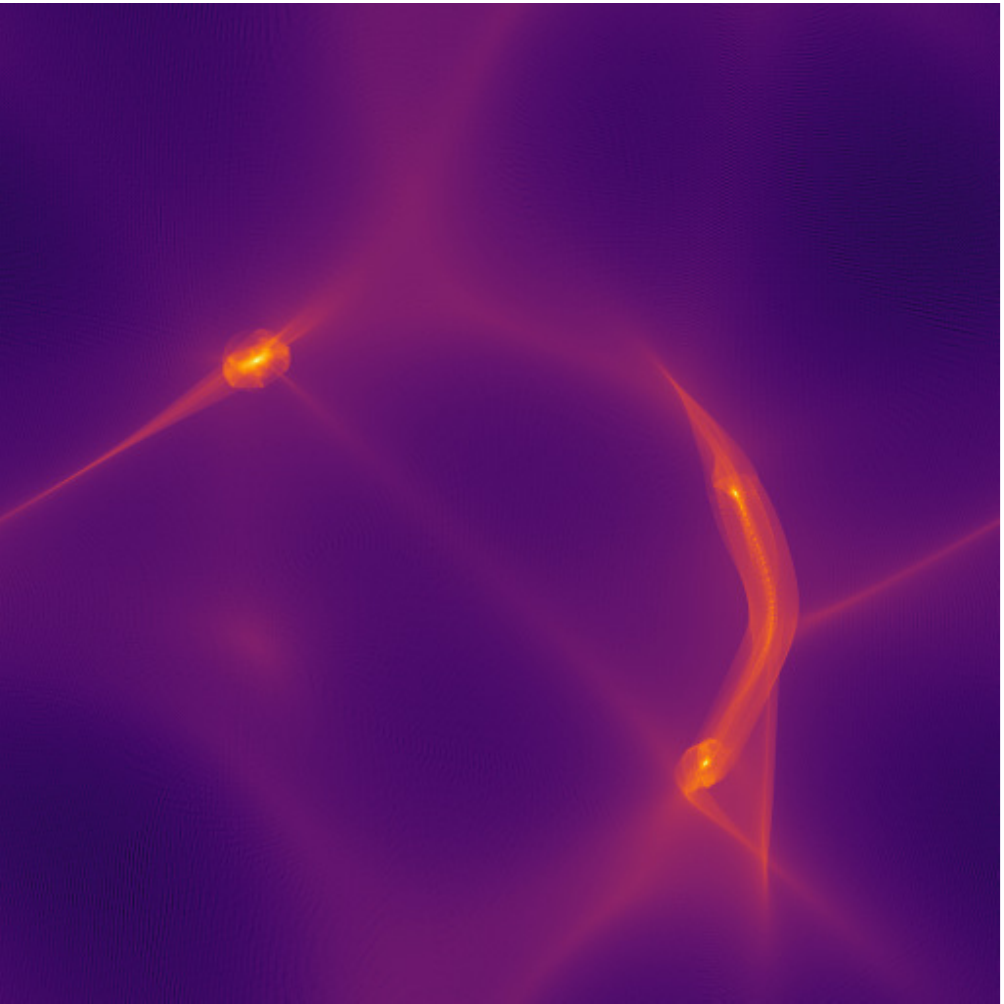}  &
\includegraphics[width=0.19\textwidth]{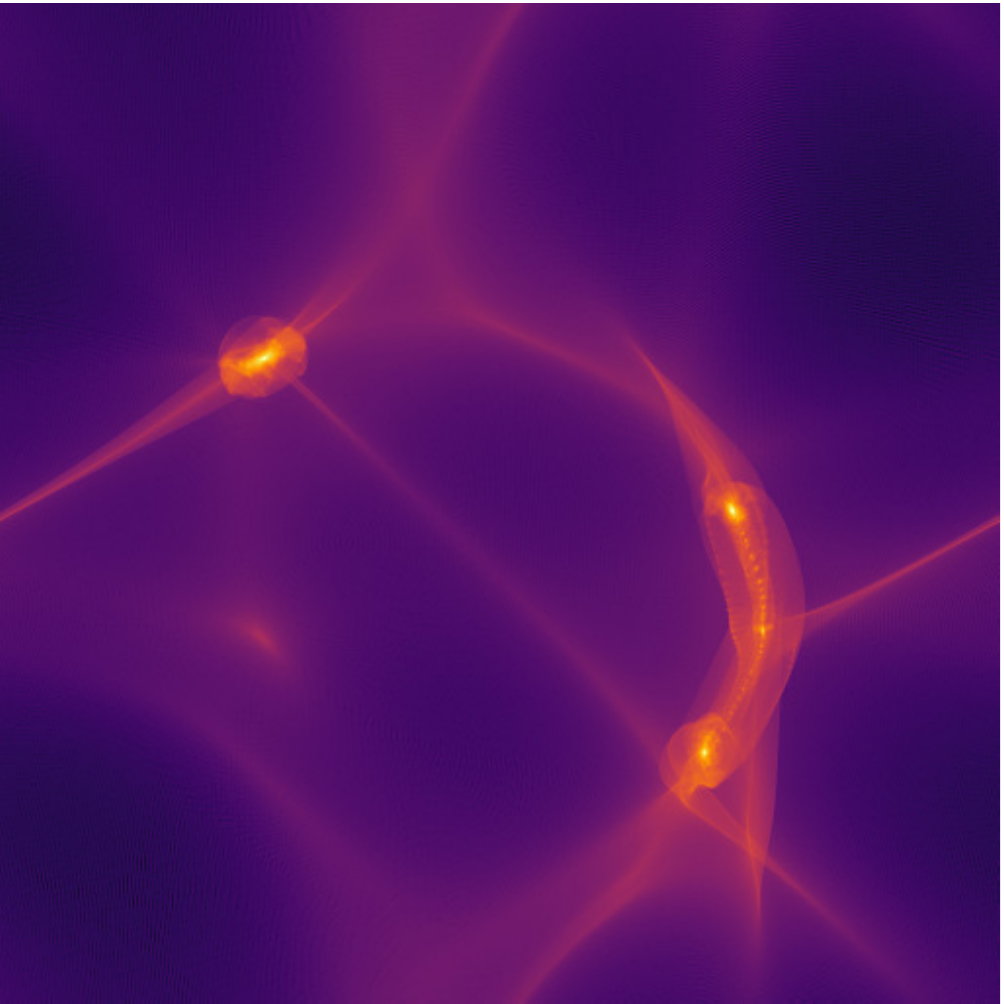}  \\
\rotatebox{90}{\,BECDM - dm} & 
\includegraphics[width=0.19\textwidth]{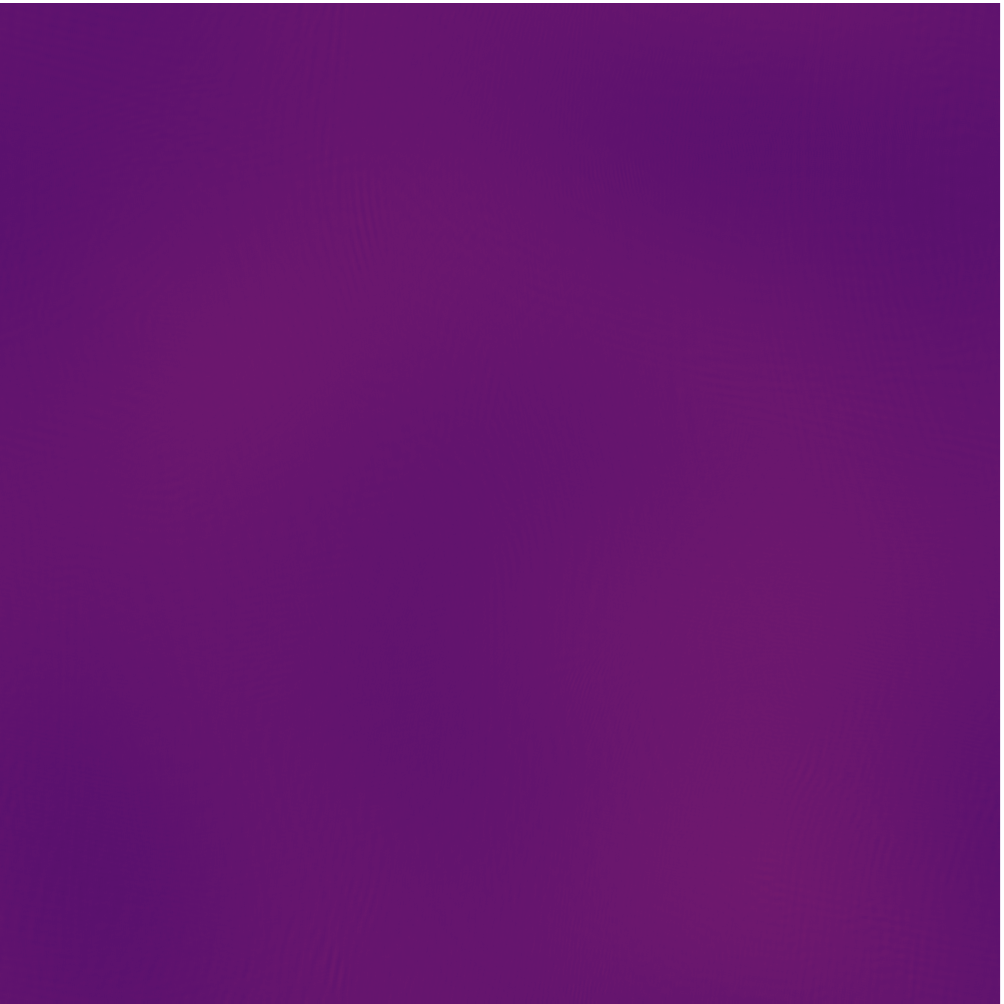}  &
\includegraphics[width=0.19\textwidth]{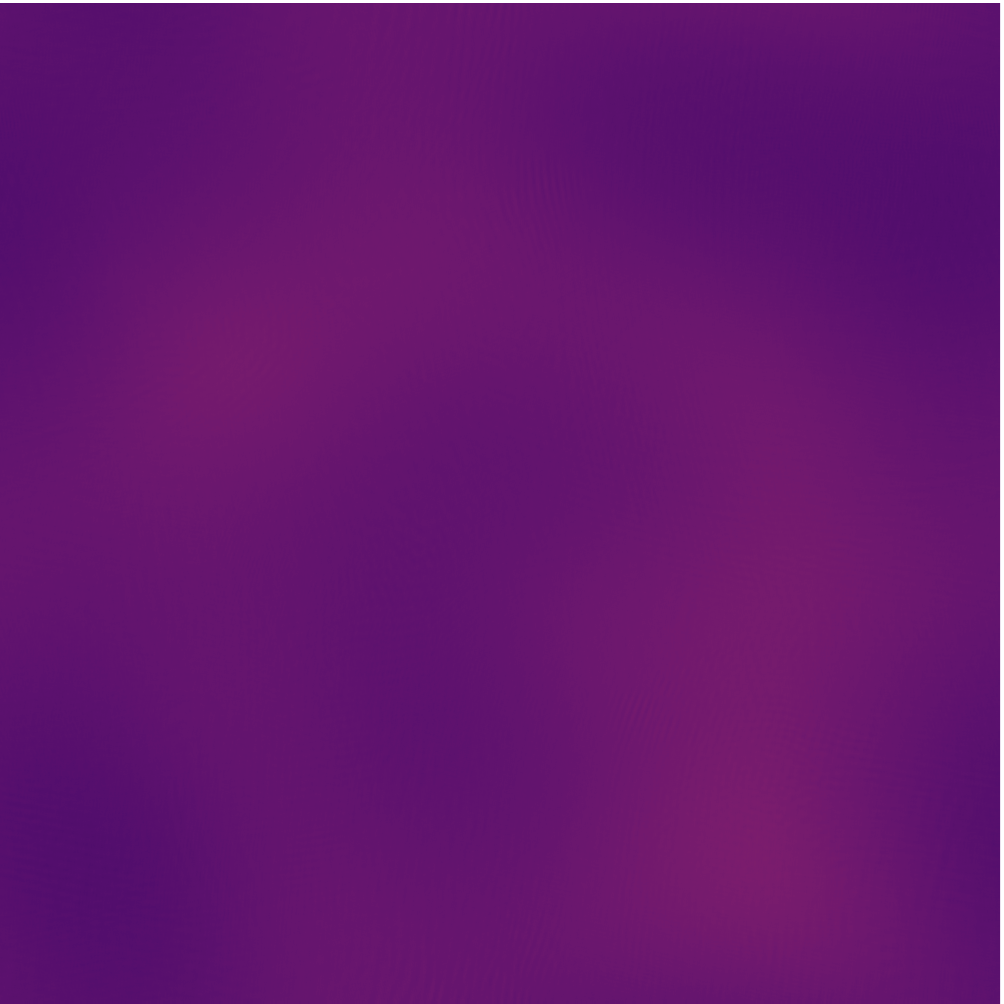}  &
\includegraphics[width=0.19\textwidth]{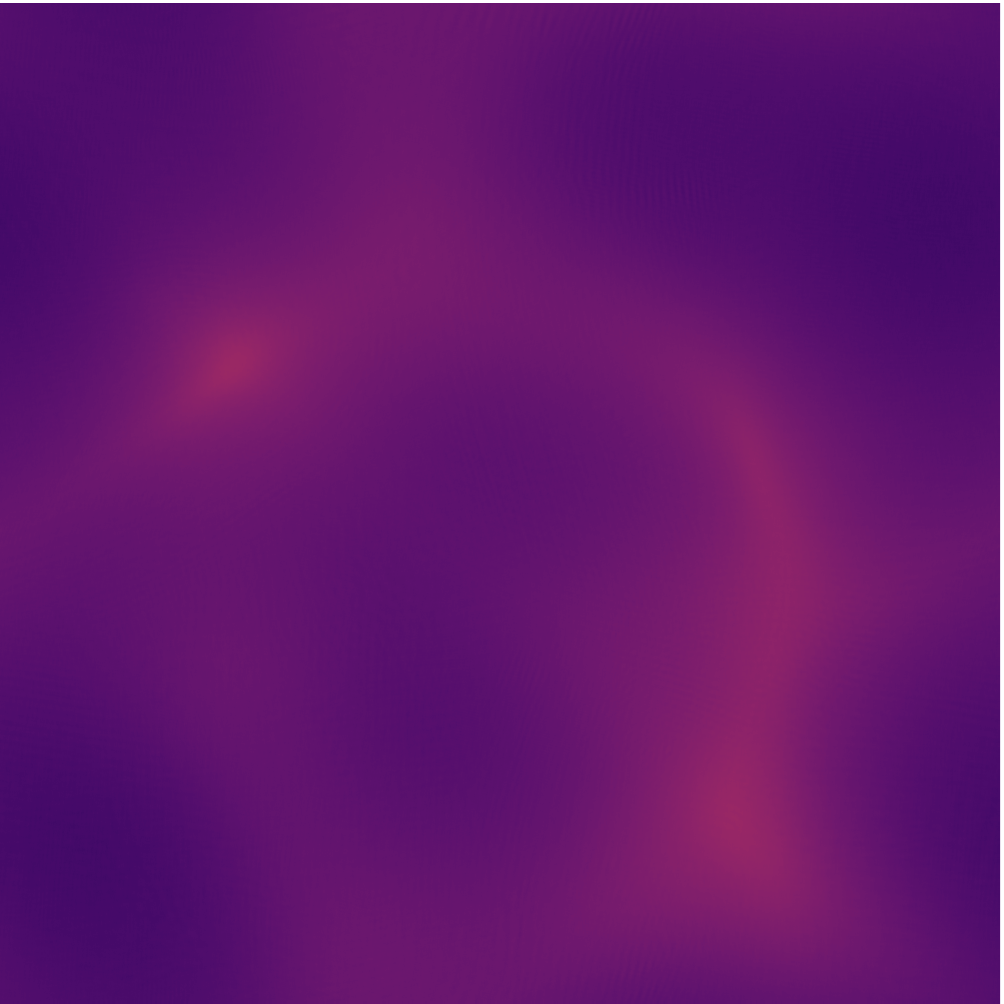}  &
\includegraphics[width=0.19\textwidth]{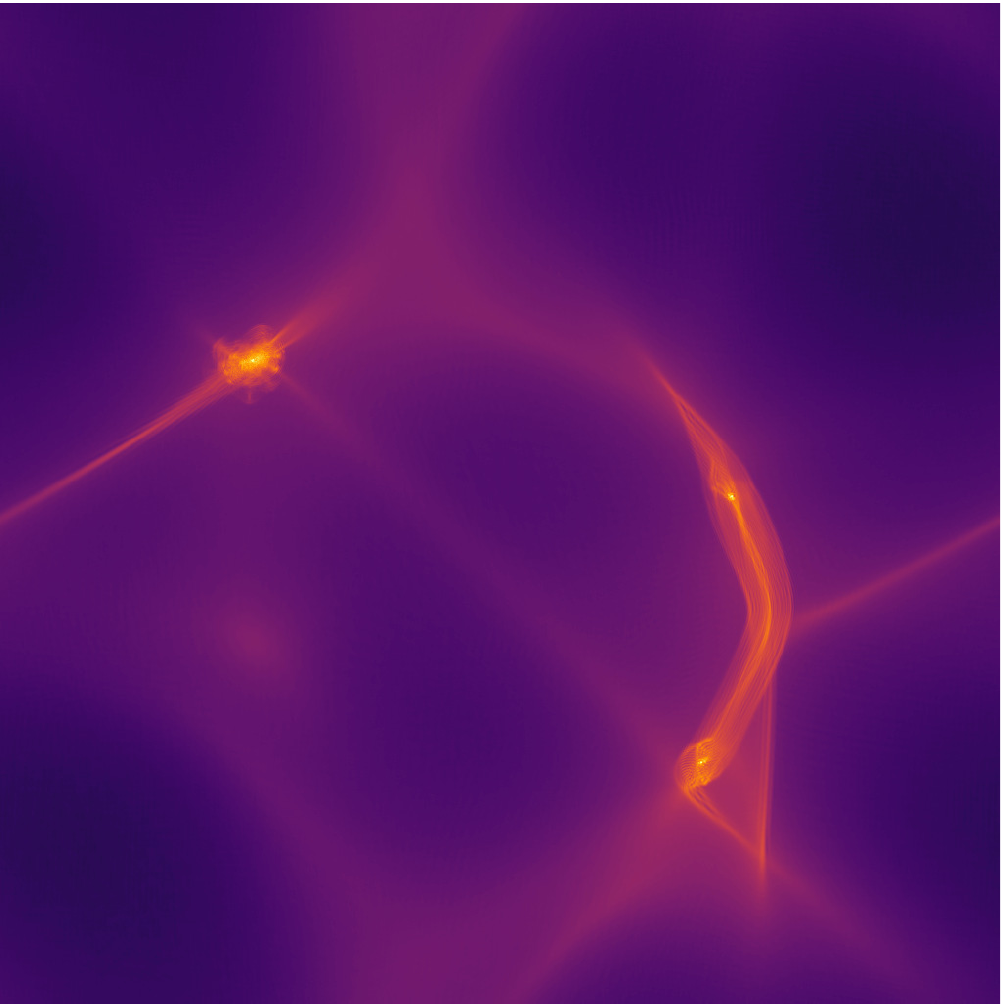}  &
\includegraphics[width=0.19\textwidth]{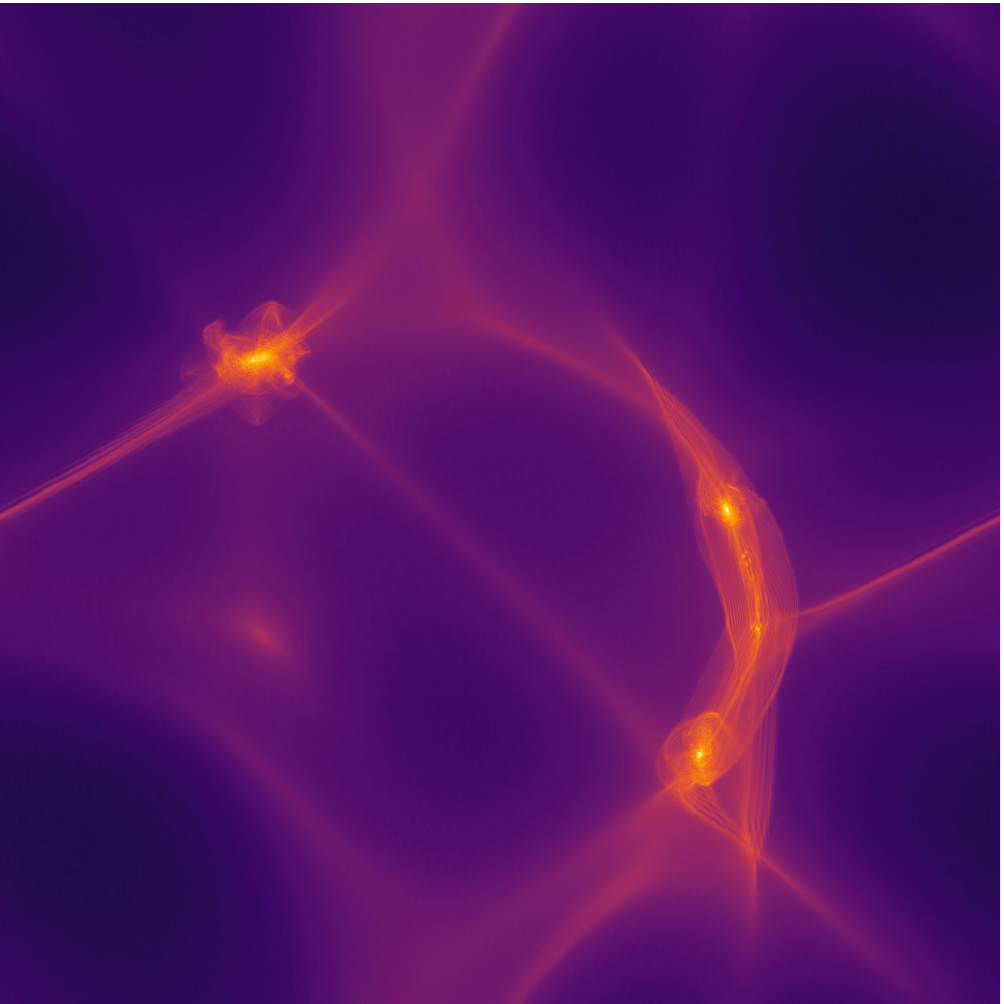}  \\
\rotatebox{90}{\,CDM - gas} & 
\includegraphics[width=0.19\textwidth]{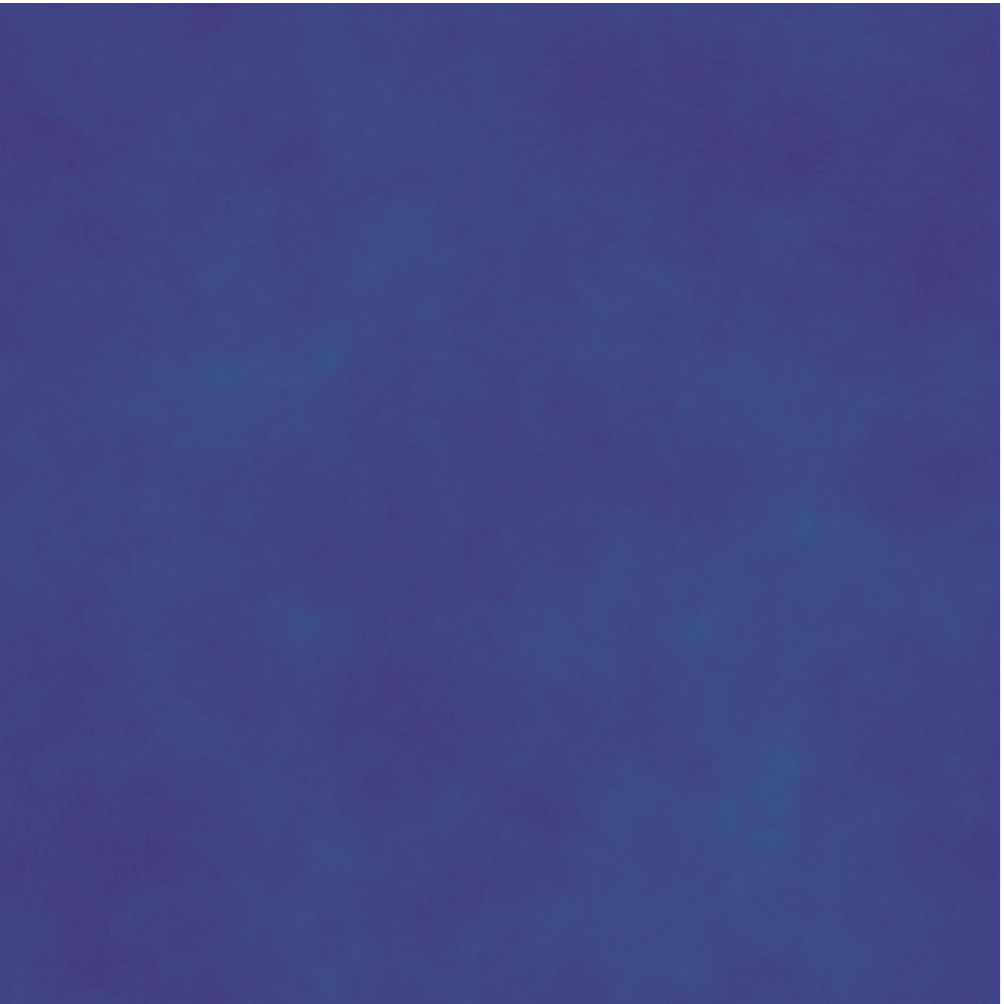}  &
\includegraphics[width=0.19\textwidth]{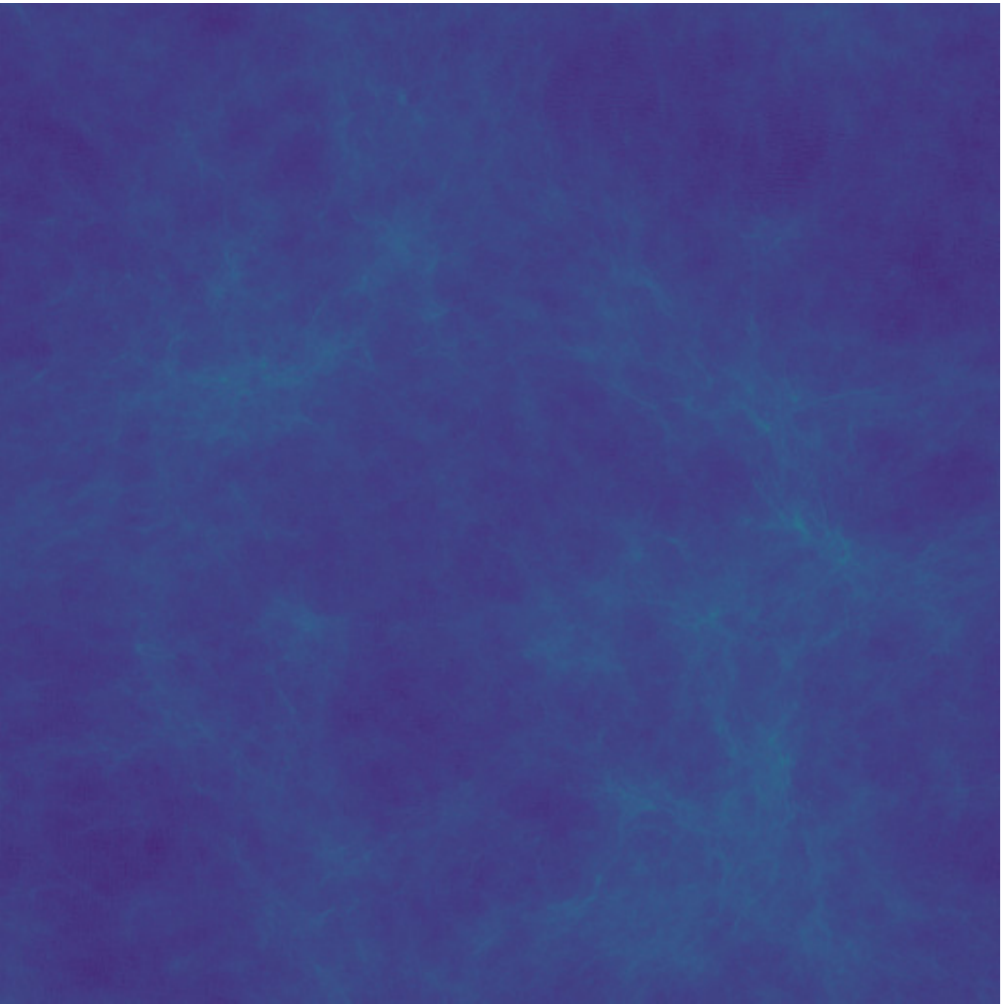}  &
\includegraphics[width=0.19\textwidth]{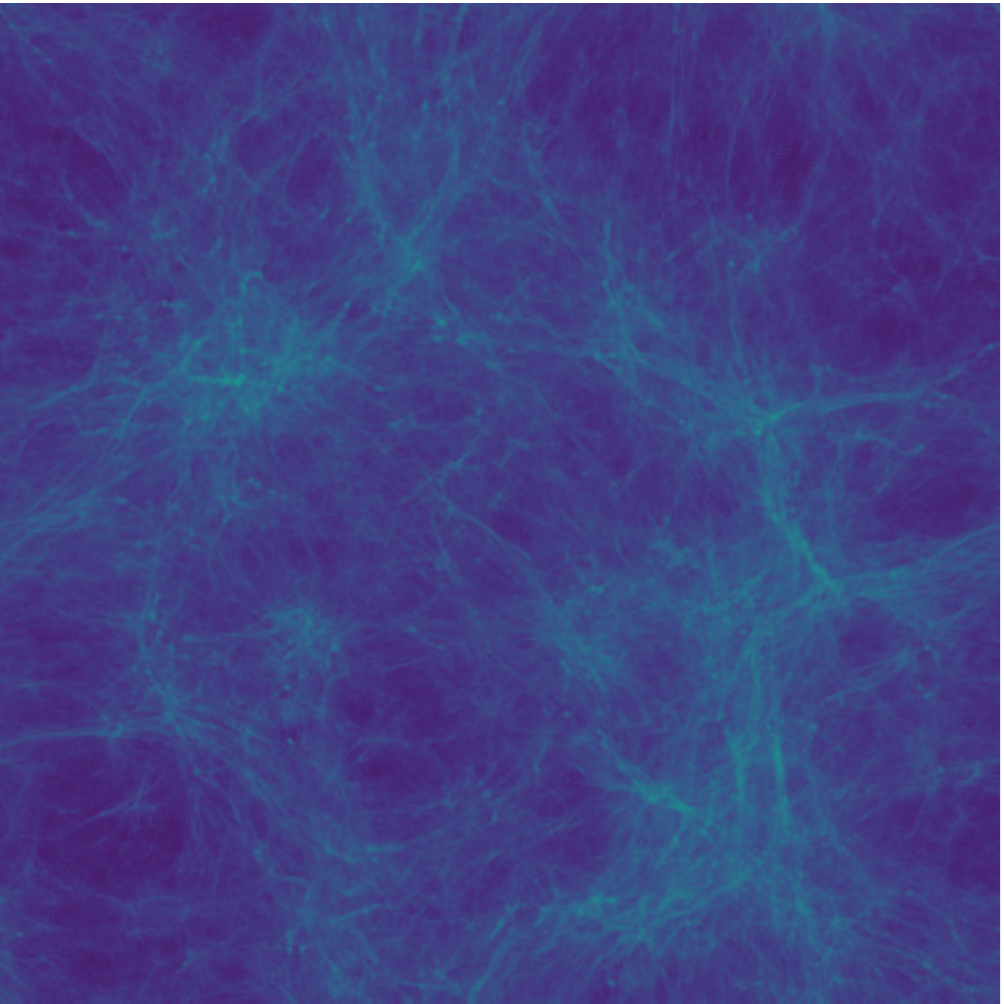}  &
\includegraphics[width=0.19\textwidth]{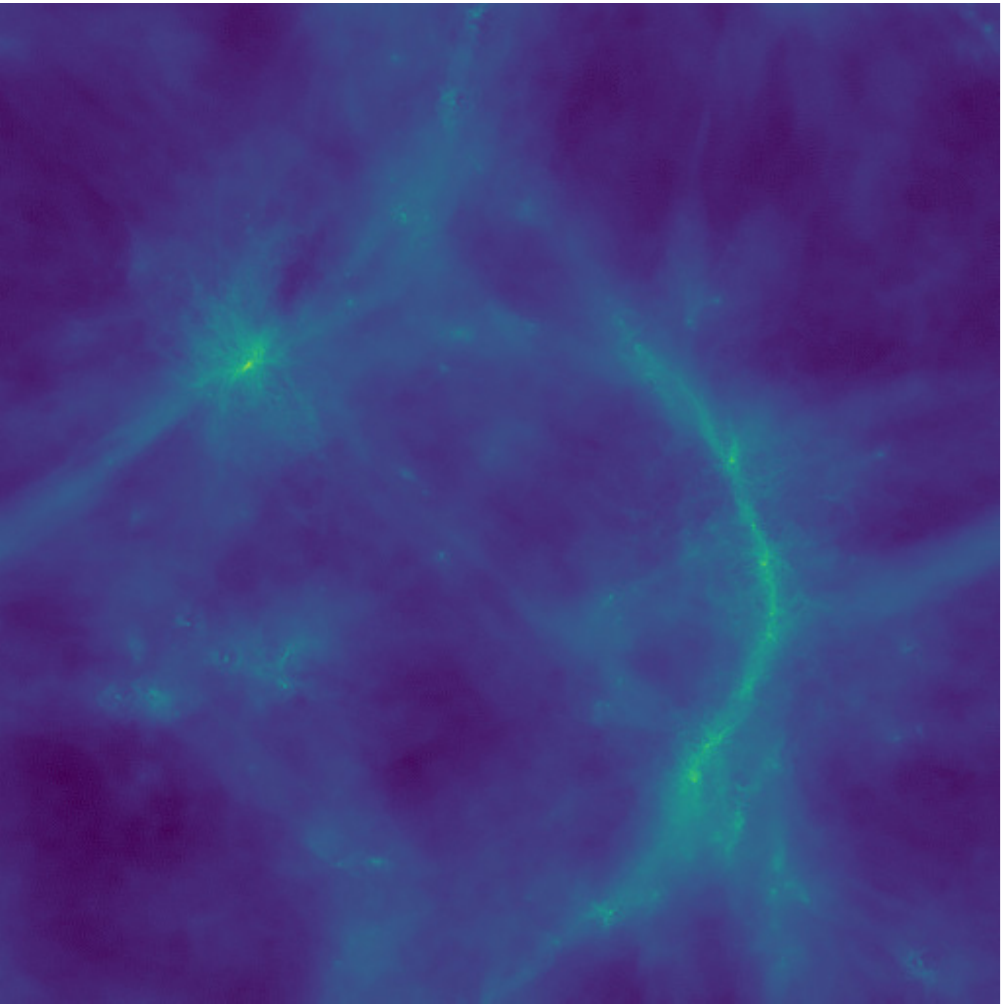}  &
\includegraphics[width=0.19\textwidth]{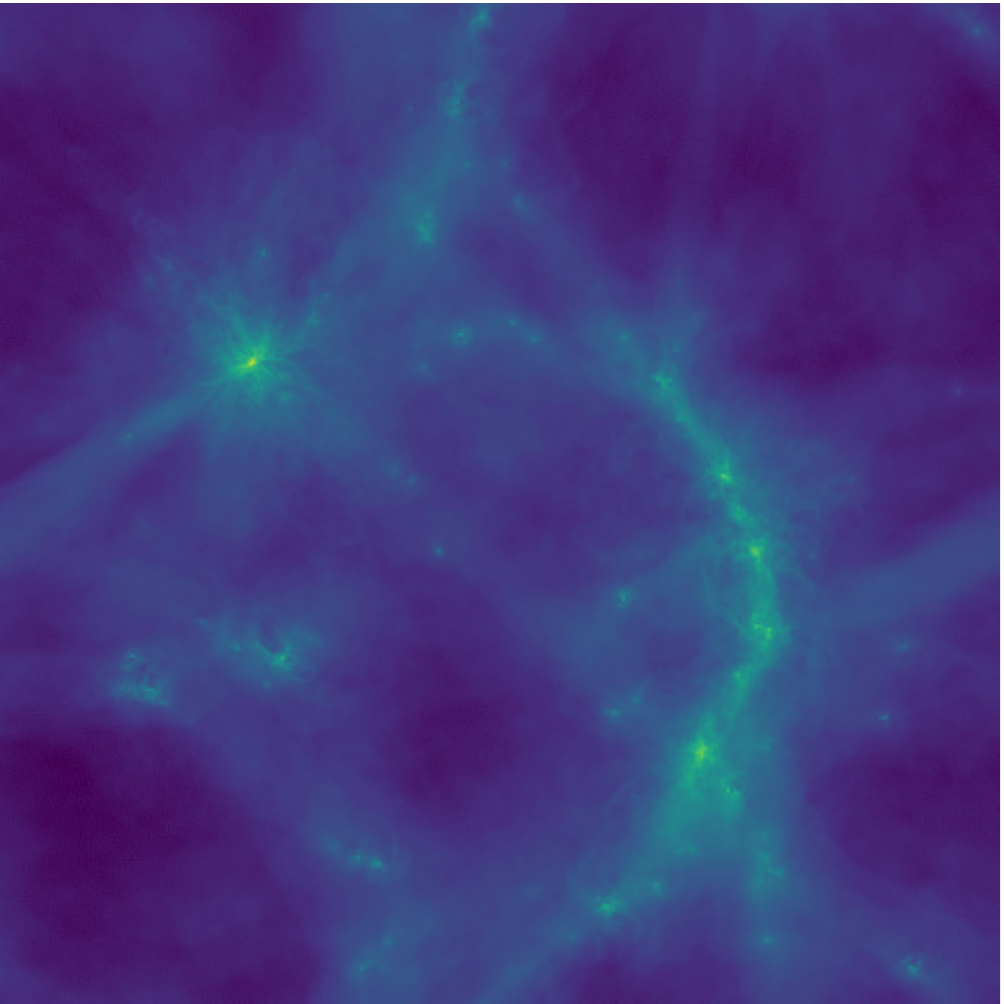}  \\
\rotatebox{90}{\,``WDM'' - gas} &                        
\includegraphics[width=0.19\textwidth]{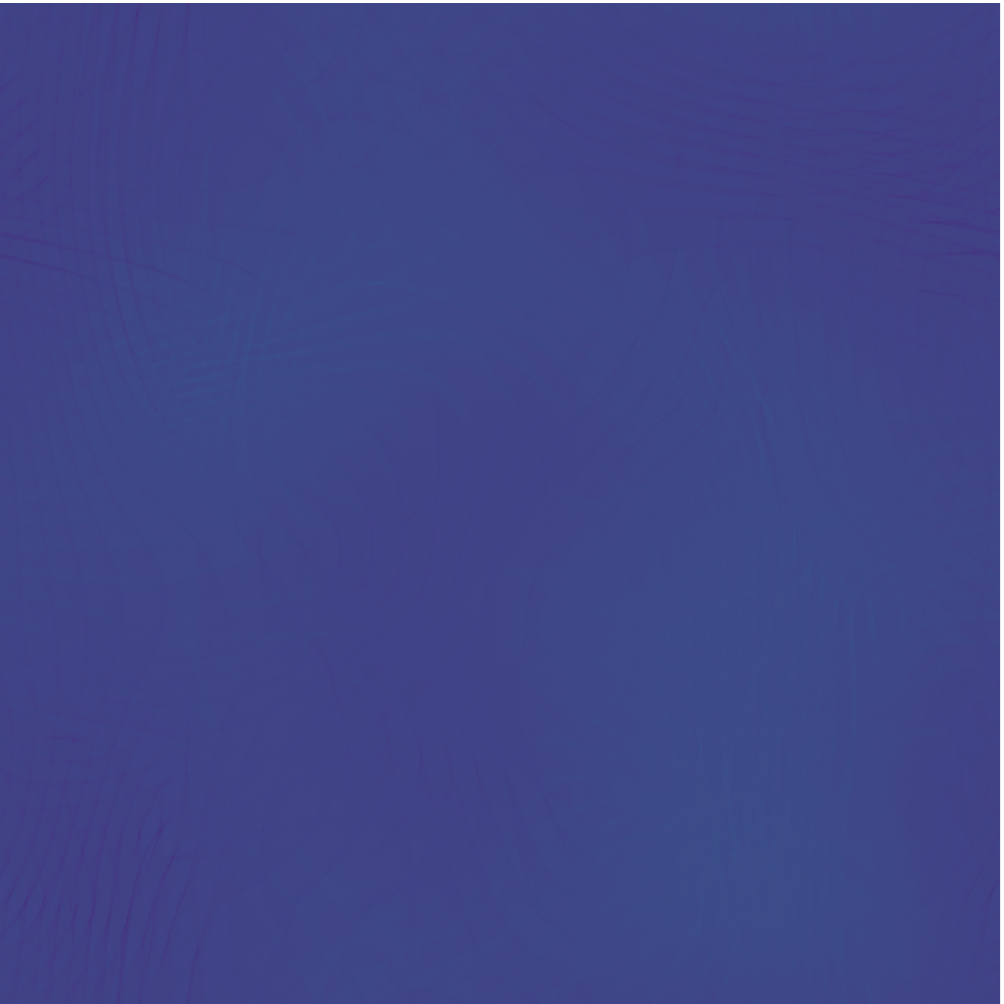}  &
\includegraphics[width=0.19\textwidth]{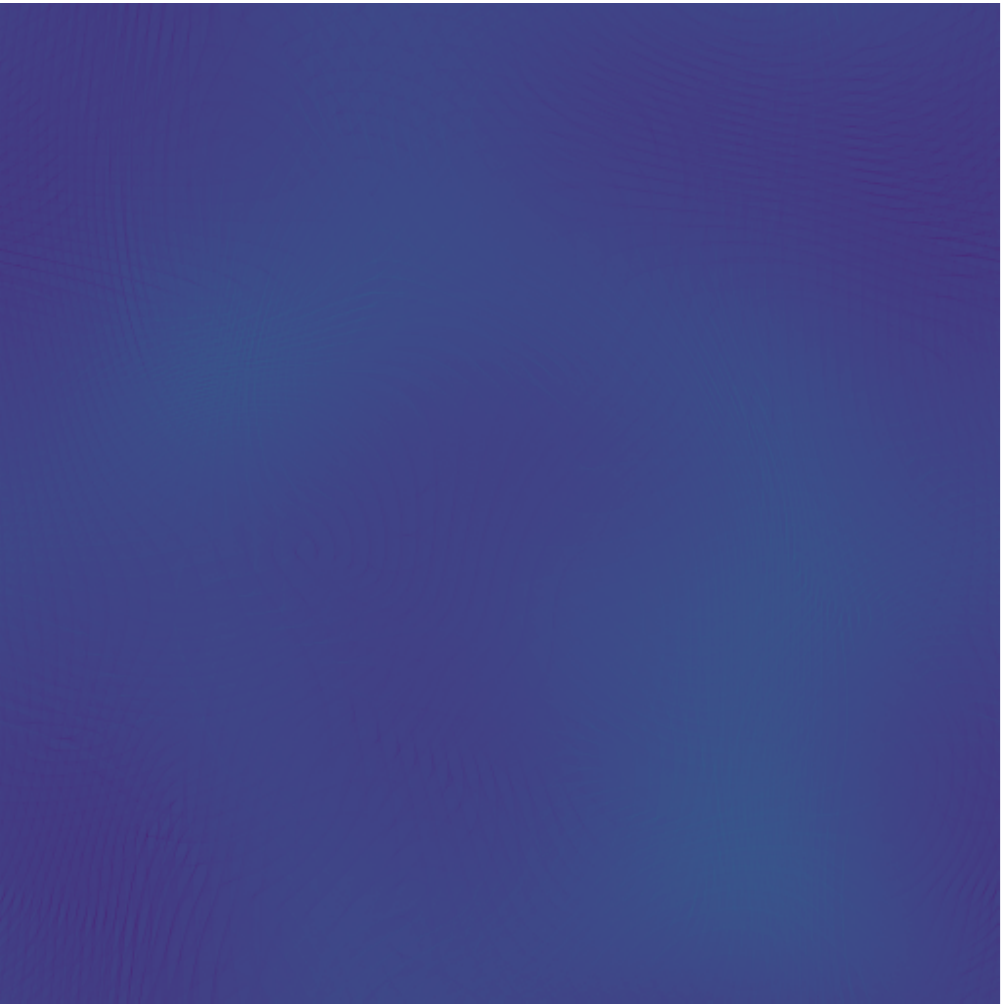}  &
\includegraphics[width=0.19\textwidth]{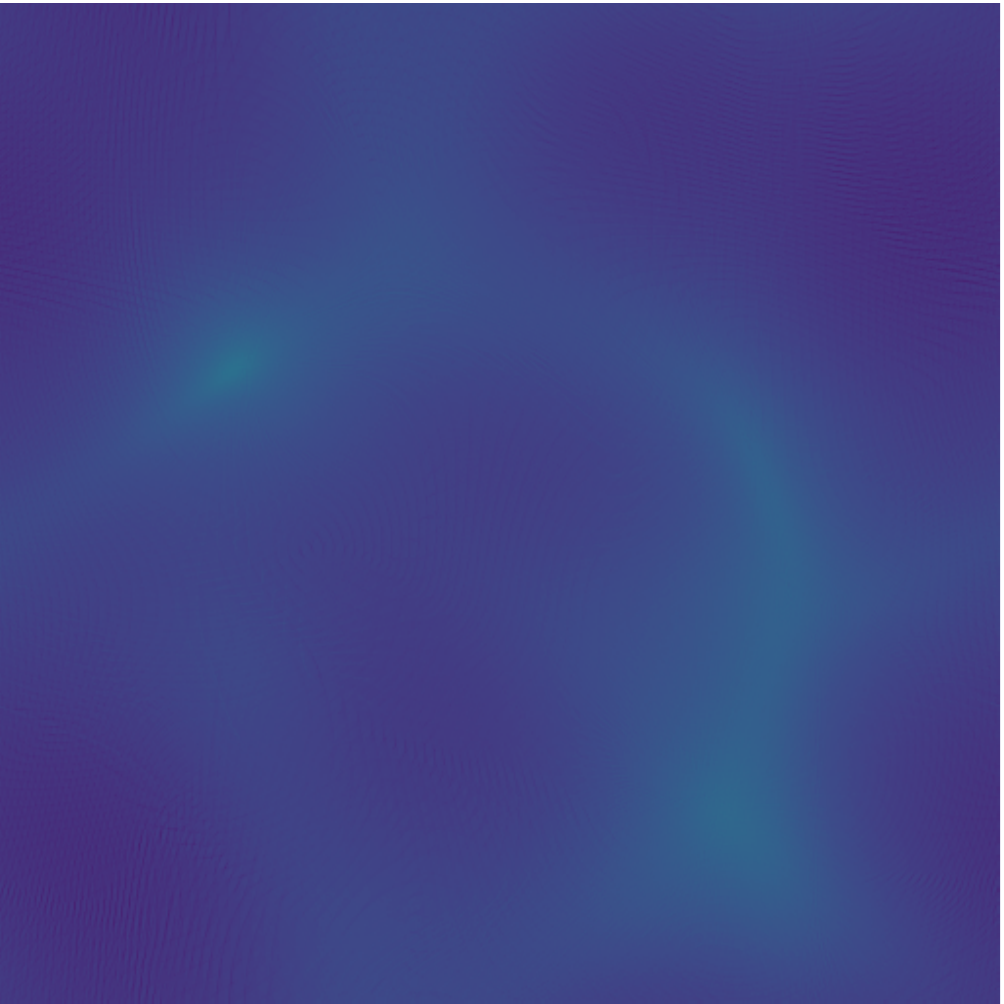}  &
\includegraphics[width=0.19\textwidth]{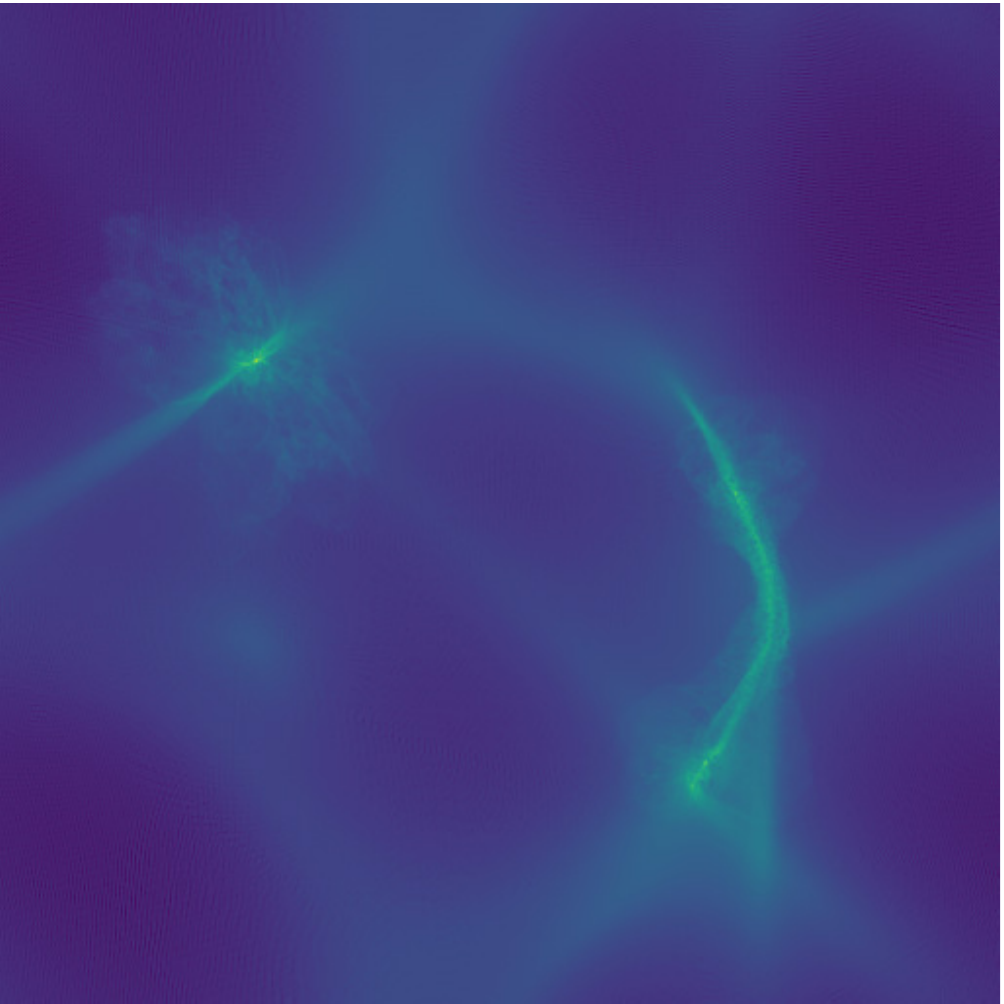}  &
\includegraphics[width=0.19\textwidth]{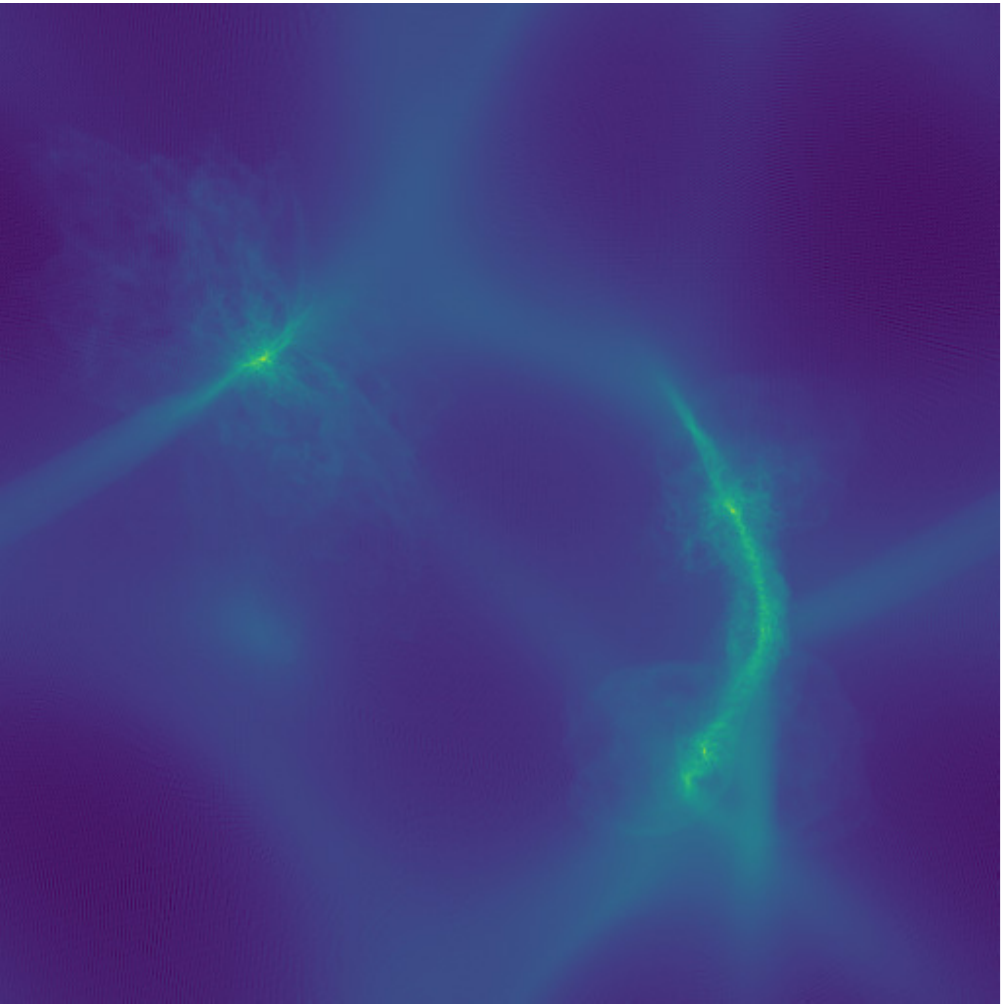}  \\
\rotatebox{90}{\,BECDM - gas} &                          
\includegraphics[width=0.19\textwidth]{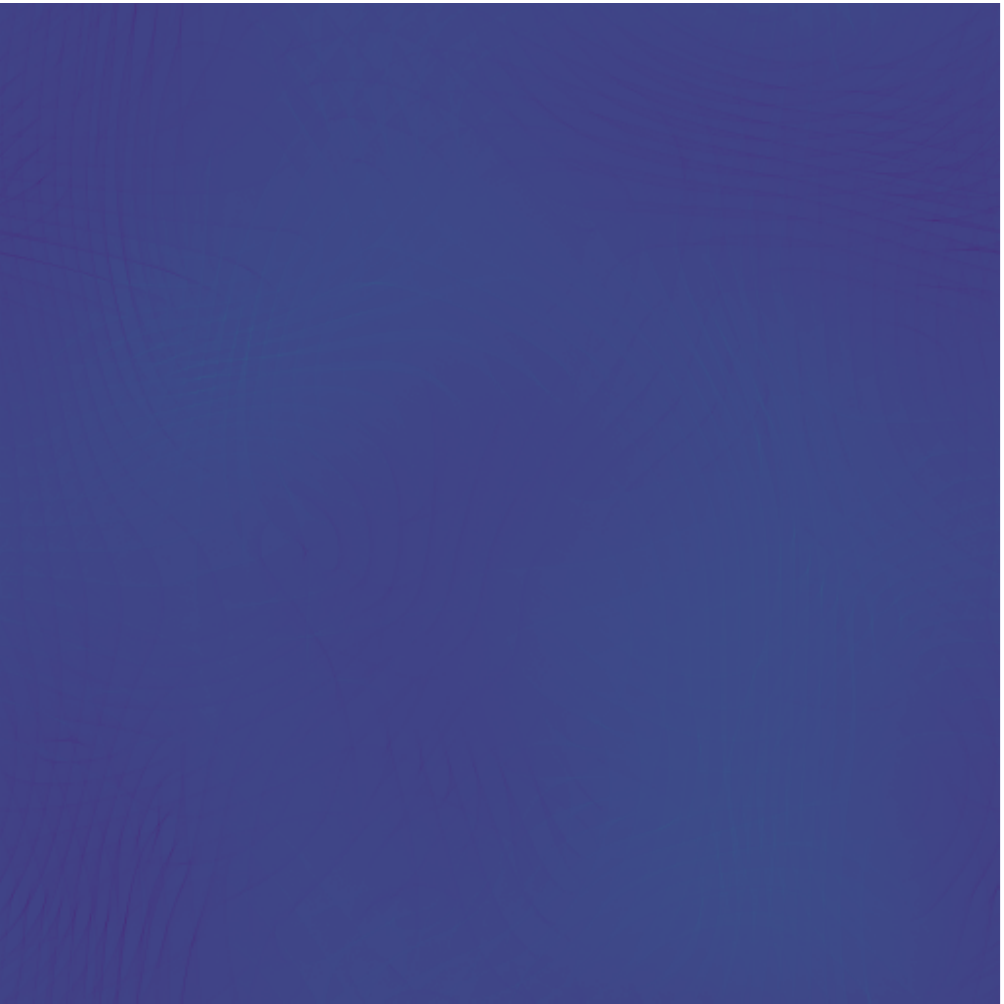}  &
\includegraphics[width=0.19\textwidth]{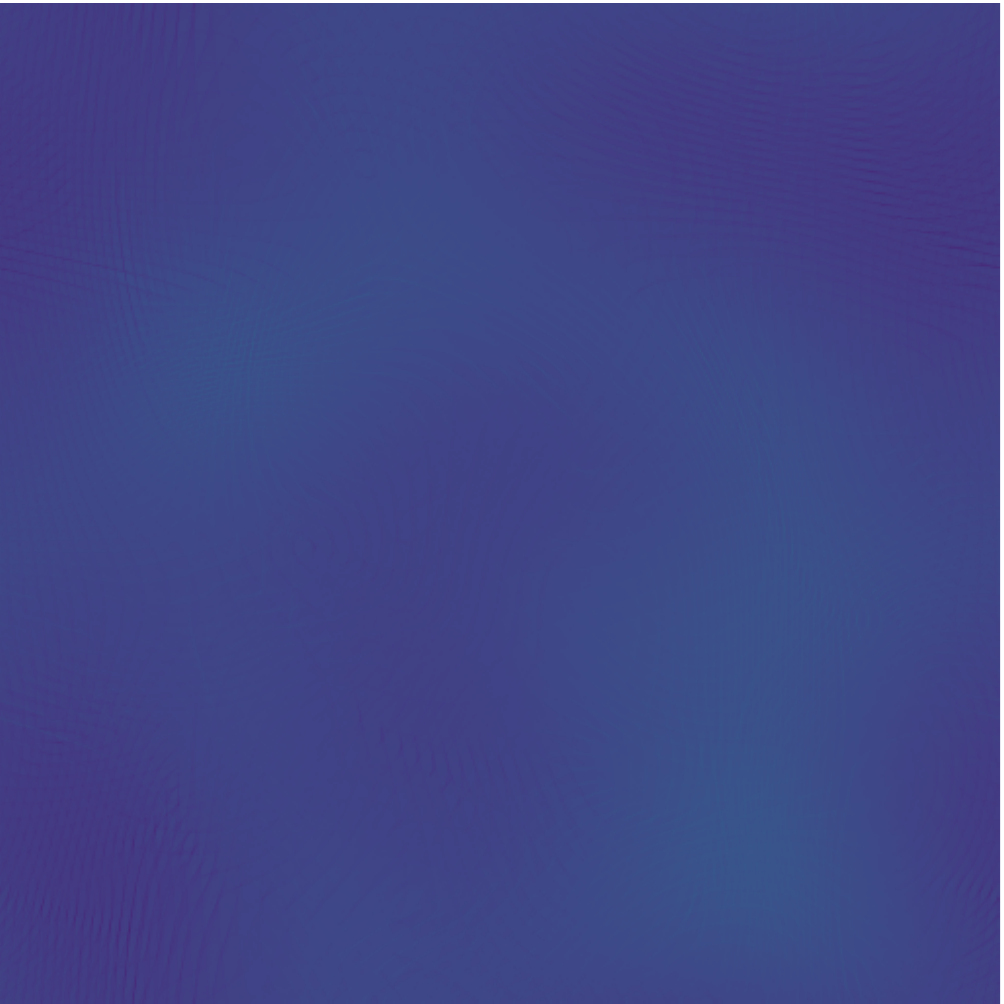}  &
\includegraphics[width=0.19\textwidth]{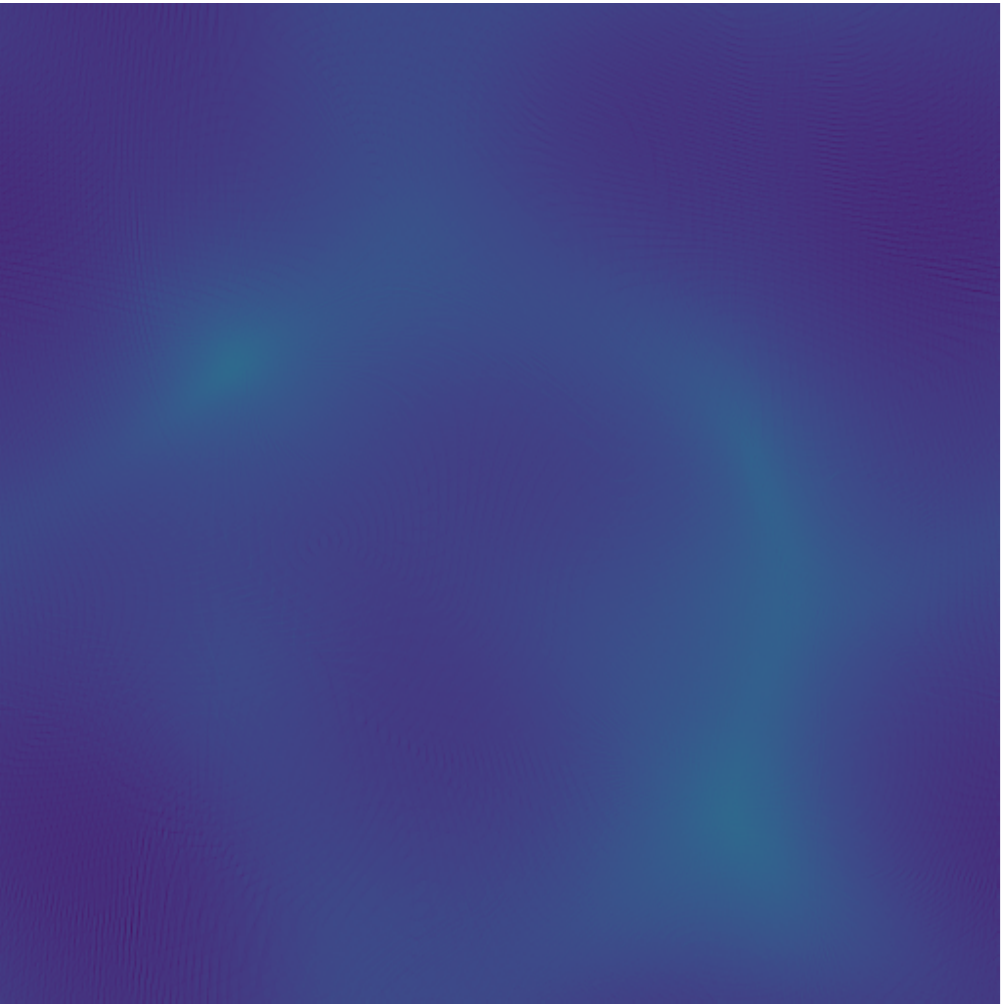}  &
\includegraphics[width=0.19\textwidth]{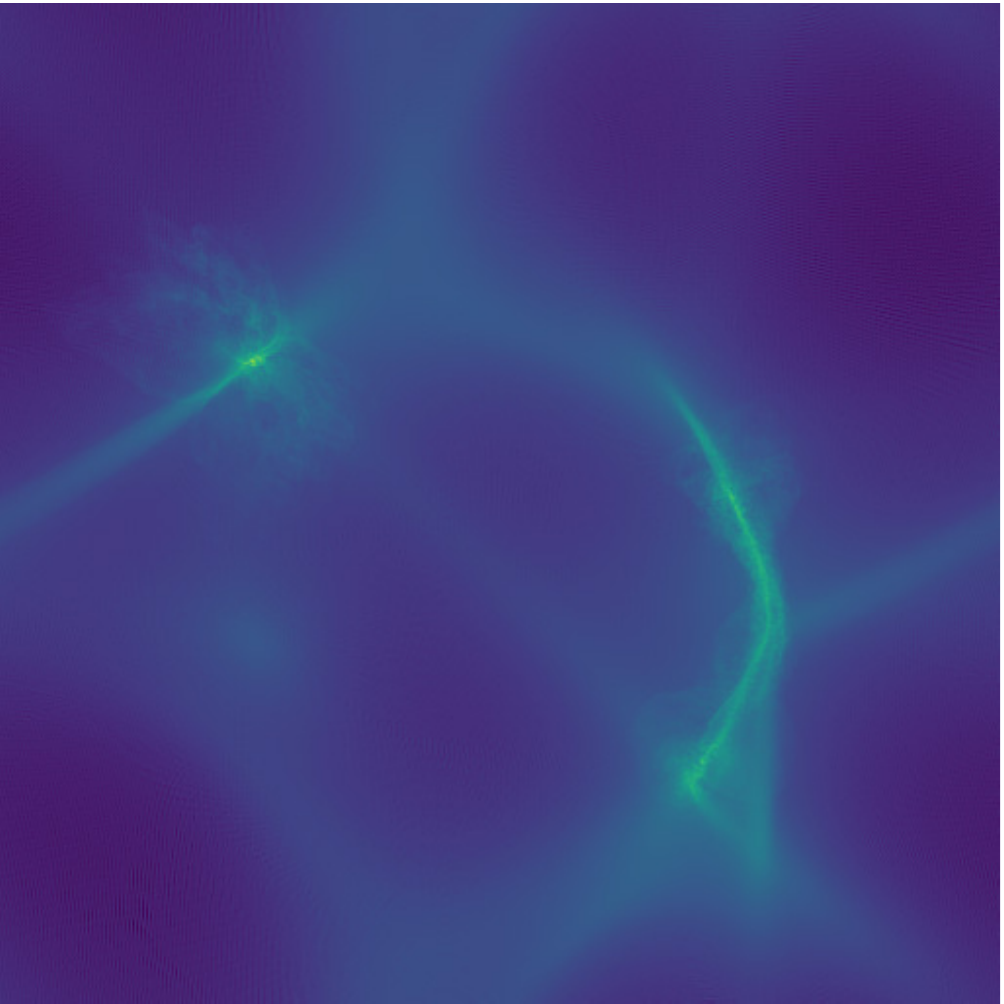}  &
\includegraphics[width=0.19\textwidth]{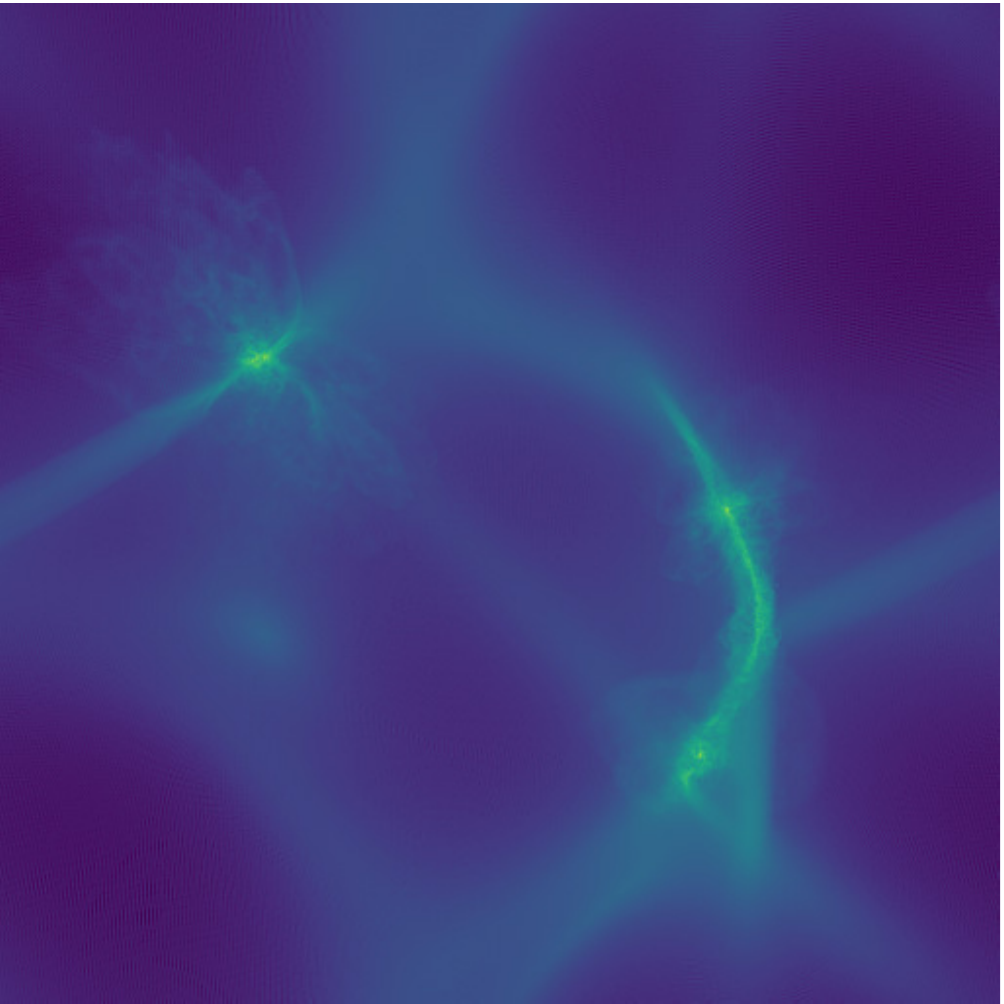}  \\
& 
&
\multicolumn{2}{c}{\includegraphics[width=0.25\textwidth]{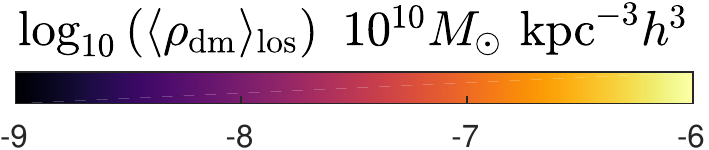}} &  
\multicolumn{2}{c}{\includegraphics[width=0.25\textwidth]{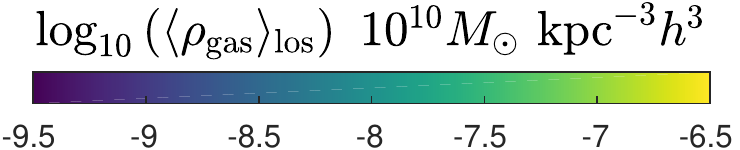}}
\end{tabular}
\caption{Structure formation of dark matter (orange/purple) and gas (green/blue) in our simulations under the 3 cosmologies studied. We plot projected (comoving) densities along the line of sight (see colorbars at the bottom of the figures for the values of the projected density field).
The snapshots are shown at intervals of the scale factor increasing by a factor of $2$, as well as the final snapshot of the simulation. The box size is $1.7h^{-1}~{\rm Mpc}$.
In these large-scale projections, gas follows the dark matter potential wells, and BECDM and ``WDM'' appear similarly filamentary, while CDM has filaments fragmented into subhaloes.} 
\label{fig:snaps}
\end{center}
\end{figure*}

We compare 3 small-box cosmological simulations (CDM, ``WDM'', BECDM), which have led to the formation of 3 $\sim 10^9$--$10^{10}~M_\odot$ haloes by $z\sim6$. 
In Fig.~\ref{fig:snaps} we show the projected densities of the dark matter and baryonic gas at $z=63,31,15,7,5.5$, that is, at the scale factor increasing by a factor of $2$ at each step. The BECDM and ``WDM'' projected densities resemble each other closely on the scales of the box, indicating that on large scales the primary effect of the quantum potential is the initial suppression of the power spectrum, while dispersion and interference have not up-scattered significantly to affect large scale modes. The truncated initial power spectrum leads to the formation of filaments as well as haloes. The CDM simulations show the same large-scale structures as well, but the filaments are now comprised of numerous subhaloes owing to the lack of a length scale in CDM (the fragmentation into subhaloes is a function of the mass resolution of the dark matter).

The baryonic gas initially traces the dark matter density very closely. The gas follows the gravitational potential on large scales, and shocks and cooling are unimportant for the dynamics. But the gas has a sound-speed/pressure, so the cosmic Jeans criterion prevents collapse of gas in the smallest substructures (the Jeans mass is about $10^5$ M$_\odot$ at $z=100$ and  $2\times 10^4$ M$_\odot$ at $z=20$). Additionally, star formation triggers reionization which, in these simulations, is driven by a uniform ionizing background that is turned on by hand at $z\sim 6$; the ionization erases small scale features. Furthermore, after stars form the gas distribution is also modified by feedback -- stellar winds from supernovae blow hot, low-density $100$~kpc scale bubbles.

Even though on large scales three simulations resemble each other, on small scales all these scenarios yield  very different structures highlighted  by \citep{moczPRL} and investigated in more details in the following Sections  (Sections~\ref{sec:dm}, \ref{sec:gas}, and \ref{sec:stars}).

\section{Dark matter structures}
\label{sec:dm}

\subsection{The dark matter power spectra}
\begin{figure}
\begin{center}
\includegraphics[width=0.47\textwidth]{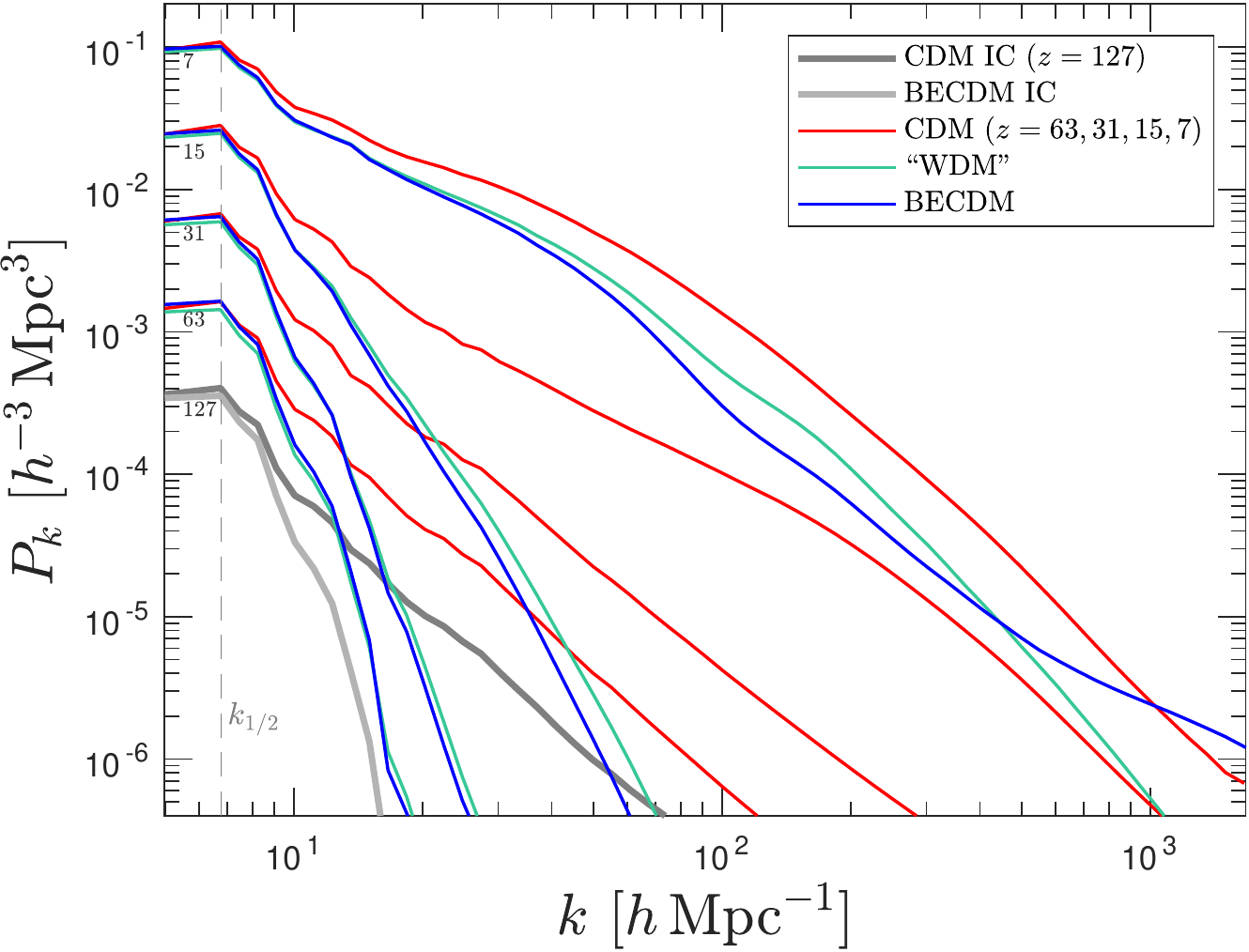}\\
\includegraphics[width=0.47\textwidth]{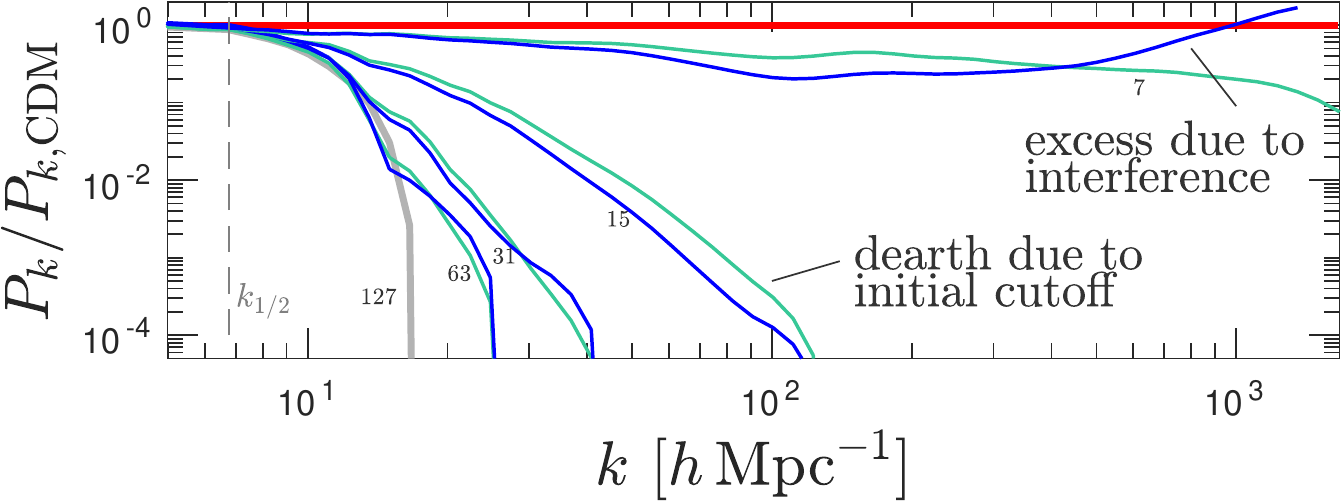}\\
\includegraphics[width=0.47\textwidth]{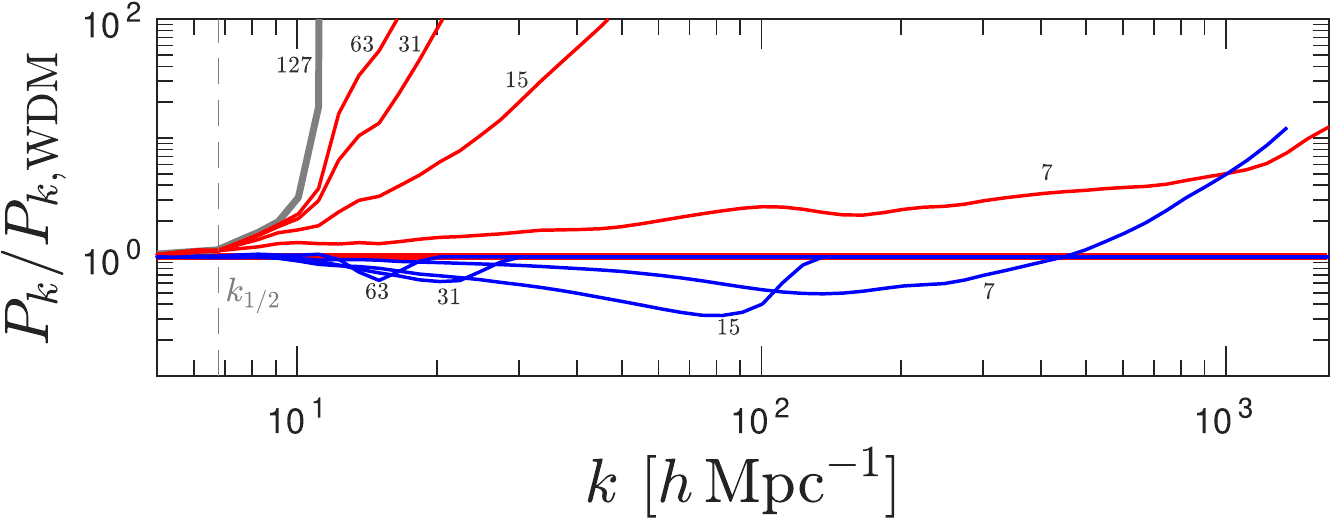}
\caption{Top panel: Evolved power spectrum of the dark matter (comoving) density field. Color-coding is indicated on the legend. Corresponding redshifts are marked below each set of lines on the left of the panel. Vertical dashed grey line marks the initial suppression scale $k_{1/2}$.
Middle panel: the ratio of the BECDM and ``WDM'' power spectra with respect to CDM (blue and green lines respectively). Grey line shows the relative initial suppression of power. Redshift of each line is marked on the plot.
Bottom panel: similar to the middle panel, but shows the ratio of BECDM/CDM to ``WDM'' (blue/red lines). Power in ``WDM'' and BECDM is exponentially suppressed on small-scales compared to CDM, but by $z\sim 7$ BECDM builds up excess kpc-scale power due to interference patterns. }
\label{fig:pk1d}
\end{center}
\end{figure}

The evolved dark matter power spectrum is shown in Fig.~\ref{fig:pk1d}. At wavenumbers smaller than  the initial suppression scale $k_{1/2}$ (shown by vertical dashed line in the figure), BECDM and ``WDM'' continue to follow CDM closely down to the end of our simulation at $z=5.5$, which means that there is no inverse-cascade of power to scales $\gtrsim 1$~Mpc due to, for example, the quantum potential. BECDM and ``WDM'' follow each other closely between the initial redshift and $z\sim 15$, indicating that the dynamically active quantum potential has not modified structure significantly. In both cases of BECDM and ``WDM'', the power spectra show a lack of power at large $k$ compared to CDM due to the initial cutoff. But by redshift $z=7$ BECDM exhibits excess power on small scales (few kpc) compared to both CDM and  ``WDM'' due to interference patterns in the simulations that have formed as a result of  shell-crossing and virialization inside haloes (more on this in our Section \ref{sec:fila} on dark matter filaments). This highlights the need to solve BECDM self-consistently, with the inclusion of the quantum potential.

\subsection{First dark matter objects}

\begin{figure*}
\begin{center}
\begin{tabular*}{0.99\textwidth}{l @{\extracolsep{\fill}} cccc}
& dm & \qquad\qquad\qquad\qquad gas & \qquad\qquad\qquad\qquad stars   \\
\end{tabular*} \\
\begin{tabular*}{0.99\textwidth}{l @{\extracolsep{\fill}} llllllllll}
&   $z=10.9$ &  $7.8$  & $5.5$ & $10.9$ &  $7.8$  & $5.5$  & $10.9$ &  $7.8$  & $5.5$   \\
\end{tabular*} \\
\rotatebox{90}{  \qquad\qquad\qquad --- halo 1 ---} 
\rotatebox{90}{\qquad BECDM   \qquad ``WDM''  \qquad  CDM} 
\includegraphics[width=0.31\textwidth]{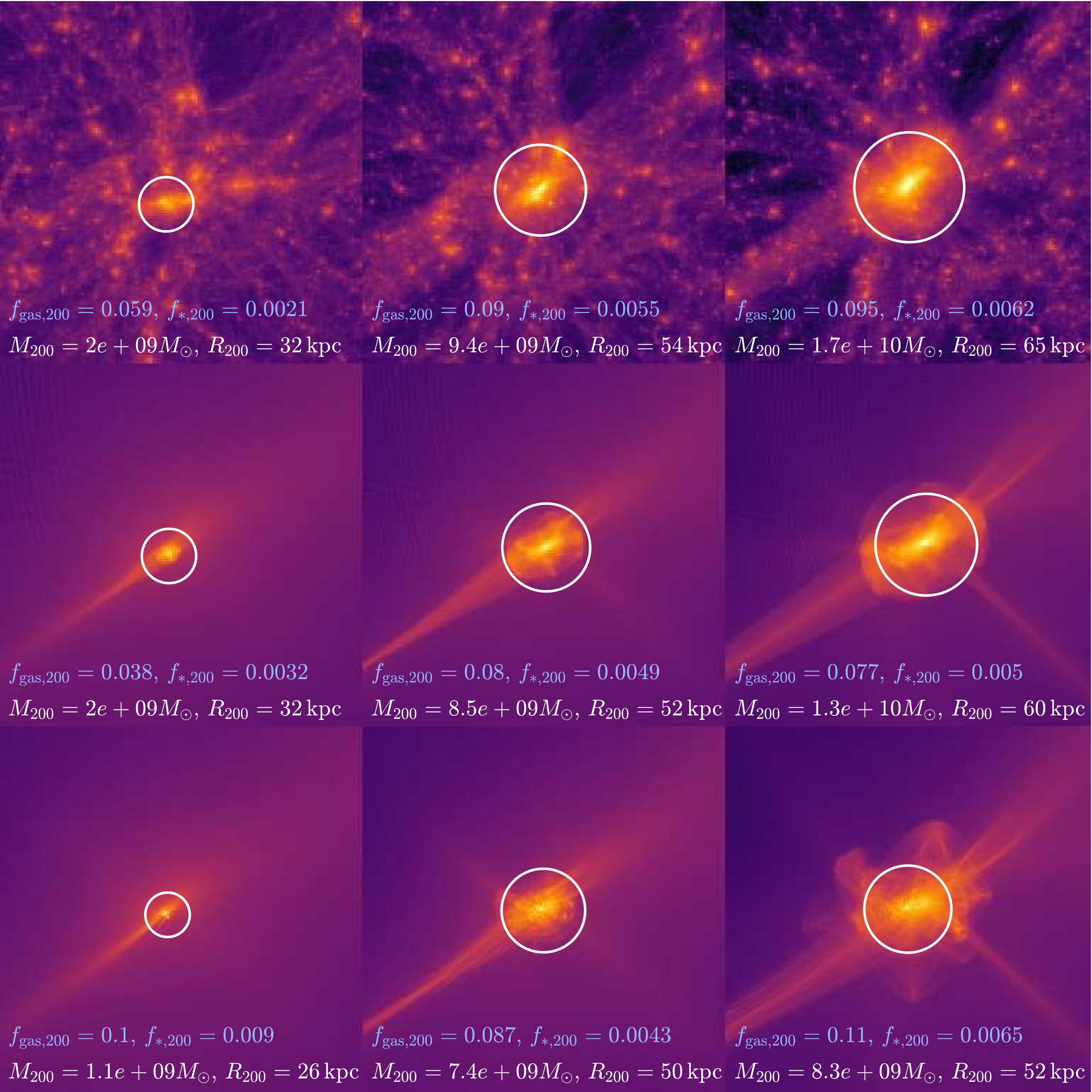} 
\includegraphics[width=0.31\textwidth]{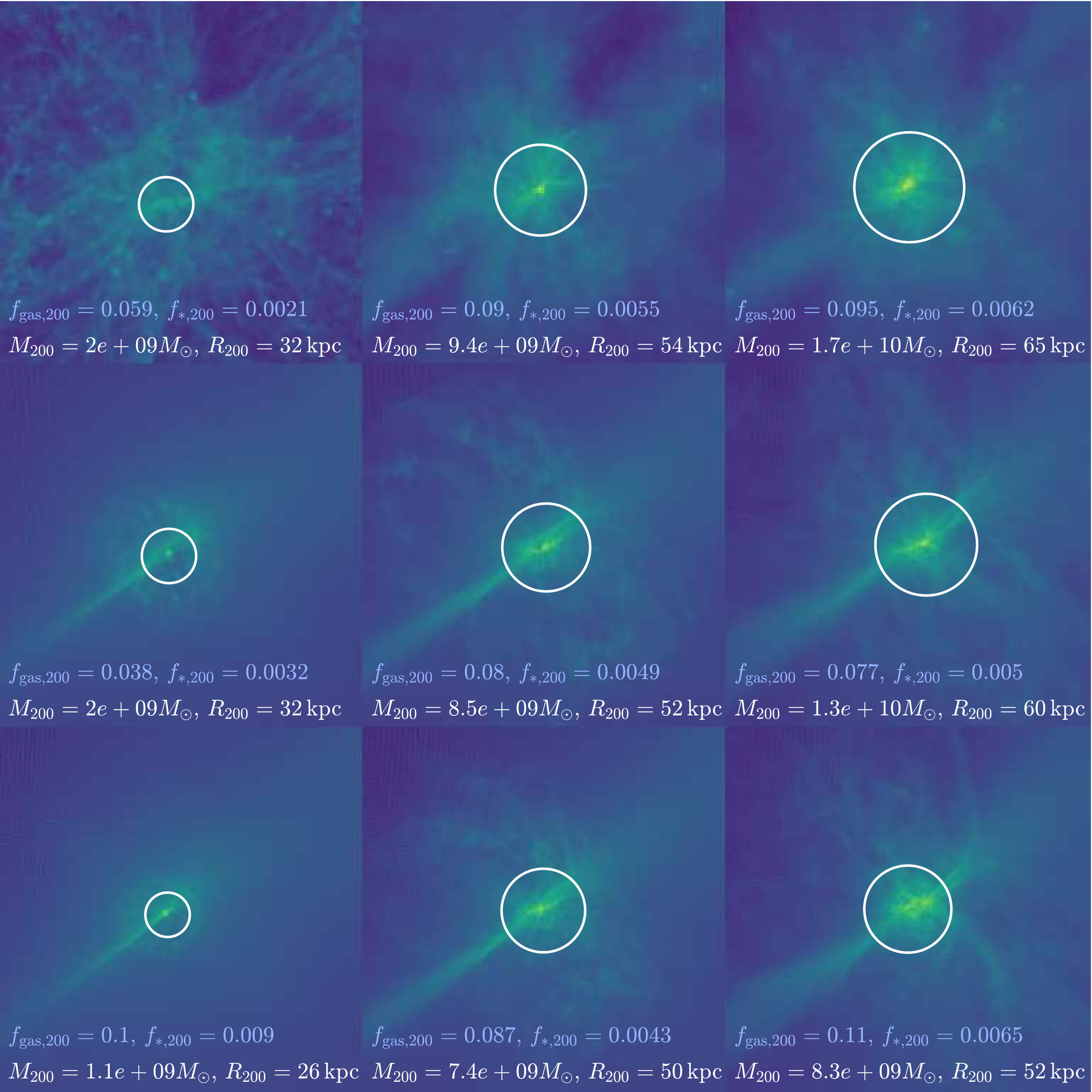}
\includegraphics[width=0.31\textwidth]{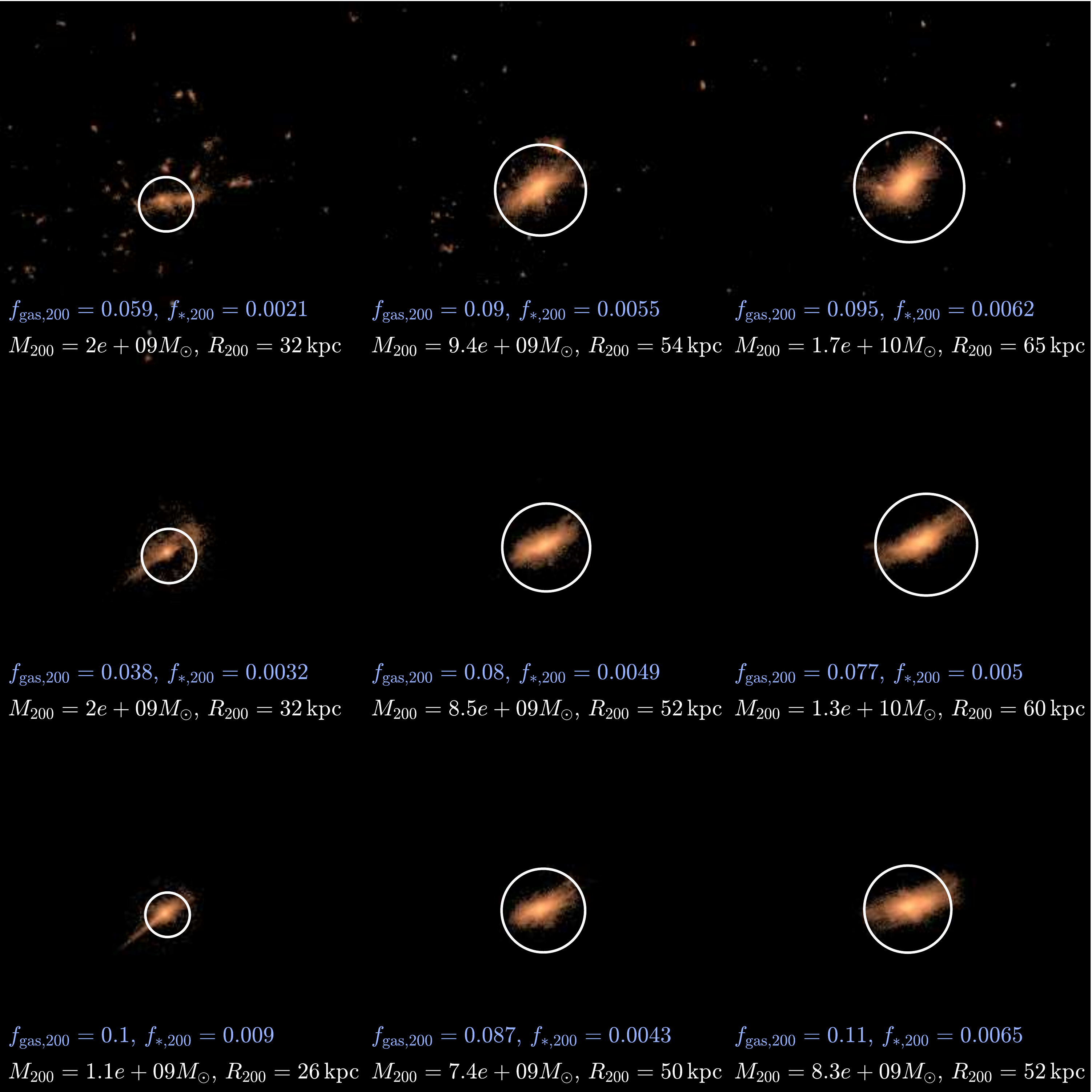} \\
\rotatebox{90}{   \qquad\qquad\qquad\qquad\qquad\qquad\qquad\qquad --- haloes 2 \& 3 ---} 
\rotatebox{90}{ \qquad\qquad BECDM   \qquad\qquad\qquad\qquad ``WDM''  \qquad\qquad\qquad\qquad CDM} 
\includegraphics[width=0.31\textwidth]{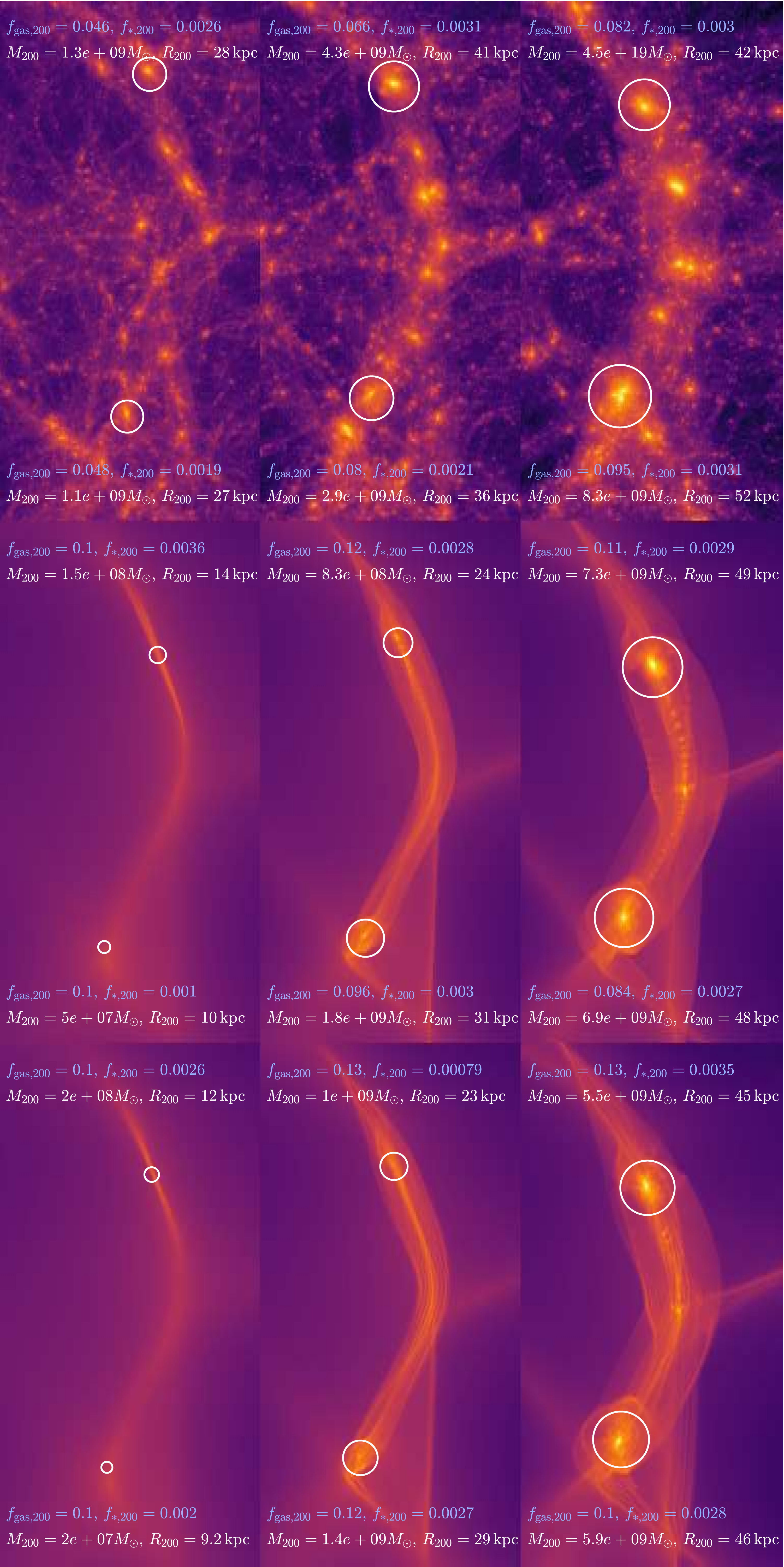} 
\includegraphics[width=0.31\textwidth]{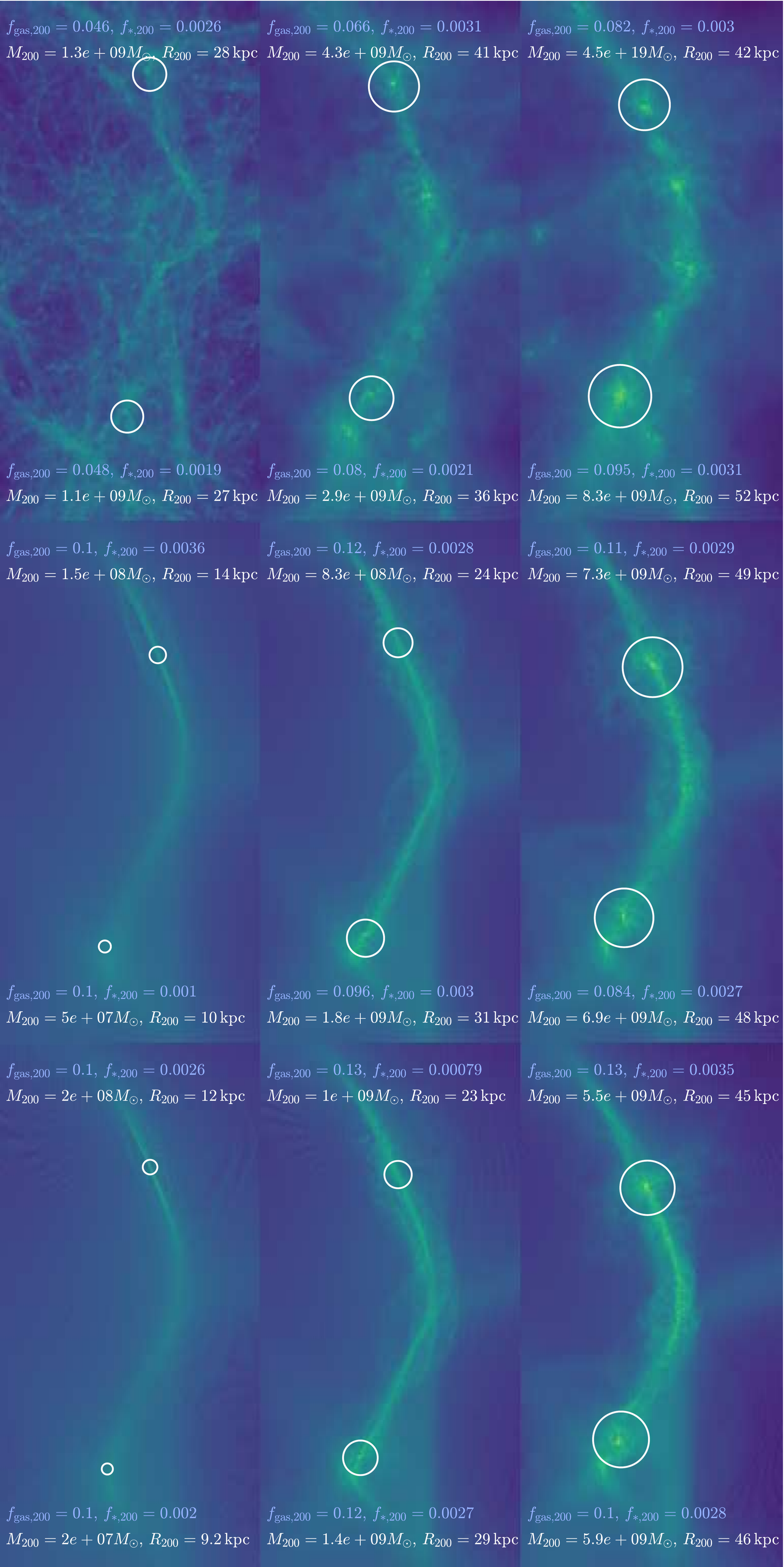}
\includegraphics[width=0.31\textwidth]{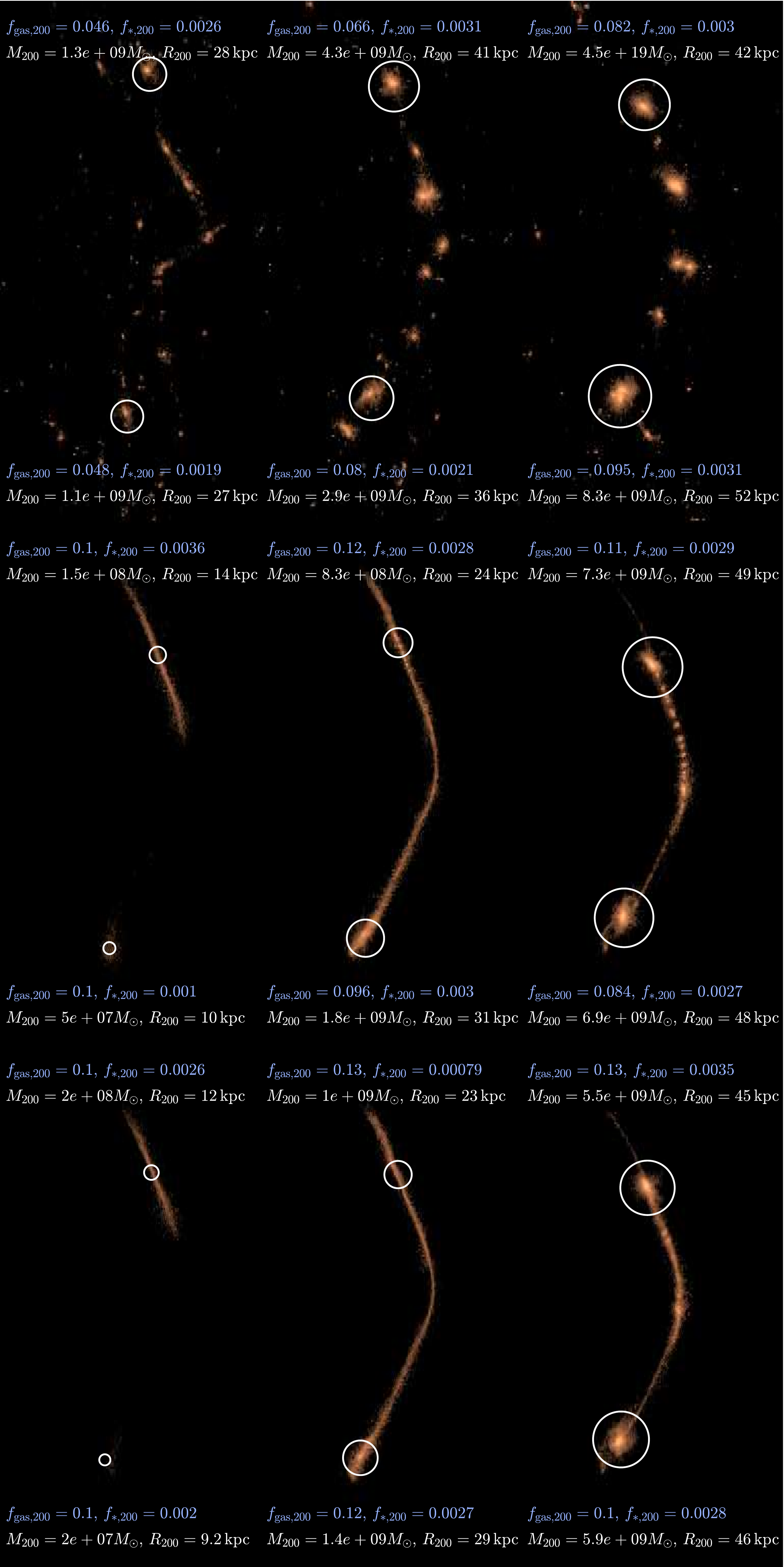} \\
\,\,\,
\includegraphics[width=0.27\textwidth]{cb-eps-converted-to.pdf}  \,\,\, \,\,\, \,\,\,
\includegraphics[width=0.27\textwidth]{cbb-eps-converted-to.pdf} \,\,\, \,\,\, \,\,\,
\includegraphics[width=0.27\textwidth]{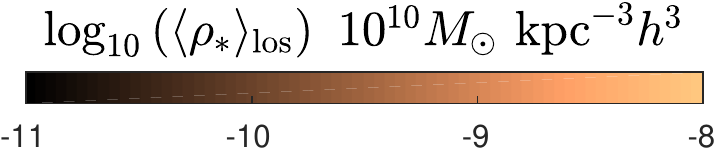}
\caption{Diversity of structures in dark matter (orange/purple), gas (green/blue) and stars (black/yellow). We show projected densities of the selected haloes simulated under the three different cosmologies.}
\label{fig:zoomD}
\end{center}
\end{figure*}

In all of the considered cosmologies, structure formation is hierarchical with smallest objects forming first. 
In CDM, the minimum mass is defined by the resolution of our simulation; while in BECDM/``WDM'' the  truncated initial power spectrum sets the minimum mass of a halo that is allowed to form \citep{2017PhRvD..95d3541H}
\begin{equation}
M_{\rm min}\simeq 4\cdot 10^6 M_\odot \left(\frac{2.5\cdot10^{-22}~{\rm eV}}{m}\right)^{3/2}.
\end{equation}
Our limited cosmological volume forms $3$ haloes of mass $M_{200}\sim 10^{10}~M_\odot$ by redshift $z\sim 6$.
Some of the main properties of the haloes (masses, sizes, fraction of mass in gas and stars, triaxiality parameters, and when stars first form), discussed here and in the subsequent sections, are summarized in Table~\ref{tbl:haloes}.
We also show zoom-in projections of the haloes (dark matter, gas, and stars) in Fig.~\ref{fig:zoomD}.
Here we discuss the leftmost panels of Fig.~\ref{fig:zoomD}, which show the halo dark matter distributions.
The structures that form are quite interesting. In BECDM/``WDM'', filaments are able to form early. They are unstable and eventually fragment to form a core / halo. In CDM (a scale-free theory) haloes form much earlier, as they are seeded by smaller-scale perturbations, and filaments themselves have substructure down to the mass resolution scale.

\begin{table*}
	\centering
	\caption{Summary of  halo properties at $z\simeq 6$ under different cosmologies. $M_{200}~[M_\odot]$ is the total mass of each halo at mean density over the critical density of $\Delta_c = 200$; $R_{200}$~[kpc] is the virial radius;  $f_{{\rm gas},200}$ is the fraction of matter in gas;  $f_{{\rm *},200}$ is the fraction of matter in stars;  $q$ is the  first triaxiality parameter defined as the intermediate-to-major
axis ratio;  $s$ is the  second triaxiality parameter defined as the minor-to-major axis ratio;  $z_{\rm stars\,form}$ is the formation redshift of the first stars  ($> 100~M_\odot$);  $z_{\rm halo\,forms}$ is the formation redshift of the halo,  defined as  the time of the last major merger for CDM; and as the first instant at which one can identify $M_{200}$ and $R_{200}$ for BECDM/``WDM'' (filament fragmentation).}
	\label{tbl:haloes}
	\begin{minipage}{\textwidth}
	\centering
	\begin{tabular}{l|cccccccc} 
		  & $M_{200}~[M_\odot]$ & $R_{200}$~[kpc] & $f_{{\rm gas},200}$ & $f_{{\rm *},200}$ & $q$ & $s$ & $z_{\rm stars\,form}$ & $z_{\rm halo\,forms}$ \\
		 \hline
		 \hline 
		  & \multicolumn{6}{c}{halo 1}\\
		\hline
		CDM     & $1.4\cdot10^{10}$ & $61$ & $0.095$ & $0.0058$ & $0.7$ & $0.5$ & 35 &    10 \\
		``WDM'' & $1.1\cdot10^{10}$ & $51$ & $0.077$ & $0.0049$ & $0.4$ & $0.3$ & 13.5 &  13.5 \\
		BECDM   & $8.2\cdot10^{9}$  & $42$ & $0.11$  & $0.0057$ & $0.4$ & $0.3$ & 13 &    13 \\
		\hline
		\hline
		  & \multicolumn{6}{c}{halo 2}\\
		\hline
		CDM     & $4.5\cdot10^{9}$  & $42$ & $0.082$ & $0.0030$ & $0.6$ & $0.4$ & 35 &   13 \\
		``WDM'' & $4.4\cdot10^{9}$  & $42$ & $0.11$  & $0.0033$ & $0.3$ & $0.2$ & 11.5 & 7.5 \\
		BECDM   & $3.9\cdot10^{9}$  & $40$ & $0.13$  & $0.0032$ & $0.3$ & $0.2$ & 11.0 & 7 \\
		\hline
		\hline
		  & \multicolumn{6}{c}{halo 3}\\
		\hline
		CDM     & $4.6\cdot10^{9}$  & $42$ & $0.095$ & $0.0027$ & $0.8$ & $0.5$ & 35   & 13 \\
		``WDM'' & $5.1\cdot10^{9}$  & $44$ & $0.084$ & $0.0025$ & $0.4$ & $0.3$ & 12.5 & 7.5 \\
		BECDM   & $4.6\cdot10^{9}$  & $42$ & $0.10$  & $0.0023$ & $0.4$ & $0.3$ & 12.0 & 7 \\
		\hline
	\end{tabular}
	\end{minipage}
\end{table*}

Table~\ref{tbl:haloes} also lists when each of the $z\sim 6$ haloes formed under the three different cosmologies. 
In BECDM/``WDM'' we quote when the initial filamentary structure fragments to form a spherically-collapsed object, which occurs well after stars form in the (cylindrical) filament that hosts haloes 2 and 3: stars form in the filaments around $z\sim 12$ but the fragmentation into haloes only occurs at $z\sim 7$, around which time central solitons start becoming visible too. Halo 1 forms out of more spherical conditions at $z\sim 13$ and also starts forming stars at that time.
In CDM the halo is a result of many smaller mergers, so we quote the redshift of the last major merger prior to $z=6$. Stars begin to form in subhaloes quite early $z>30$ that then merge to form the $z=6$ halo. Last major mergers for these CDM haloes occurs at $z\sim 10$.

\subsubsection{Triaxiality}

There is a stark contrast in the triaxiality of first haloes in CDM vs BECDM/``WDM''. Triaxiality is an important measure for dark matter haloes, because non-sphericity affects weak and strong lens statistics, the nonlinear clustering of dark matter, and the orbital evolution of satellites galaxies \citep{2002ApJ...574..538J}, as well as the orbits of stars in the stellar halo of our own Milky-Way \citep{2018MNRAS.474.2142I}. Our results indicate that triaxiality could be used as a probe of dark matter nature,  and, therefore,  it is important to  quantify systematic differences in triaxialities for  different dark matter models.
We estimate the triaxiality of the haloes at $z=6$ and list   $q$ and $s$ triaxiality measures in  Table~\ref{tbl:haloes}.
The triaxiality parameters of the halo are computed at the radius $R_{200}$ following the procedure in
\cite{2019MNRAS.484..476C}. haloes are characterized by $0<q\leq 1$, the intermediate-to-major axis ratio, and $0<s\leq q$, the minor-to-major axis ratio. In CDM, haloes may have typical intermediate-to-major axis ratios of $q\sim 0.7$ \citep{2019MNRAS.484..476C} which is what we observe in our CDM simulation as well (our haloes have  $q =0.6-0.8$ at $z\sim6$). For comparison, BECDM/``WDM'' haloes are significantly more triaxial than in CDM with $q = 0.3-0.4$.
Furthermore, BECDM/``WDM'' low-mass haloes are much more elongated than those of CDM with $s = 0.2-0.3$, while for CDM this parameter measures $s=0.4-0.5$.


\subsubsection{Halo radial profiles}

The shape of the dark matter gravitational potential wells determines the distribution and motion of the observed luminous objects. The structure of virialized DM haloes in the case of BECDM is predicted to be very different from the cuspy Navarro-Frenk-White (NFW) profile found for pure CDM \citep{1996ApJ...462..563N}, as well as cuspy WDM haloes \citep{2014MNRAS.439..300L}. Simulations of merging DM haloes show that solitonic cores are formed on scales of the de Broglie wavelength after a free-fall time of the halo and that the resulting DM profile is cored \citep{2014NatPh..10..496S, Schwabe:2016, 2017MNRAS.471.4559M}. Here we study the radial profile of first objects at early redshifts, where BECDM, ``WDM'' and CDM may have somewhat more similarities, because we are comparing them right around a free-fall time, which occurs at $z\sim 5$; the CDM simulations themselves are not fully NFW-like.

\begin{figure}
\begin{center}
\includegraphics[width=0.45\textwidth]{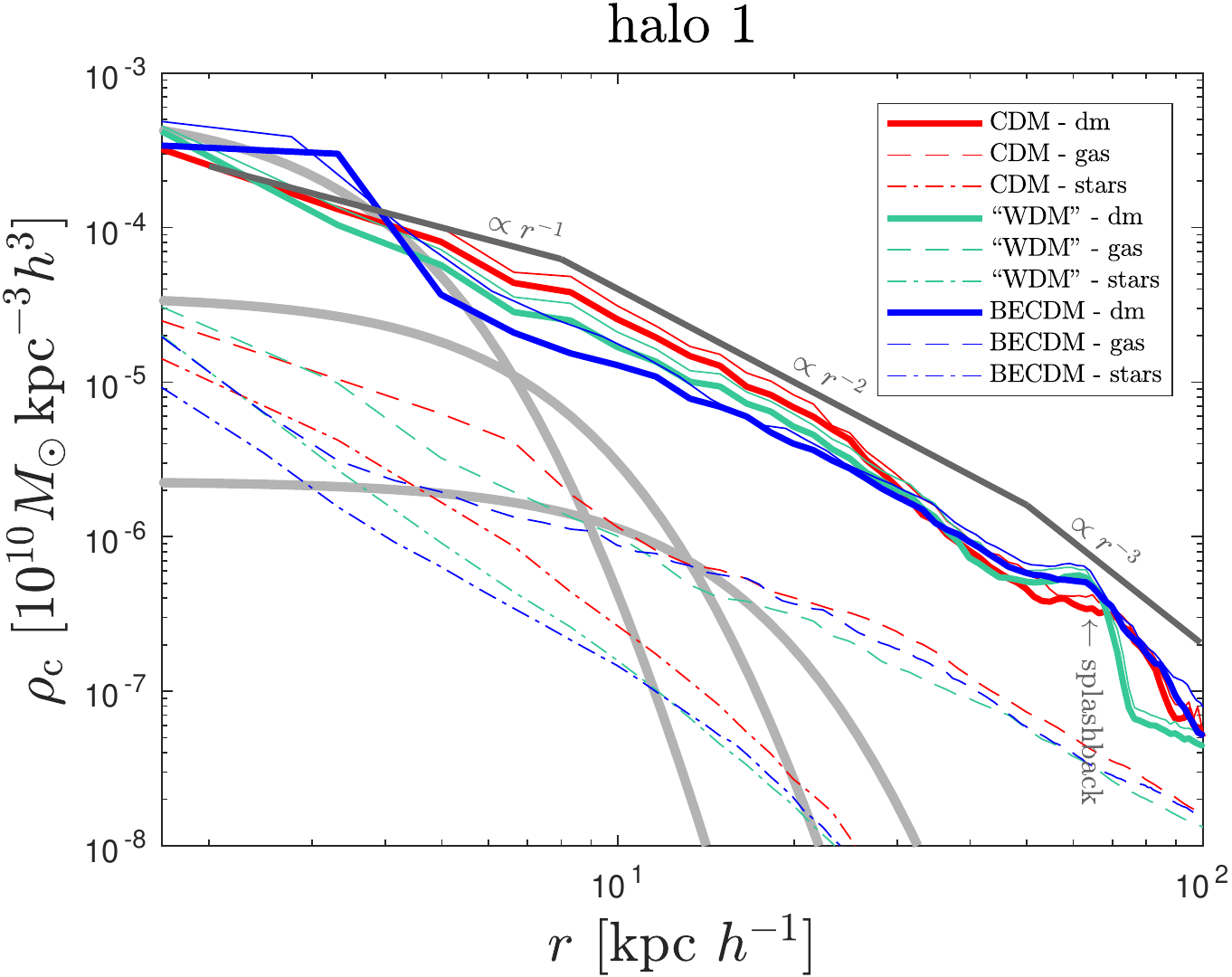} \\
\includegraphics[width=0.45\textwidth]{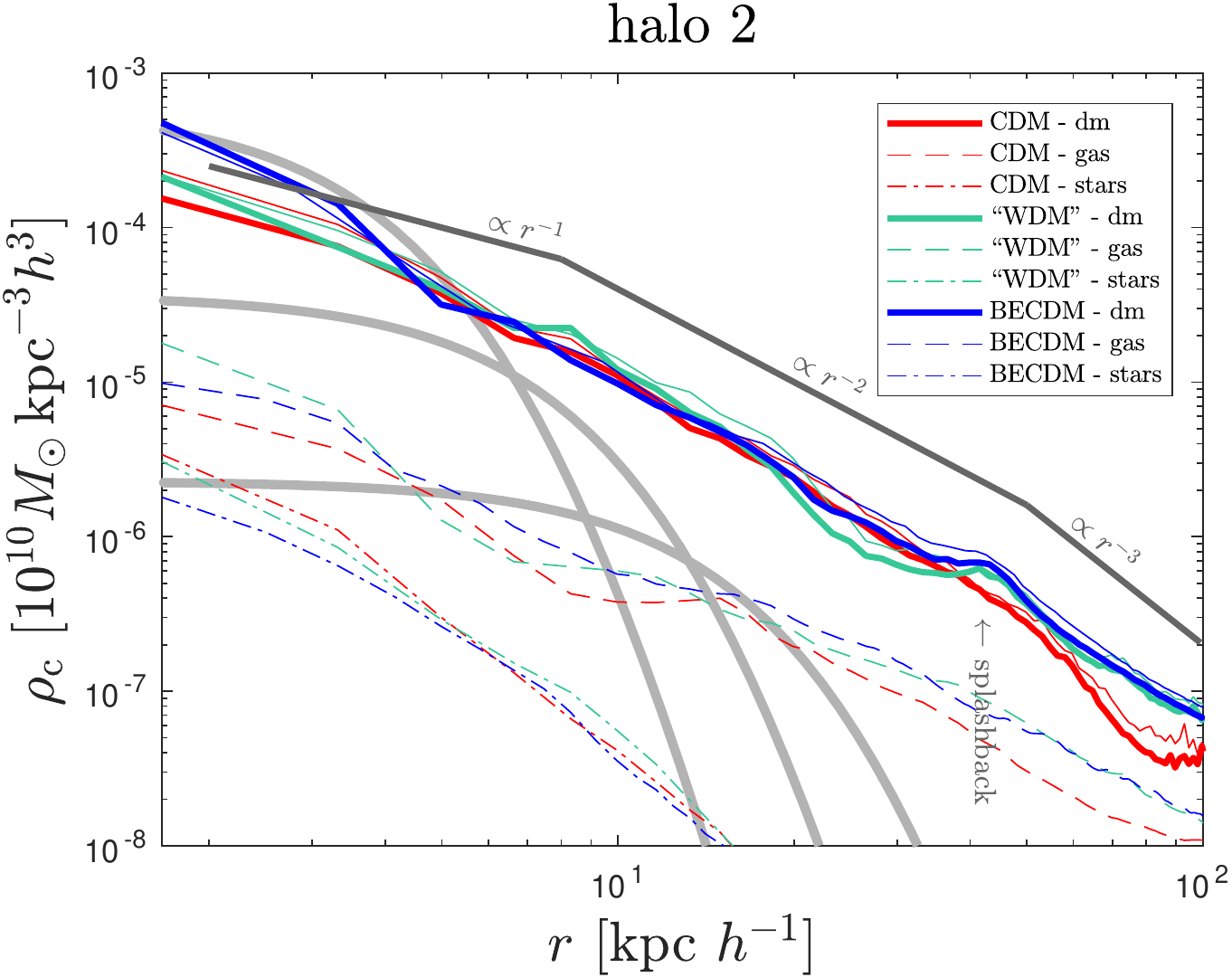} \\
\includegraphics[width=0.45\textwidth]{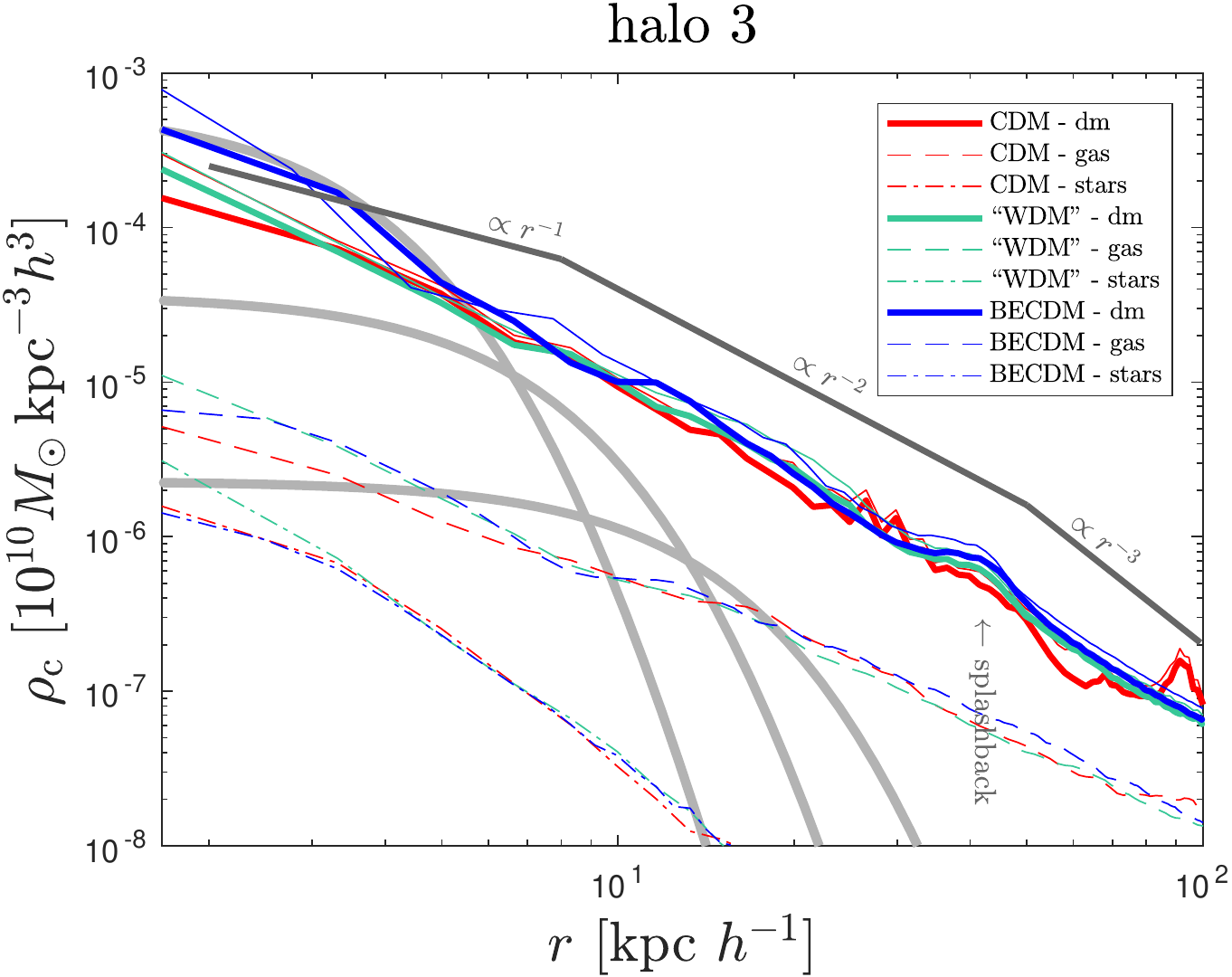} 
\caption{Radially averaged (comoving) density profiles for the dark matter, gas, and stars  for 3 haloes in our simulations under different cosmologies are shown at $z=6$. The thick solid lines are dark matter density in the baryon full-physics run, and we also show corresponding thin lines in the dark matter only runs, which are similar and show that the baryons have not strongly modified the dark matter potential wells for these low mass haloes in the early universe. Thick grey lines show where soliton profiles of various mass/size lie, which are just marginally resolved by our simulations. The smallest, densest, most massive soliton profile approximately matches the simulations. }
\label{fig:profilesH}
\end{center}
\end{figure}

Even though there is less spherical symmetry observed in BECDM/``WDM'' cosmologies at high redshifts, as we go lower in redshift, large, more-spherically symmetric cores are formed embedded in filaments. 
In contrast, in CDM filaments fragment into subhaloes which can merge into the central halo.
We plot radial density profiles for dark matter, gas, and stars of our three haloes in Fig.~\ref{fig:profilesH}. Halo centres are identified by gravitational potential minima. We also compare the dark matter profiles to corresponding dark matter only runs  finding that baryons are not significant at affecting the profiles at these halo masses ($M\sim 10^9$--$10^{10}~M_\odot$) and redshifts ($z\geq 6$), and do not soften the initial cuspy profile of collapse. The relative change in the density introduced by baryons to the DM profiles is only $\sim 20$ per cent, consistent with the fact that in the dark matter only simulation the total dark matter density is larger by that amount. Feedback is not found to alter the dark matter significantly.  
The radially averaged inner profiles of CDM and ``WDM'' are similar. They are cuspy, and follow a $r^{-1.5}$ power-law.
Thus these high redshift first structures follow a Moore profile 
\citep{1998ApJ...499L...5M,2001ApJ...557..533F},
and have not yet evolved into an NFW ($r^{-1}$) power-law -- typical profiles of massive haloes at lower redshift  which have formed though hierarchical merging after the free-fall time  \citep{1996ApJ...462..563N}.
In contrast, the BECDM dark matter profiles have central soliton cores that form on around a halo free-fall time. Due to limited resolution of our simulations, the cores are only marginally resolved. In dark matter-only simulations, the cores are slightly more compact because they are more massive.

The ``splashback'' radius -- the radius ($r\sim 50$--$70$~kpc for our haloes) at which accreted matter reaches its first orbital apocenter after turnaround and creates a sharp density drop in the halo outskirts \citep{2015ApJ...810...36M} -- is also a clearly identifiable feature in the radial profiles. The splashback radius is seen clearly in our most massive halo, which is the most spherically symmetric, and it is especially sharp in the ``WDM'' simulation, since it is due to a sharp caustic in the dark matter distribution. In the other two haloes the splashback feature is still there, but in WDM/BECDM these haloes are embedded in a filament which smooths out the radially-averaged profile, i.e., the asphericity of haloes washes out the radially-averaged splashback feature. The splashback radius has not been studied previously in BECDM/``WDM'' and it may be possible that systematic differences in splashback sharpness can be used to observationally identify cosmology. It is possible that in BECDM this caustic feature would  appear more smoothed-out due to the density structure being fuzzy on the de Broglie wavelength scale. However, with our current simulations we have very limited statistics and we defer this study to future work.

\subsection{Dark matter filaments}\label{sec:fila}

\begin{figure}
\begin{center}
\includegraphics[width=0.47\textwidth]{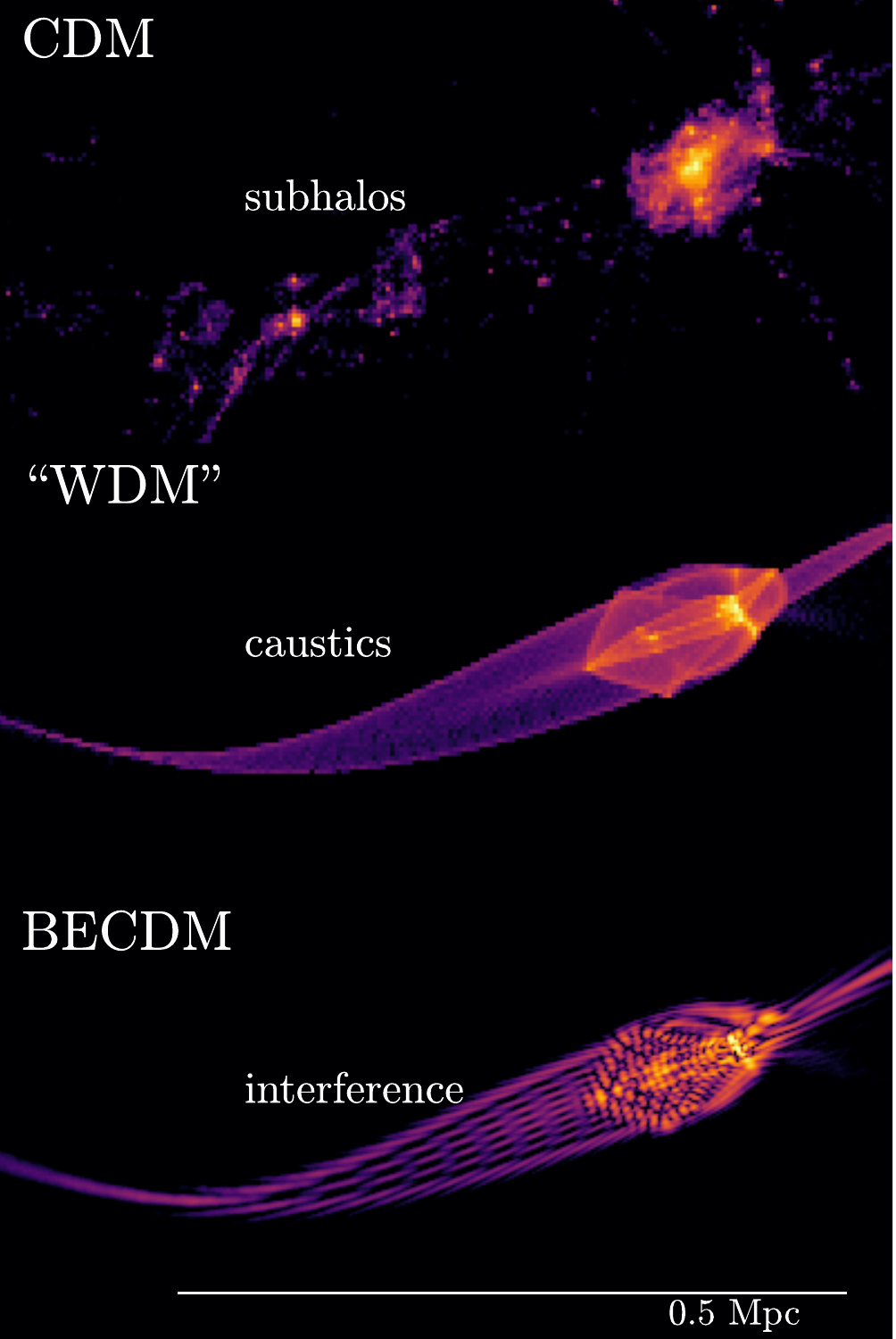} \\
\includegraphics[width=0.47\textwidth]{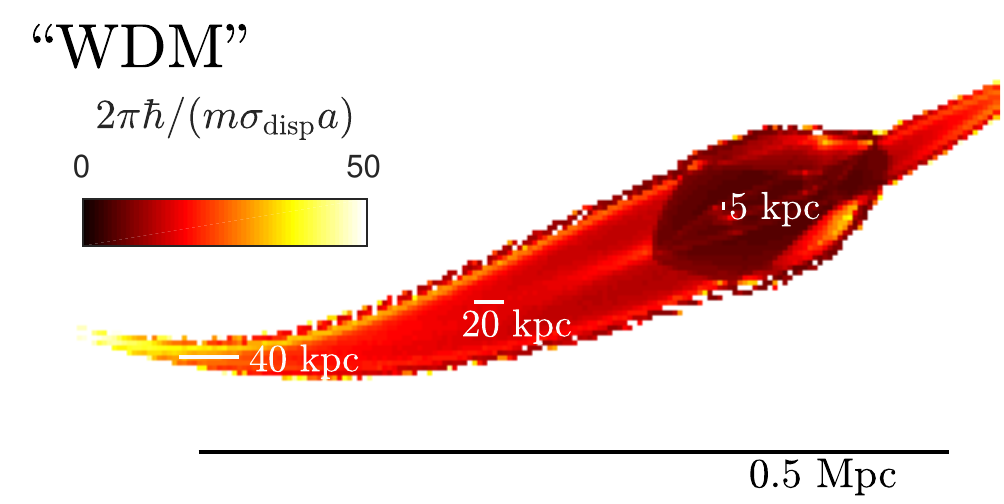} \\
\includegraphics[width=0.47\textwidth]{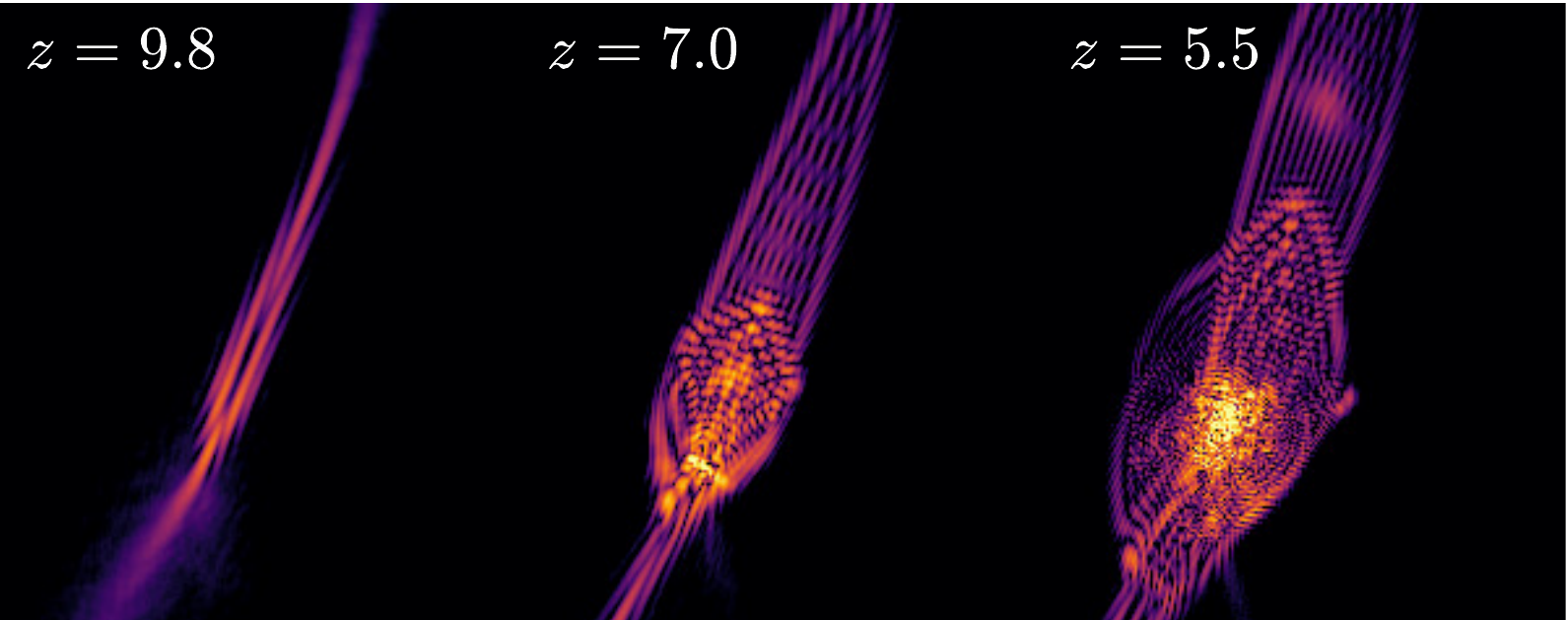}
\caption{Anatomy of a cosmic web dark matter filament. Three upper panels show a density slice through a filament at $z=7$. CDM has subhaloes on all scales. ``WDM'' shows caustic structures. And BECDM has large-scale coherent interference patterns due to converging flow towards the filament, and a coarse-graining of caustics on the local de Broglie length scale. The forth sub-panel shows the estimated sizes of BECDM interference patterns (at $z=7$) by taking $\lambda_{\rm dB}$ of the velocity dispersion of  ``WDM'', which are in good agreement with the actual BECDM simulation.
Bottom panel shows redshift evolution of the interference pattern in the BECDM filament (middle snapshot is the same as BECDM case in panel above, just rotated).
} 
\label{fig:filament}
\end{center}
\end{figure}

Even though large scale structure above $\sim 1$ comoving Mpc is the same  under the different cosmologies, the small scale structure is strikingly varied. To illustrate this, we consider a density slice through a cosmic filament, shown in Fig.~\ref{fig:filament}. In CDM the filament is comprised of low mass subhaloes. In the ``WDM'' simulation the same filament is formed, but there are no small scale initial perturbations and, hence, no fragmentation. Instead, dark matter is distributed continuously along the ``WDM'' filament.   Caustic structures are seen as material converges towards the filament. 
The filament formed with BECDM is distinct in that it displays interference patterns due to the relative velocity of the matter converging onto the filament. The interference in the filaments stays regular/coherent over time as there are just a small number of phase-sheets overlapping (see the lowest panels of the figure which exemplifies the temporal evolution of the interference profile). Inside the virial radius of the halo, the density structure appears much more `turbulent' \citep{2017MNRAS.471.4559M} as it is the superposition of a large number of plane waves that encode the velocity dispersion in the halo.   These features are linear in the sense that they are not held together by self-gravity but arise simply from the superposition of modes, as in the linear Schr\"odinger equations.

We point out that the scale of the interference patterns in the BECDM simulations can be estimated (to within a factor of $2$) as a function of location from ``WDM'' simulations by taking the de Broglie wavelength of the measured velocity dispersion of the dark matter particles that sample the 6D distribution function (see Fig.~\ref{fig:filament}).
This is expected by the Vlasov/Schr\"odinger-Poisson correspondence \citep{2018PhRvD..97h3519M}, as the interference features are just linear features (i.e., self-gravity is unimportant) and the BECDM equations approximate collisionless dynamics. 
Thus, it may be possible to conduct ``WDM'' simulations where quantum effects are post-processed, as a way to simulate larger box sizes. This method would not be exact, however, because it ignores areas in the superfluid where the quantum pressure support is important (e.g. support in soliton cores and the spines of filaments), which can feed back to reshaping larger scales of the internal structure of the haloes.

Even though discreteness noise is well-known to sometimes artificially collapse structure in such WDM filaments \citep{2007MNRAS.380...93W}, it only marginally affects our simulation and is not apparent in Fig.~\ref{fig:filament}. This is  due to the high particle resolution used (the cores formed in ``WDM'' are physical and not due to numerical effects because the same cores form in BECDM, which is unaffected by discreteness noise). 

\cite{moczPRL}  showed that unstable cylindrical soliton-like core can be found in  the center of dark matter filaments. This cylindrical structure is unstable and can fragment and form spherical solitons. The solitonic structures are unique to BECDM, and detecting them would be a smoking gun of such cosmologies. Therefore, of interest is the core/filament mass per unit length relation along the cylinder, and whether the normalization is similar to the relation found for spherical solitons embedded in haloes (e.g. \citealt{2014NatPh..10..496S}). \cite{2018PhRvD..97b3529D} have carried out an analytic calculation of the structure and stability of such cylindrical cores, including the presence of an axion self-interacting force, and check whether such structures would be visible in the Lyman-$\alpha$ forest power spectrum. They find that there would be a detectable impact on the distribution of Lyman-$\alpha$ lines if the core/filament mass per unit length relation is different from the relation of haloes by a factor of $A_{\rm c}\gtrsim 100$. However, for our object we find $A_{\rm c} \simeq 0.3 \sim 1$, in agreement with the spherical relation.


\section{Gas} \label{sec:gas}

In this section we discuss the distribution of gas in the simulated cosmological volume. The middle panels of   Fig.~\ref{fig:zoomD} show projected densities of the gas, zoomed in on our 3 haloes.

\subsection{Density distribution and power spectra}

The distribution of baryonic gas in the intergalactic medium is quite similar in BECDM and ``WDM'', as can be seen readily in the projected densities (Fig.~\ref{fig:zoomD}),
despite the small-scale disparities in the dark matter density field (there is order unity differences due to quantum interference patterns).
However, baryons are only coupled to dark matter via the long-range gravitational force, and, thus,  in the dark matter force field, the small scale structures (e.g. interference patterns) are suppressed by a factor of $k$ , the wave number, relative to the density field, and the coupled baryonic motions in BECDM is expected to approximate that of ``WDM'', converging as $\mathcal{O}(m^{-1})$ \citep{2018PhRvD..97h3519M}.

\begin{figure}
\begin{center}
\includegraphics[width=0.47\textwidth]{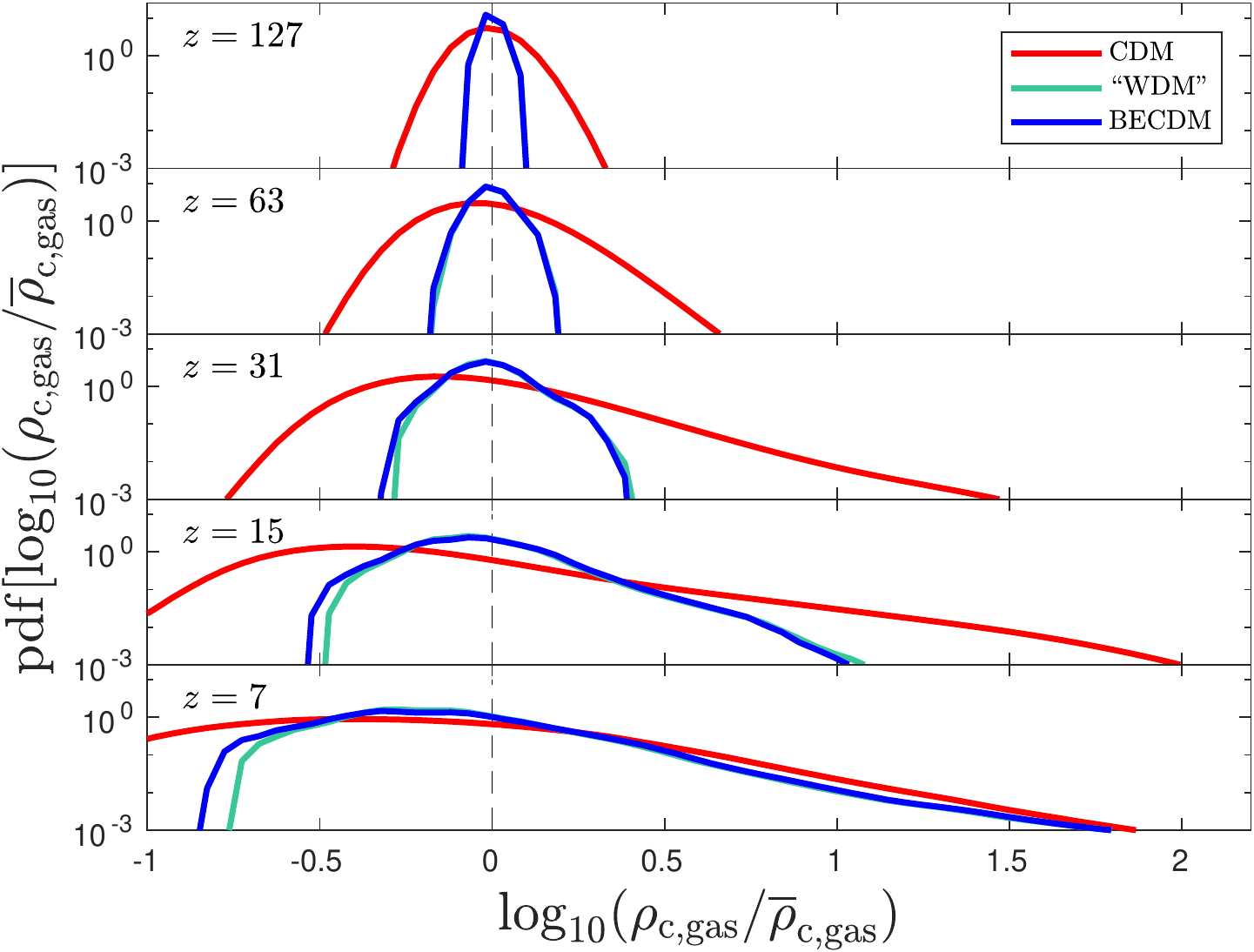}
\caption{The PDF of (comoving) baryonic densities in the cosmic volume for the three different types of simulations shown at various redshifts (indicated in each panel). The broader distribution of CDM is explained by the abundance of over- and under-dense regions which are smoothed in ``WDM''/BECDM.} 
\label{fig:densitypdf}
\end{center}
\end{figure}

Fig.~\ref{fig:densitypdf} shows the probability density distribution (PDF) of the baryon gas density in the cosmic volume. We see that the distribution of gas in BECDM/``WDM'' is significantly narrower in these cosmologies than in CDM. 
This is because in ``WDM''/BECDM cosmologies, structure is smoothed and there is a dearth of both over- and under-dense regions compared to CDM. 
At lower redshifts ``WDM''/BECDM  catch up with CDM in terms of the abundance of overdense regions; However, CDM voids are still much emptier than those in ``WDM''/BECDM. Exploring the contrast of matter density in voids might be an interesting route to constrain ``WDM''/BECDM  cosmologies.

\begin{figure}
\begin{center}
Baryon gas power spectrum:
\includegraphics[width=0.47\textwidth]{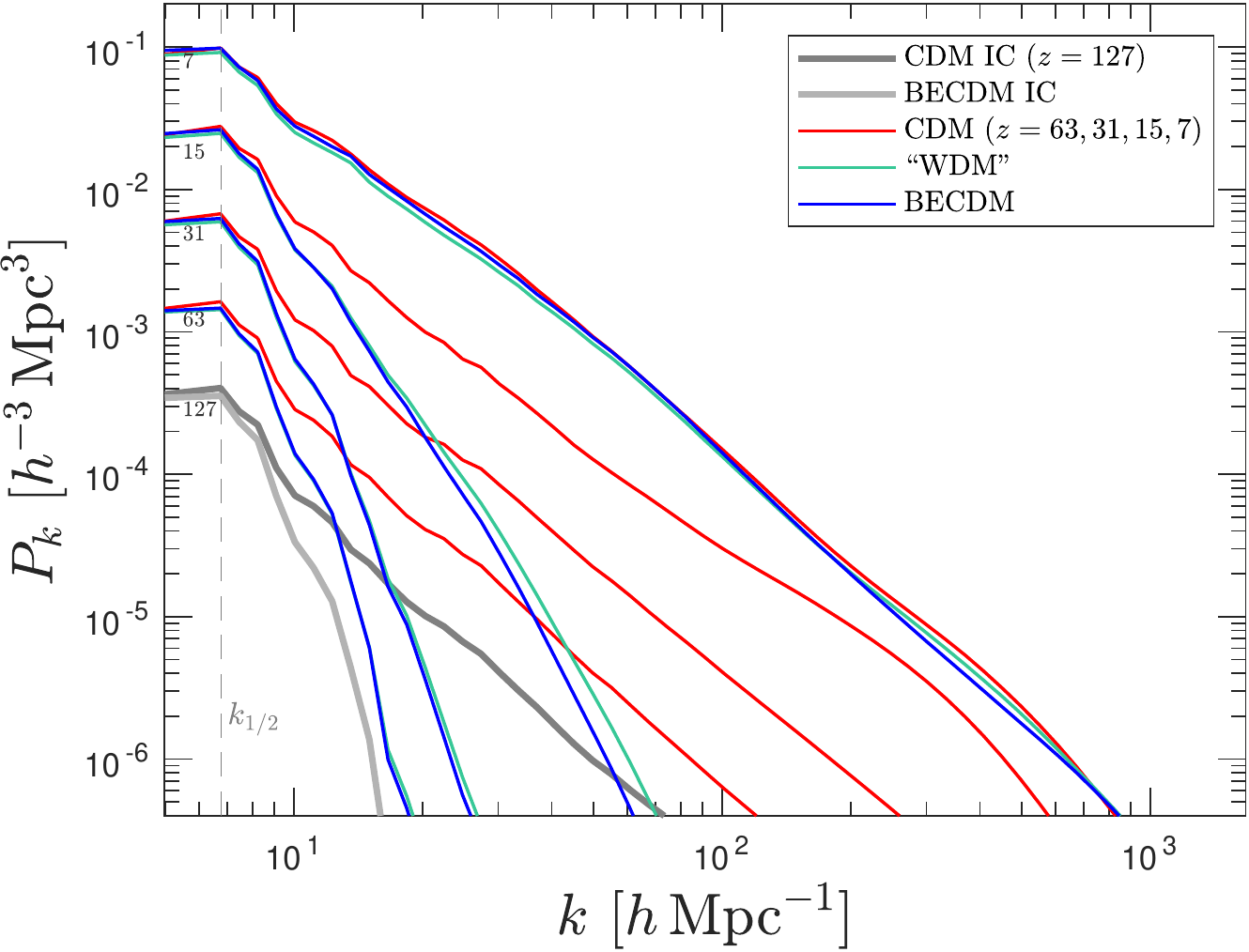}\\
\includegraphics[width=0.47\textwidth]{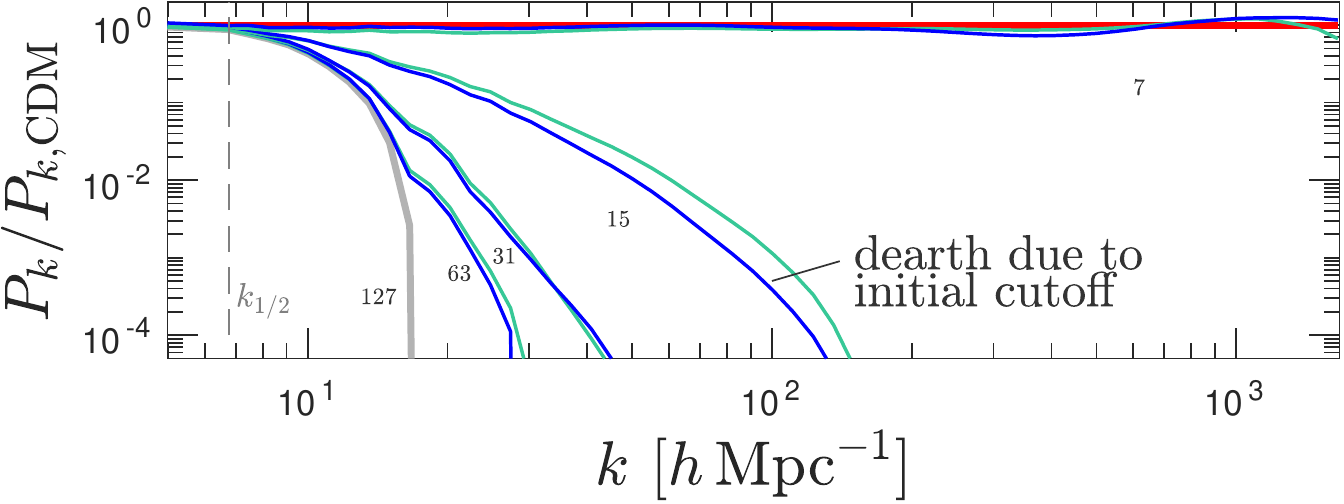}\\
\includegraphics[width=0.47\textwidth]{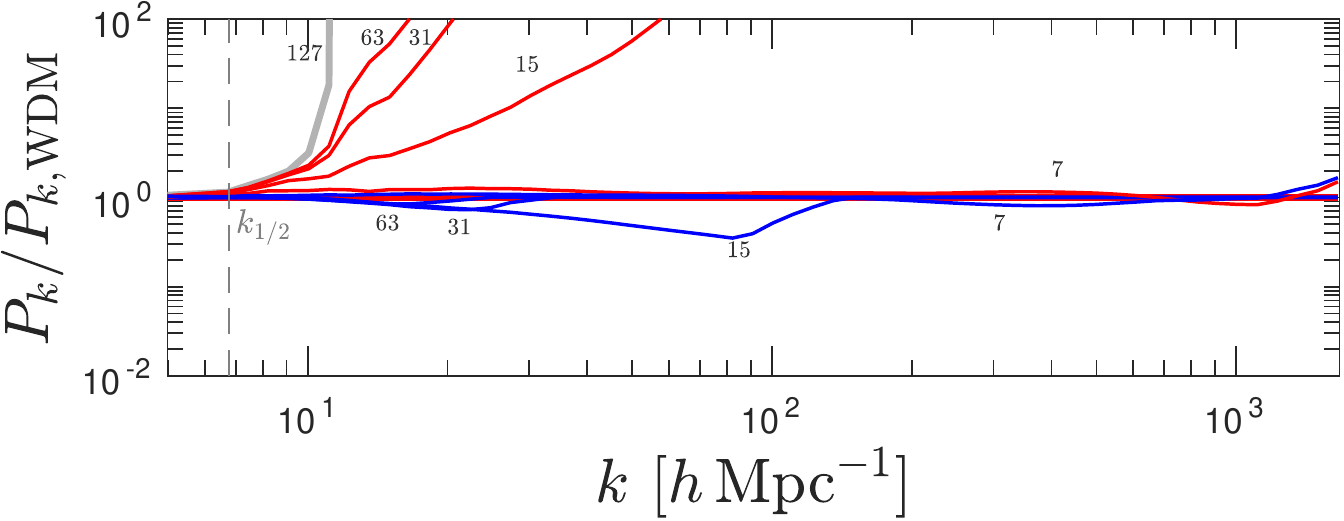}
\caption{Baryon gas (comoving) density evolved power spectra. Two bottom panels show the  ratio between the BECDM power spectra and CDM (middle panel),  ``WDM'' (bottom panel).}
\label{fig:bpk1d}
\end{center}
\end{figure}

\begin{figure}
\begin{center}
\includegraphics[width=0.47\textwidth]{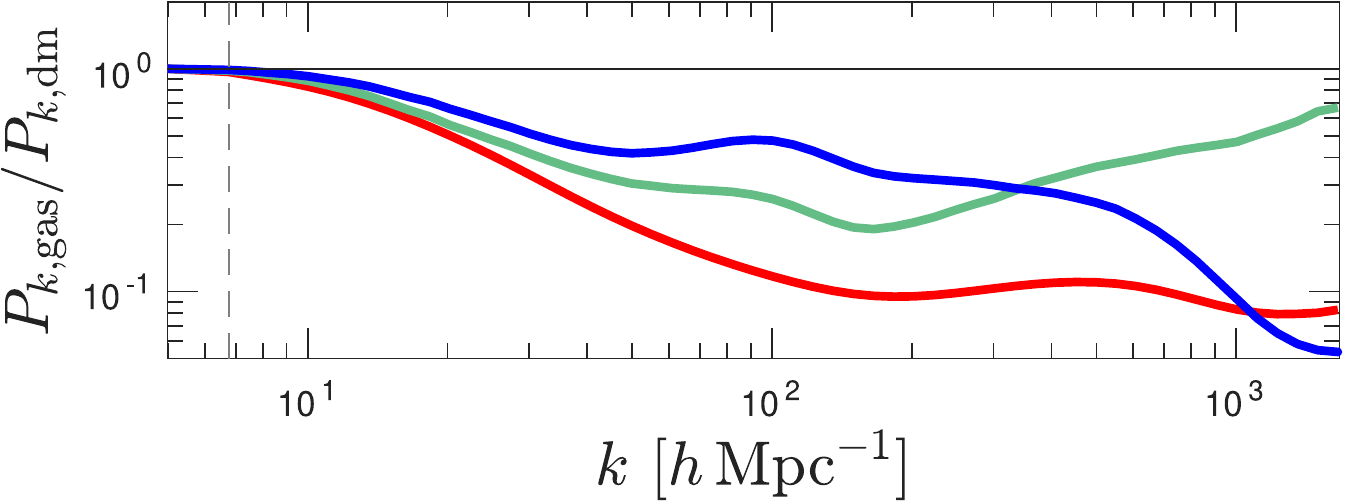}
\caption{Ratio  between gas and dark matter power spectra in CDM (red), ``WDM'' (green) and BECDM (blue) highlighting that on small scales baryons can be an imperfect tracer of dark matter structure at $z=7$.}
\label{fig:bdmpk1d}
\end{center}
\end{figure}

The baryonic power spectrum is shown in Fig.~\ref{fig:bpk1d}. The baryons initially follow the dark matter quite closely from $z=127$ to $z\sim 10$. Following the dark matter, baryons evolved under BECDM/``WDM'' show the same lack of power at large $k$   compared to CDM as is seen in dark matter due to the initial exponential cutoff. However, after about a free-fall time, by $z\sim 7$, the baryons start feeling their own pressure as well, and their distribution can become different from the dark matter in this non-linear regime. 
We highlight this fact by showing the relative power between gas and dark matter in Fig.~\ref{fig:bdmpk1d} at $z=7$. The ratio between gas and dark matter power spectra in ``WDM'' is closer to unity at small scales (large wavenumbers) than in CDM reflecting the fact that the  dark matter distribution is smoother. Similarly, in BECDM, the ratio is close to unity at small values of $k$; however, it  drops to very low values at  large wavenumbers owing to the small scale dark matter structures (interference profile) which are not imprinted in the gas spectrum. Even though in CDM the baryons have less small-scale power than the underlying dark matter (we discuss the reasons for this -- baryons feel gas pressure and are no longer as tightly coupled to dark matter below filtering scale -- in the next Section~\ref{sec:acc}),  the gas distribution  in BECDM/``WDM'' is smoother compared to CDM  even after reionization and feedback have affected  it. This is  because  for our choice of the axion mass,  the minimum dark matter halo mass is slightly above the filtering scale in BECDM/``WDM'' (small scale structures were never formed). But, importantly, the baryonic power spectra between the different cosmologies agree to a much better extent than the dark matter ones by $z\sim 7$. We will investigate the implications of this agreement for the Lyman-$\alpha$ forest, which depends on the full phase distribution of the gas (density, temperature, ionization), in future work.

\subsection{Accretion of Gas onto Collapsed Objects} \label{sec:acc}

Here we discuss the formation of baryonic objects in our simulations  -- the accretion of gas onto collapsed DM structures (haloes, filaments and sheets) -- under the different cosmologies. 
In CDM, despite dark matter structure forming at all scales, baryons do not cluster below the filtering scale \citep{Gnedin:1998}.
This scale relates to the amount of gas available for cooling and star formation inside collapsed objects.
At low redshifts, where large haloes dominate, there may be no large difference in the accretion process in BECDM/``WDM'' compared to CDM. 
However, in the high-redshift domain there is an important difference: in the case of CDM, baryonic objects form on all scales above the filtering scale; however, in BECDM/``WDM''
the suppression in the initial power spectrum of BECDM/``WDM'' defines the minimum dark matter halo mass that can form, which in our case is above the filter scale. Therefore, BECDM/``WDM'' baryonic objects look ``fuzzier''/more smoothed than those in CDM. 
To give a quantitative comparison, we list  fraction of mass in stars and gas inside $R_{200}$ at $z\sim 6$  in Table~\ref{tbl:haloes} (and shown in Fig.~\ref{fig:zoomD}). Despite different values of the total stellar masses in haloes (total  mass of gas and stars in CDM haloes is larger than in ``WDM'', which is larger than in BECDM), the fractions of gas and stars are comparable across cosmologies $f_{{\rm gas},200}\sim 0.1$,
$f_{*,200}\sim {\rm few} \times 0.001$, despite differences in halo shapes, although some systematic differences are observed: at $z=6$  BECDM/``WDM'' may have slightly larger gas fractions and smaller stellar fraction than CDM.  

\subsection{Baryon feedback}

We point out that stellar winds / supernovae feedback is active in our simulations. By $z\sim 7$ this leads to $\sim 100$~kpc scale winds, which adds to small-scale power in the baryon distribution, and may help resolve some of the observational tension with constraints on the axion particle mass from the Lyman-$\alpha$ forest. The differences in the gas distribution across our three cosmological models is also less significant compared to the underlying dark matter distribution, in part due to how feedback reshapes power on small-scales.

\section{First star formation}
\label{sec:stars}

Chemically pristine gas heats as it falls into dark matter potential wells, cools radiatively due to the formation of molecular hydrogen, and becomes self-gravitating collapsing to form first stars. The Rees-Ostriker-Silk \citep{1977MNRAS.179..541R} cooling criterion sets the minimum dark matter halo mass which is able to host star formation. In the CDM cosmology, first sites of star formation are in mini-haloes ($M_{\rm halo}\sim 10^5$--$10^6M_\odot$) at $z=50$--$60$ \citep{2013RPPh...76k2901B,Bromm:2013, Naoz2006}; while a more significant amount of stars can be built up by $z=20$--$30$. Formation of first stars and black holes provides a sensitive probe of the small-scale nature of dark matter \citep{Gao:2007, Yoshida:2003}, and changing the properties of dark matter might have  a strong effect on the formation of first objects. For instance,  \citet{Hirano:2015} assumed initial matter power spectrum with a blue tilt and found that in this case stars form early ($z\sim 100$) and are very massive. The case of BECDM is expected to be similar to the WDM scenario in which  star formation starts in filaments \citep{Gao:2007,Hirano:2017}. In the WDM scenario the way filaments fragment affects the  initial spatial distribution and amount of first stars, and, as a result, the rate of supernovae explosions and early metal enrichment. The likely very different initial mass function of stars and the rapid formation of massive black holes in a WDM scenario, as opposed to CDM, implies very different reionization, thermal and metal enrichment histories, greatly affecting galaxy formation and predictions for the James Webb Space Telescope (JWST) and Square Kilometre Array (SKA).  Some of these effects of the change in abundance of low-mass halos have been explored in the context of the ETHOS model with a cutoff in the power spectrum similar to WDM \citep{2018MNRAS.477.2886L,2019MNRAS.485.5474L}.

\begin{figure}
\begin{center}
\includegraphics[width=0.47\textwidth]{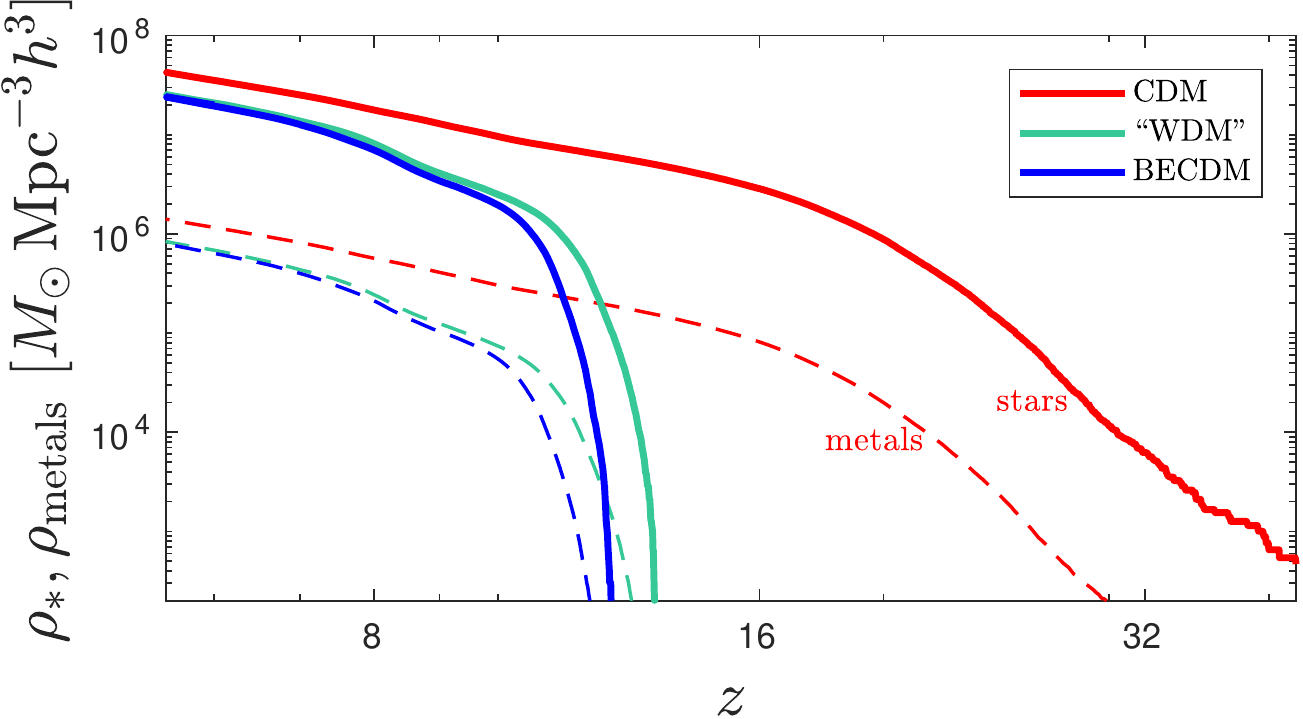}
\caption{The comoving density of stars (solid) and metals (dashed) as a function of redshift in our $1.7h^{-1}~{\rm Mpc}$ box. Star formation and metal enrichment are  found to be delayed in the ``WDM'' and BECDM simulations. The amount of metals trace the stellar mass.} 
\label{fig:sfr}
\end{center}
\end{figure}

In our simulations, under BECDM/``WDM'', the lack of small scale power elevates the minimum mass of star forming haloes leading to a delay in star formation.
On average in our simulated volume we find star formation is delayed due to the suppression in the initial power spectrum in BECDM/``WDM'' relative to CDM (see Fig.~\ref{fig:sfr}). Additionally, the quantum effects further delay star formation in BECDM, compared to the ``WDM'' case. For instance, in the most massive halo (\#1), first stars are formed at $z=35$, $13.5$ and $13$  in CDM, ``WDM'' and BECDM respectively (see details in Table~\ref{tbl:haloes}). The simulations reach a stellar density per cosmic volume of $10^6~M_\odot~{\rm Mpc}^{-3}$
at $z\sim 20$ in CDM, $z\sim 11$ in ``WDM'', and $z\sim 10.5$ in BECDM.  
Contrary to ``WDM'' where the distribution of gas and stars in the galactic centers are cuspy, in BECDM first galaxies can develop a solitonic core,  as was discussed in detail by \citet{moczPRL}. 
Furthermore, we point out here that only half as many stars as in CDM form per unit volume in BECDM/``WDM'' by $z\sim 6$ in the low mass haloes probed by our simulations.
The suppressed star formation would have an important impact on delaying reionization. In this paper we had included, by hand, a particular UVB background model that completed reionization by redshift $z\sim 6$ inferred from CDM simulations. However, for self-consistency, it should be modified by taking into account the differences in cosmic star formation history.

\begin{figure*}
\begin{center}
\begin{tabular}{cccc}
CDM & ``WDM'' & BECDM & \\
\includegraphics[width=0.3\textwidth]{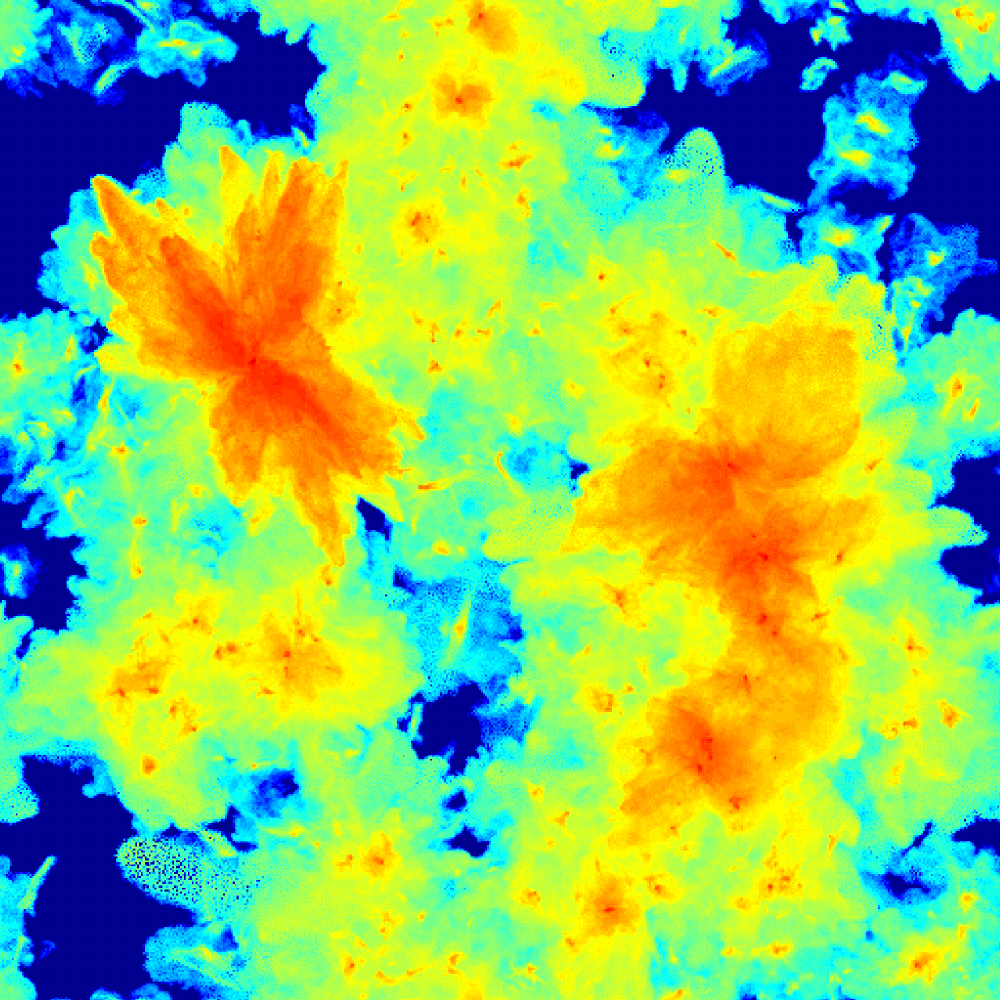} &
\includegraphics[width=0.3\textwidth]{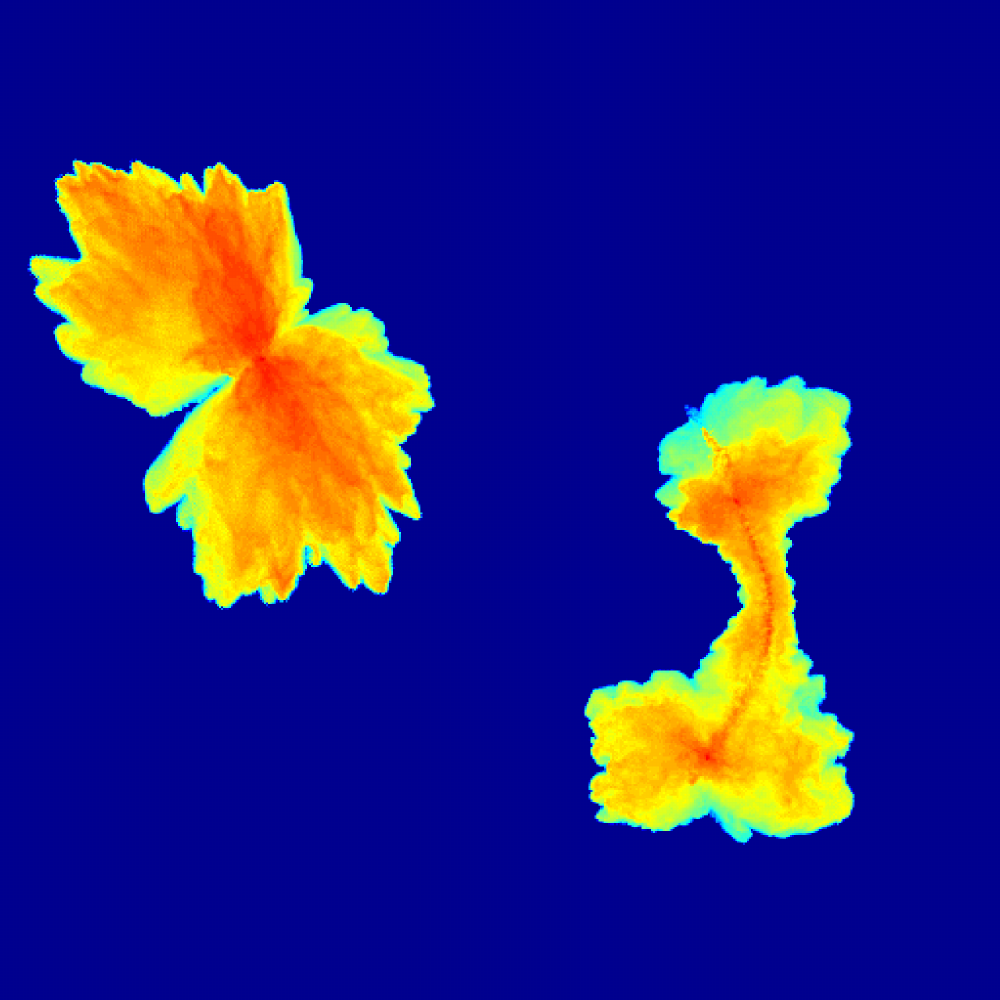} &
\includegraphics[width=0.3\textwidth]{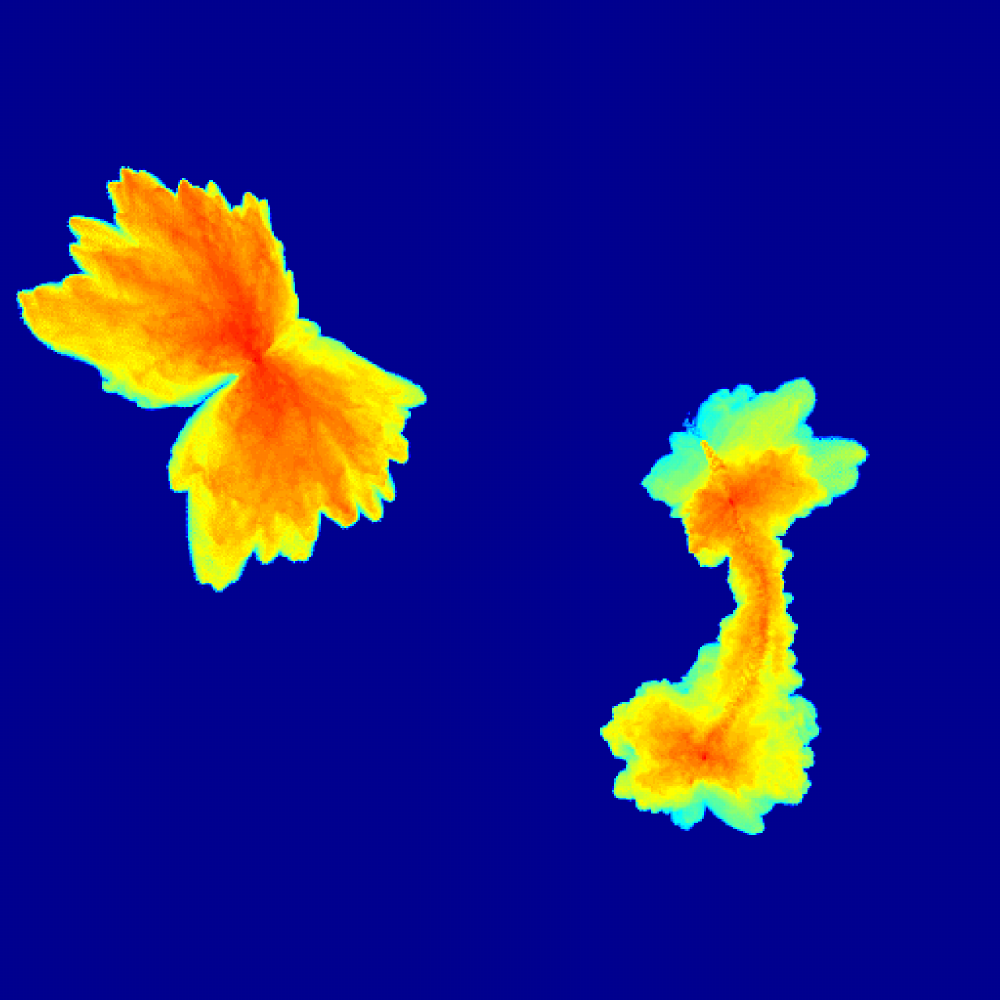} &
\includegraphics[width=0.08\textwidth]{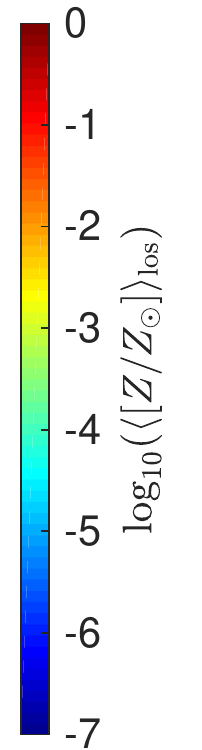} 
\end{tabular}
\caption{Mass-averaged line-of-sight metallicities in the cosmic volume at $z=6$ (see colorbar on the right). BECDM/``WDM'' shows a significantly more pristine intergalactic medium due to the lack of subhaloes.} 
\label{fig:Z}
\end{center}
\end{figure*}

 We briefly explore the metallicity distribution under the three different cosmologies. Fig.~\ref{fig:sfr} shows that the time evolution of the global mass in metals follows that of stars. Mass-averaged line-of-sight metallicities in the cosmic volume at $z=6$ is shown in Fig.~\ref{fig:Z}.
We see that BECDM/``WDM'' leaves much of the cosmic volume pristine at this redshift, compared to CDM.
This can be attributed to the absence of subhaloes below a critical mass and the delayed star formation and consequently reduced wind feedback.

We point out again that the baryonic modules have been tuned using CDM simulations.
It is possible that is the  efficacy of, e.g., the feedback or indeed the star formation law was modified so as to promote an earlier onset of  star formation, some of the differences between BECDM and CDM may be reduced.
This is left for future work. It will also be of interest to consider whether first stars that form in filaments under BECDM/``WDM'' could fragment and become globular clusters formed outside of galaxies.

\subsection{Cosmic Diversity}

In our simulations with dark matter fully coupled to the baryonic physics we observe filamentary star formation in BECDM and ``WDM'' cosmologies \citep[in agreement with][]{Gao:2007, Hirano:2017}. In filaments, we find, stars start forming much earlier than virialized haloes can be identified (Table~\ref{tbl:haloes}).  For instance, at the same location, where halo \#2 appears at $z = 7.5$/7 in ``WDM''/BECDM, star formation along the 2-D potential well starts $\sim 0.31$  Gyrs earlier, e.g., at at $ z = 11.5/11$. This filamentary mode of star formation is clearly seen in Fig.~\ref{fig:zoomD} (right column), where we show the projected stellar density of the three haloes  at $z=10.9$, 7.8, 5.5 under the three different cosmologies. 
 
However, the filamentary mode is reserved for small-mass objects (haloes \# 2 \& 3 in our simulated box, Fig.~\ref{fig:zoomD}), while in more massive haloes (such as halo \# 1, Fig.~\ref{fig:zoomD}) stars are distributed in a similar way to the CDM case -- in an isolated halo -- even at high redshifts. In halo \# 1 the redshifts of star formation and halo formation are the same ($13.5$ in ``WDM' and 13 in BECDM).

This dependence of the shape of the first galaxies on the initial conditions and the environment suggests that there is larger diversity of first star forming objects in BECDM/``WDM'' cosmologies compared to CDM. 
We note again that our simulations are not statistically representative, and the relative abundance of the extended  filamentary galaxies compared to the isolated CDM-like galaxies cannot be inferred from our simulations. Our finding may have implications for the `diversity problem' of observed dwarf galaxies
if the formation history of these dwarfs affects their late-time structure as well \citep{2015MNRAS.452.3650O,2016MNRAS.462.3628R}. 

\begin{figure*}
\begin{center}
\includegraphics[width=0.97\textwidth]{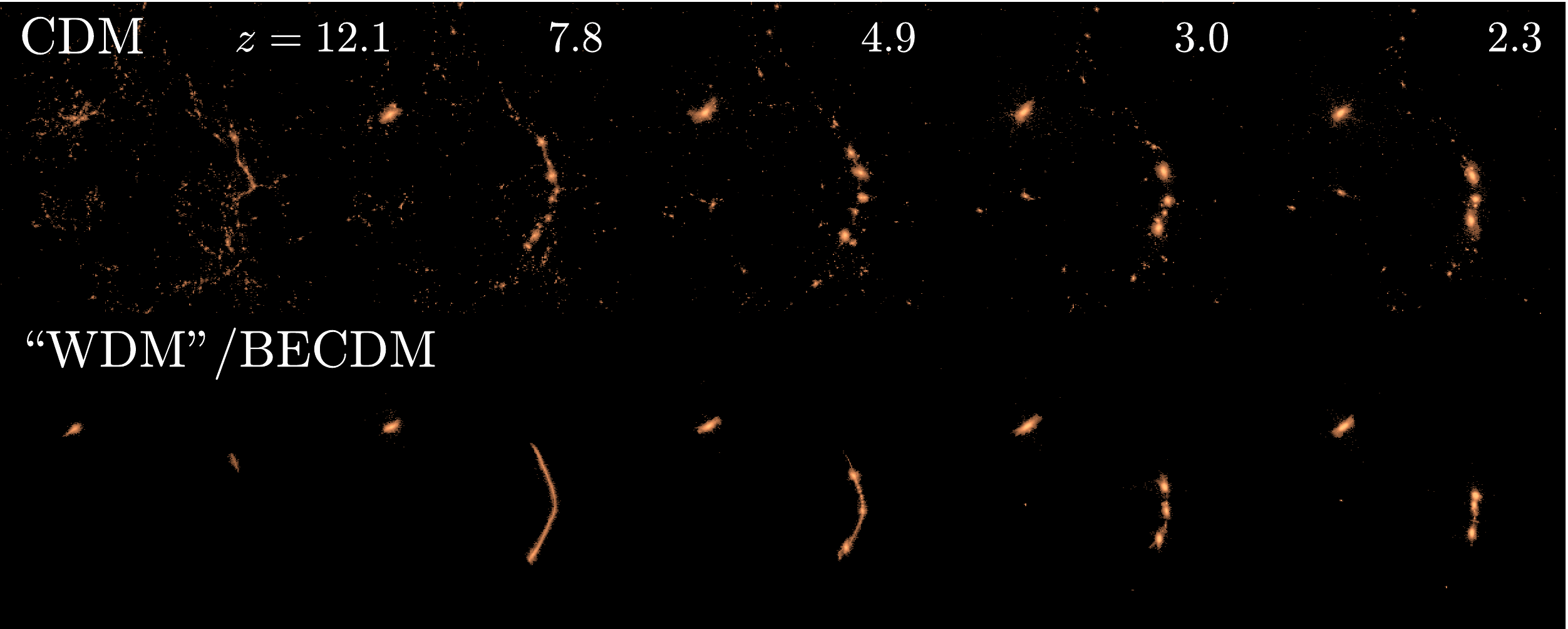}
\caption{Ultimate fate of filamentary first ``galaxies''. We show  the projected stellar densities in CDM (top row)``WDM'' (bottom row). Stars that in ``WDM''  form along cosmic web filaments before $z\sim 6$ accrete into more familiar-looking galaxies by lower redshifts. } 
\label{fig:starsW}
\end{center}
\end{figure*}

Finally, we also find that in ``WDM''/BECDM cosmology the morphology of the first small galaxies can be very different from low-redshift massive and evolved structures.   To illustrate this aspect, we show the evolution of the stellar content of a  filament in ``WDM'' down to $z = 2.3$  in  Fig.~\ref{fig:starsW} (unfortunately our full BECDM simulation lacks the resolution to go to such low redshifts, but the evolution is expected to be qualitatively similar). 
The ultimate fate of the filamentary first galaxies is that stars do end up being accreted into the few haloes that have formed along the filament, thus shaping more familiar-looking galaxies at low redshifts.

\section{JWST mock images}
\label{sec:JWST}

If the filamentary first galaxies (and not the spherical CDM  galaxies) are realized in nature, next-generation telescopes such as JWST could see bright a filamentary cosmic web illuminated by the first stars.

\begin{figure*}
    \includegraphics[width=0.325\textwidth]{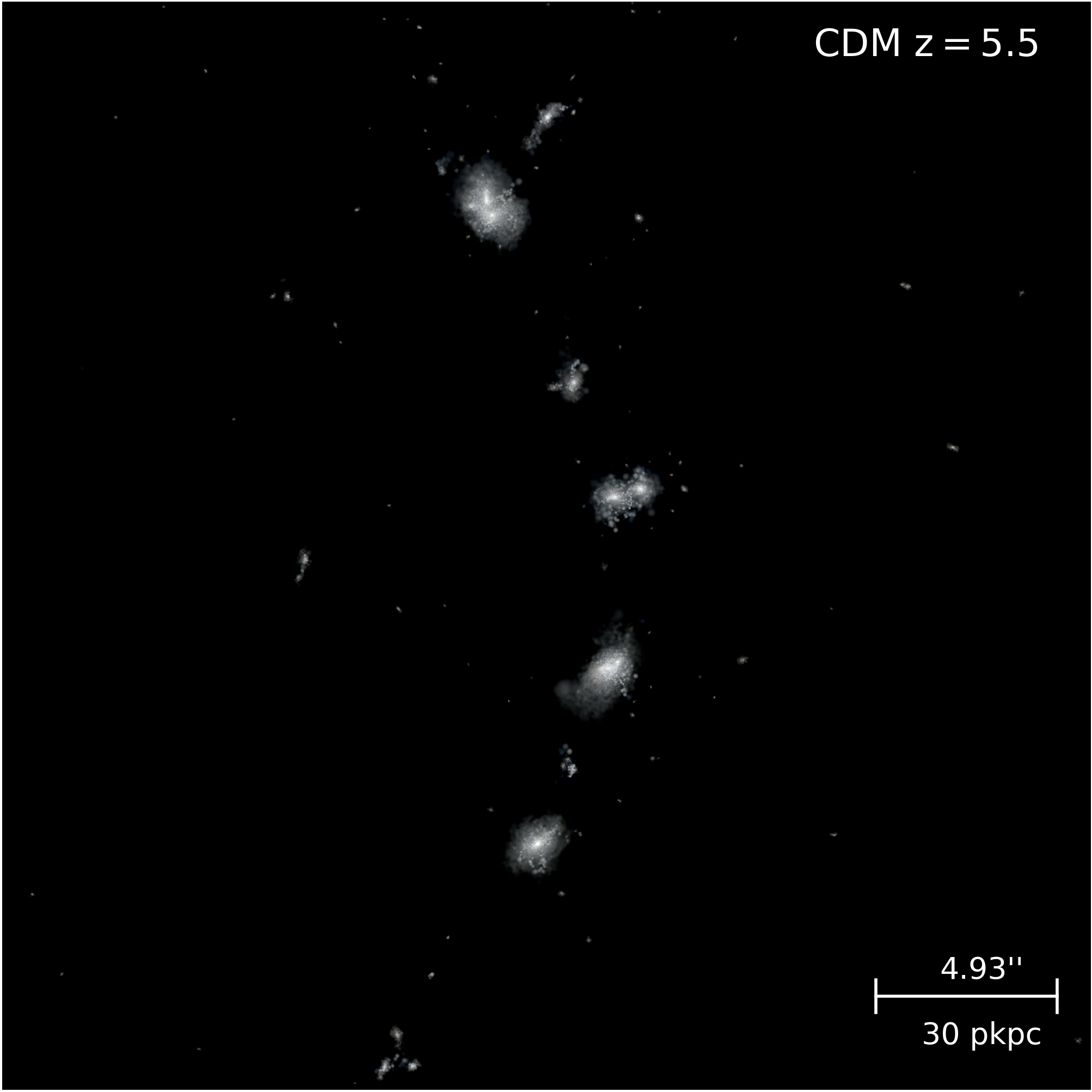}
    \includegraphics[width=0.325\textwidth]{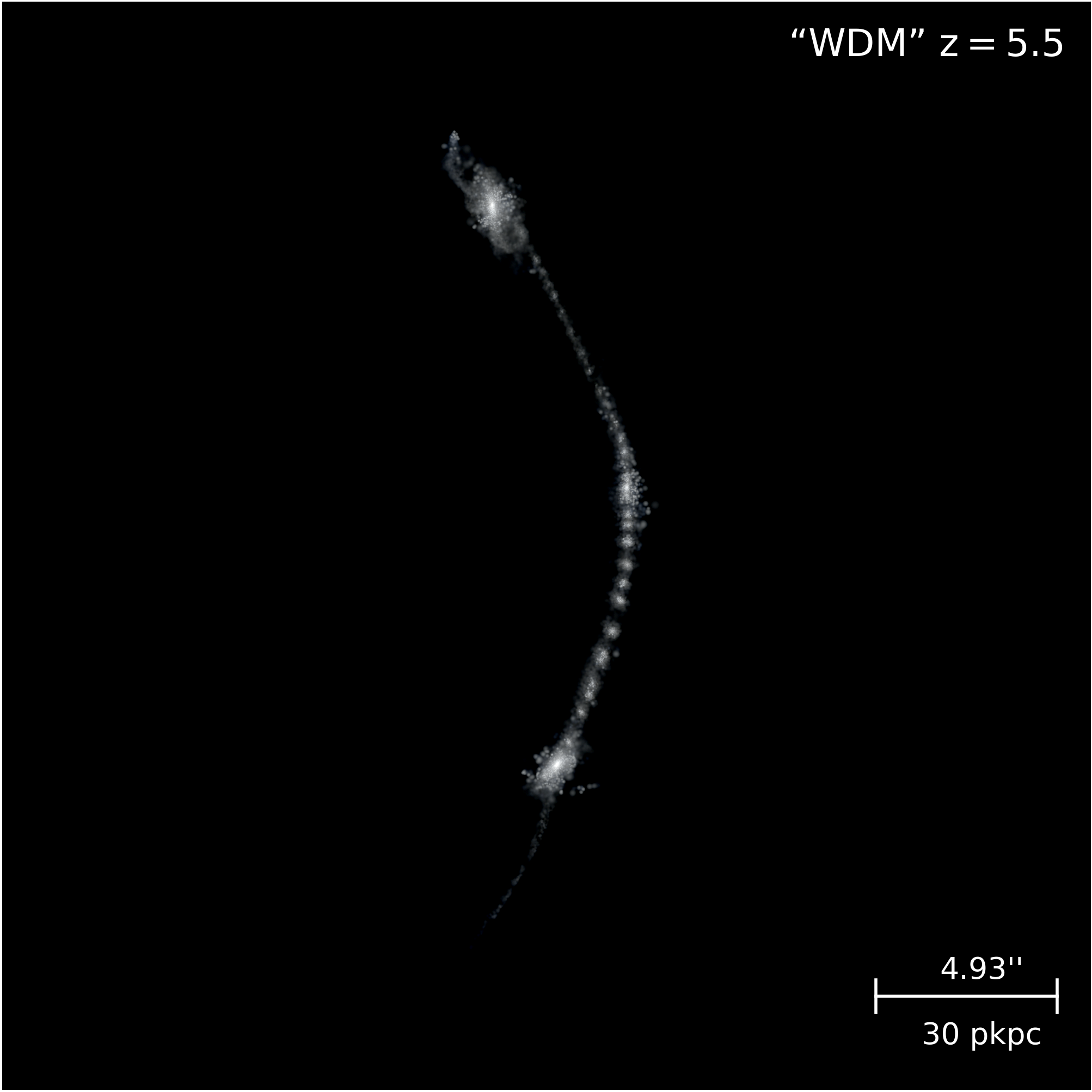}
    \includegraphics[width=0.325\textwidth]{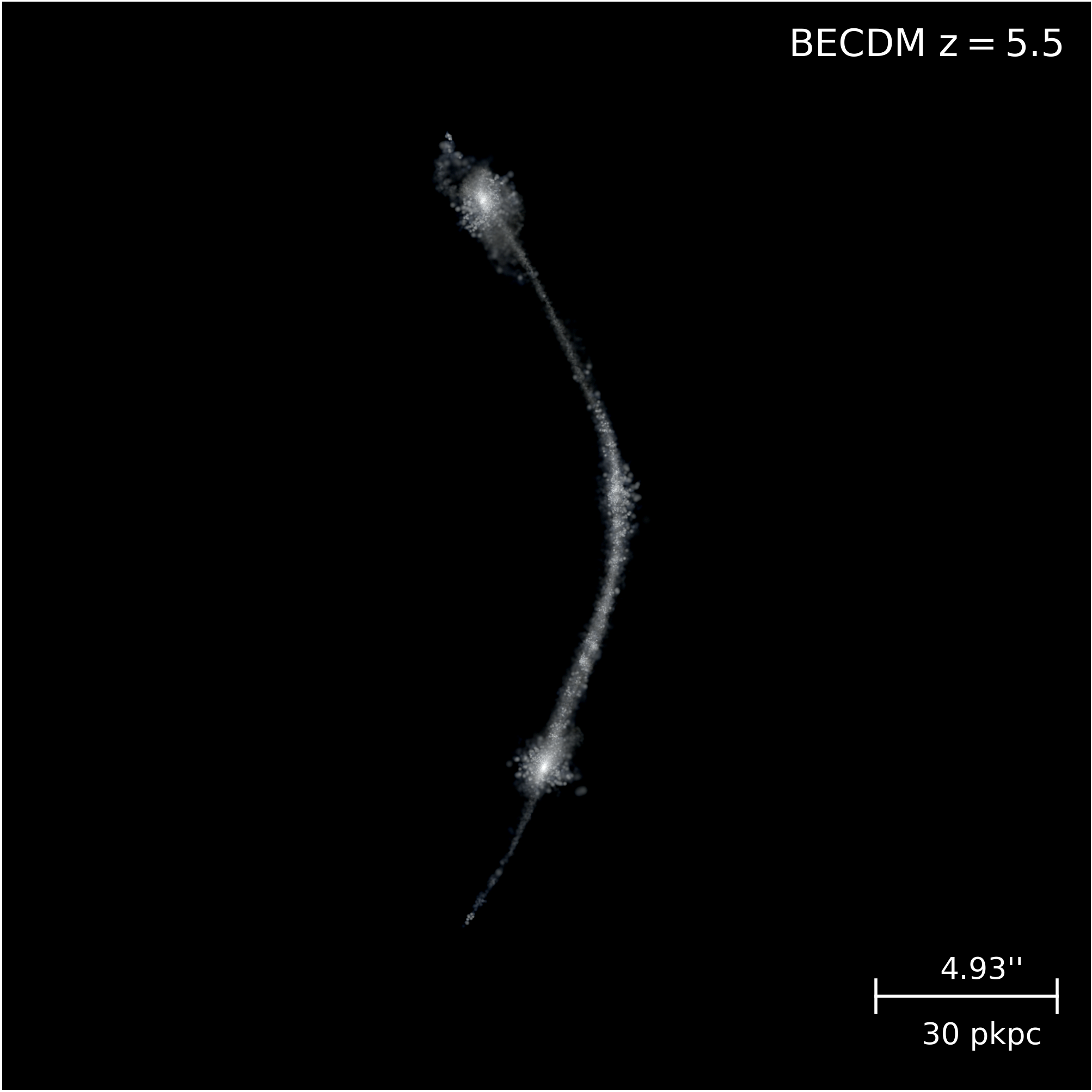}
    \includegraphics[width=0.325\textwidth]{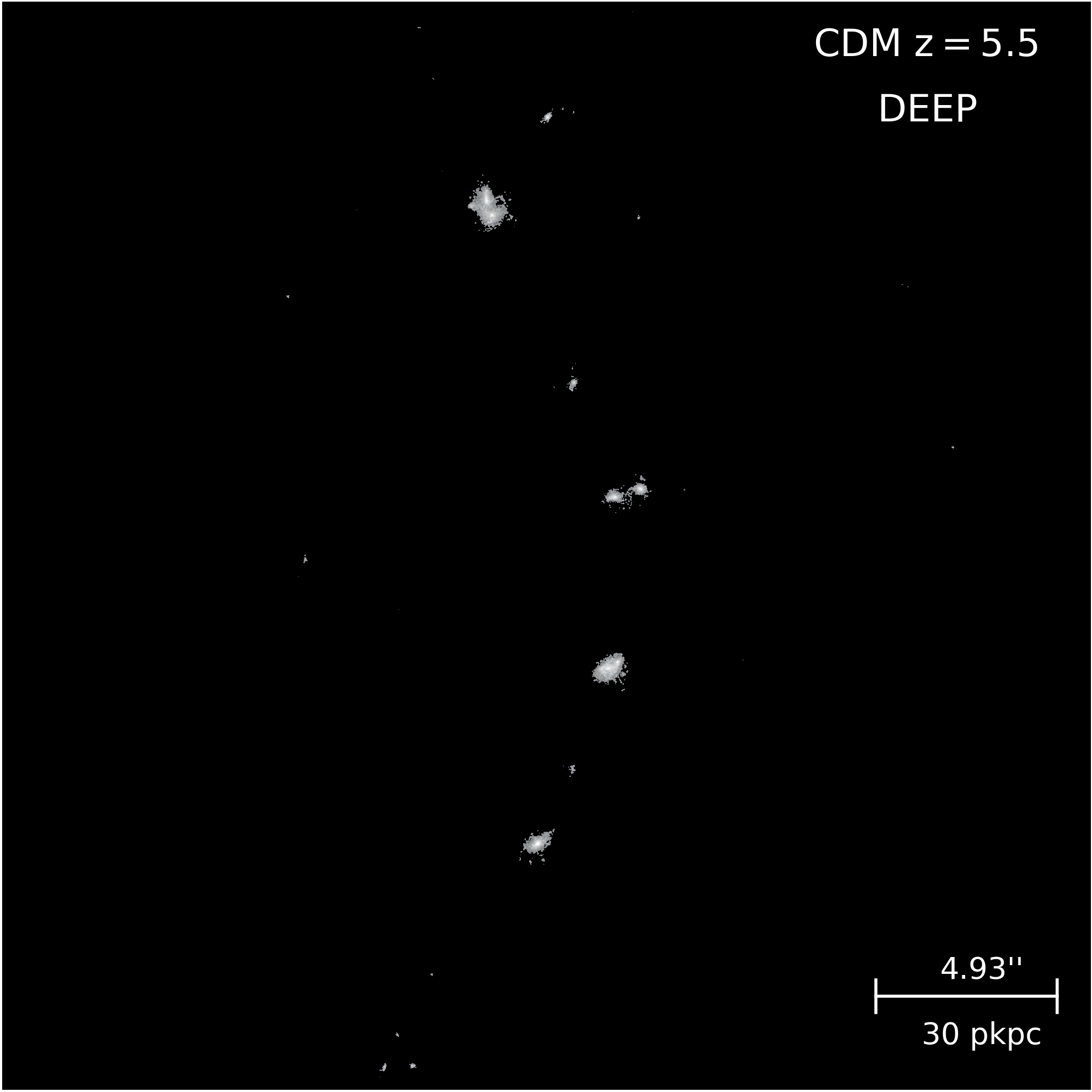}
    \includegraphics[width=0.325\textwidth]{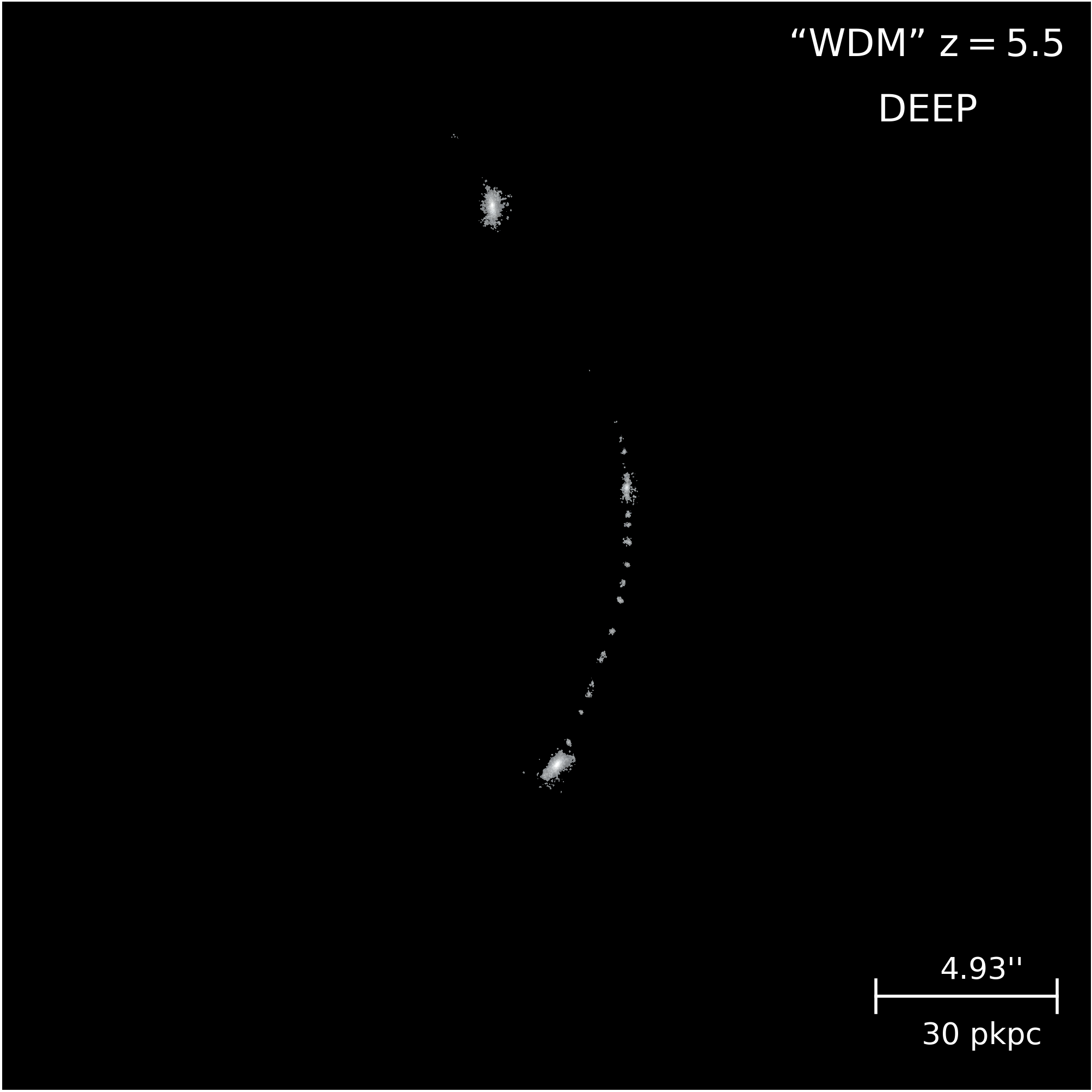}
    \includegraphics[width=0.325\textwidth]{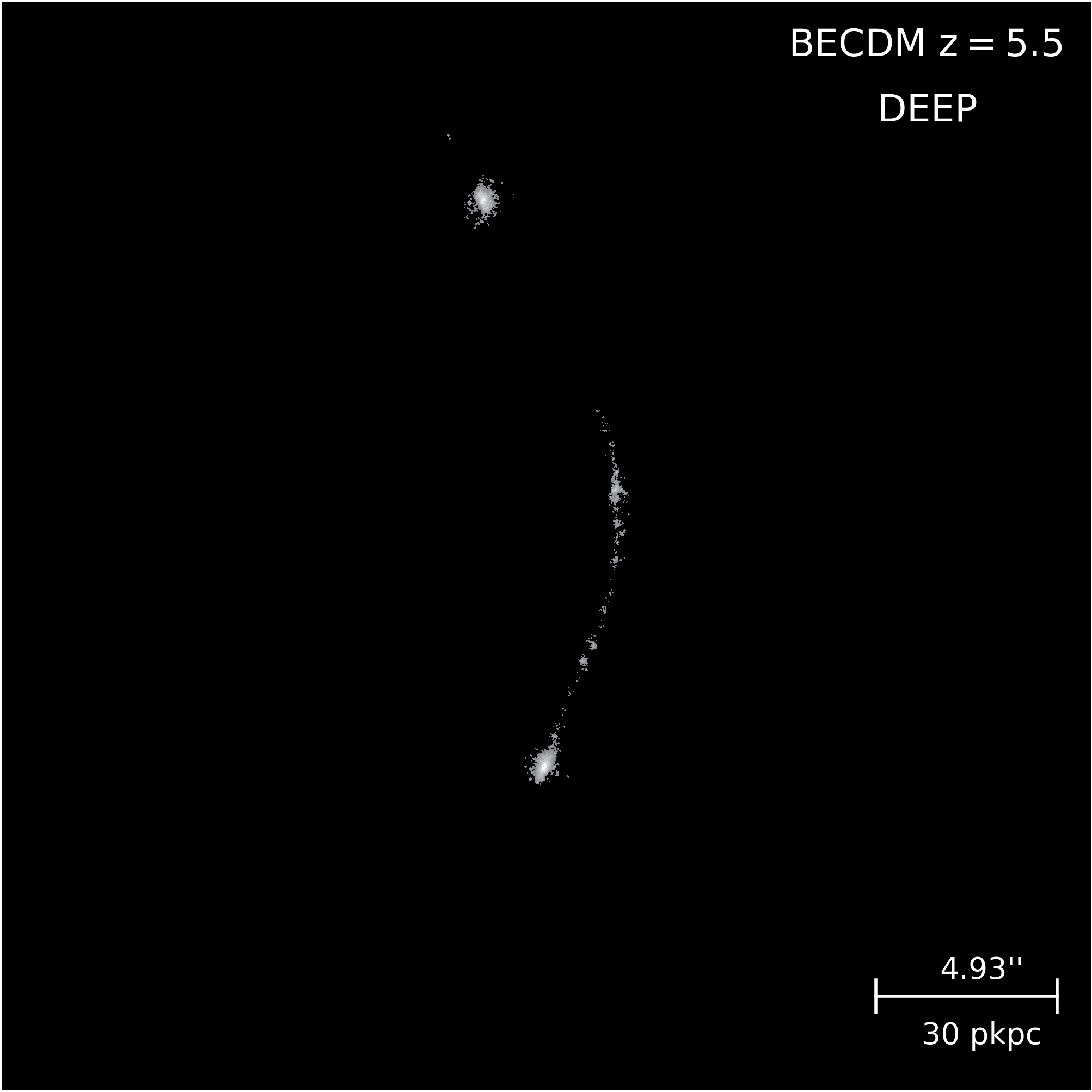}
    \caption{JWST NIRCam mock images of the first filamentary galaxy in the simulations at $z\sim5.5$. The images are synthesized based on the apparent surface brightness in F277W, F356W and F444W bands. The first row shows images without any surface brightness limit. The second row shows images with a surface brightness limit that is $50$ times deeper than the detection limit of JWST.}
    \label{fig:jwst}
\end{figure*}

To demonstrate what telescopes such as JWST would actually see, we generate mock JWST images at $z=5.5$  for the filamentary structure (stretching between haloes 2 and 3). To this end we adopt the Monte Carlo radiative transfer code {\sc Skirt}~\citep{Baes2011,Camps2013,Camps2015,Saftly2014}. Details of the parameter setup for {\sc Skirt} were introduced by \citet{Vogelsberger2019}. The images are synthesized based on the apparent surface brightness in F277W, F356W and F444W bands. We note that we do not include dust attenuation in generating these images. A radiative transfer calculation including dust attenuation performed on the biggest galaxy of this cluster has shown that dust attenuation has very limited influence on the broadband photometry of the galaxy. The raw images without any surface brightness limit are shown in the first row of Fig.~\ref{fig:jwst}. The camera is set in the positive-z direction of the simulation coordinates and the field-of-view is roughly $180\times 180$ physical kpc (pkpc). On the image we see that, as expected, at $z=5.5$ stars are distributed along the entire filament in both BECDM and WDM, while they are grouped into distinct galaxies in CDM. 

The detection limit in surface brightness of the JWST NIRCam is calculated with the {\it JWST} Exposure Time Calculator~\footnote{\url{https://jwst.etc.stsci.edu/}}~(ETC; \citealt{Pontoppidan2016}) with the following configuration details. Mock source is treated as extended with a flat surface brightness profile with a radius of $0.25''$. The aperture radius is set to $0.5''$. The target signal-to-noise ratio (SNR) is set to $5$ and the exposure time is set to $10^{5}~\rm s$. The readout pattern is set to DEEP8, which yields a high SNR and can efficiently reach a maximum survey depth. We employ $20$ groups per integration, $1$ integration per exposure and $24$ exposures per specification. For the background configuration, we choose the fiducial background at RA = 17:26:44, Dec = -73:19:56 on June 19, 2019~\footnote{\url{https://jwst-docs.stsci.edu/display/JTI/NIRCam+Imaging+Sensitivity}}. The surface brightness limit we derive for the F277W band is $\sim 0.0013\, {\rm MJy}/{\rm sr}$ which is equivalent to $27.69\, {\rm ABmag}/{\rm arcsec}^{2}$. Unfortunately, we find that these particular galaxies and the filamentary structure shown in the raw images cannot be detected under this detection limit. Therefore, in the second row of Fig.~\ref{fig:jwst}, we show the images with a surface brightness limit that is $50$ times deeper than the detection limit of the actual JWST.
We also point out that limited box statistics may mean there could be more visible cosmic filaments as well, than the one simulated.
Under this super-JWST detection limit, the filamentary structure can be revealed and the differences between three simulations are striking.
Interestingly, some amount of clumping of the star particles is observed in ``WDM'' and to a lesser extent in BECDM. Even though the clumping in ``WDM'' is a known numerical issue of ``WDM'' simulations (due to dark matter clumping) and is purely artificial, this is not the case in BECDM which is immune to the ``WDM'' numerical issues in the dark matter. Instead, minor clumping in BECDM might be a result of self gravity of stellar particles distributed along a smooth potential well, which is not accurately resolved due to the addition of a particle smoothing length. We will investigate this issue further in future work.

\section{Summary of differences} 
\label{sec:sum}
Comparing BECDM to the same-seed cosmological simulation of CDM and ``WDM'', we summarize the several qualitiative differences between these cosmologies:

\subsection{Between BECDM and CDM}

\begin{itemize}
\item BECDM appears filamentary while CDM has filaments fragmented into spherical subhalos.

\item BECDM filaments appear striated due to quantum interferences on the boson de Broglie scale while CDM has subhalos on all astrophysical scales.

\item BECDM forms smoother structures than CDM on kpc scale due to quantum pressure.

\item BECDM shows soliton cores while CDM shows much denser cusps.

\item In the BECDM scenario stars form along dense filaments (before they fragment) while in the CDM scenario they form in nearly spherical  halos (after the filaments have fragmented). In the BECDM scenario stars are distributed along the entire filaments while they are grouped into 
distinct galaxies in the CDM scenario.

\item The dark matter filaments develop cylindrical soliton-like cores that are unstable 
under gravity and collapse into kpc scale spherical solitons. There is no lower scale of collapse in CDM.

\item The distribution of gas and stars, which do form along the entire filament, exhibit central cores imprinted by dark matter. There is no equivalent in CDM.

\item BECDM halos are more triaxial than CDM halos.

\item BECDM halos shows granules (incoherent interference patterns) contrary to CDM halos.

\item Stars form later in BECDM than in CDM, and star formation is reduced in BECDM as compared to CDM. These two points should have significant consequences on the reionization history and signatures of the Universe.

\item BECDM galaxies are dimmer than in CDM model.

\item The splashback radius of halos is more distinct in BECDM than in CDM.
\end{itemize}

\subsection{Between BECDM and ``WDM''}

BECDM and ``WDM'' show similarities on large scale (e.g., both cosmologies feature dense star forming filaments), but the small scale structure is very different.

\begin{itemize}

\item BECDM filaments show  quantum interferences on the boson de Broglie scale while distribution in ``WDM'' is smooth, except for sharp caustics.

\item The dark matter density field in BECDM features cylindrical (in filaments) and spherical (inside halos) solitonic cores, while ``WDM'' profiles are cuspy.

\item The dark matter cores are imprinted in the  distribution of gas and stars  in BECDM (in both filaments and halos), while the profiles of gas and stars are  cuspy in  ``WDM''.

\item Due to the effect of the quantum potential, star formation and metal enrichment are slightly delayed (by $\Delta z \sim 0.5$) in BECDM compared to ``WDM'', and star formation is slightly systematically reduced in BECDM as compared to ``WDM''. 

\end{itemize}

\section{Conclusions}
\label{sec:conc}

In this work we have explored high-redshift galaxy formation in BECDM using first-of-the-kind cosmological simulations with BECDM fully coupled to baryons. We used the {\sc Arepo} code, a state-of-the-art high-performance parallel code for solving gravity and (magneto)hydrodynamics   \citep{2010MNRAS.401..791S}, in tandem with a newly developed but well tested module that solves the Schr\"{o}dinger-Poisson (SP) equations for BECDM \citep{2017MNRAS.471.4559M}. In addition, we ran simulations with CDM and ``WDM'' (WDM-like) cosmologies for comparison. ``WDM'' simulations are often used as a proxy for  BECDM on large cosmological scales and ignore the effect of the quantum potential on the evolution of the BECDM by only implementing the initial cutoff in the power spectrum. ``WDM'' simulations are much more economical compared to BECDM. One of the goals of this paper was to test whether or not ``WDM'' is a good approximation to BECDM when full baryonic physics is taken into account.
 

We compared BECDM to the same-seed cosmological simulation of ``WDM''. We find that, even with the baryonic feedback included, on large scales and at high redshifts ($z>5.5$) BECDM is well approximated by ``WDM'' on scales of above a few $100$~kpc. The evolved baryonic power spectrum agrees well with that of ``WDM'' on all scales and at all redshifts that we have explored. We also find that in ``WDM'' and BECDM (compared to CDM) haloes are much more triaxial (e.g., $q\sim 0.3-0.4$ instead of $0.6-0.8$ as is expected in CDM); star formation and metal enrichment are delayed from $z=35$ to $z\sim13$; the fraction of gas and stars inside virialized haloes is similar to that of CDM; depending on the initial conditions, stars can form both in isolated massive three-dimensional regions (haloes) and in more extended deep two-dimensional potential wells (along the cosmic web filaments). However, there is a systematic delay in star formation in BECDM compared to the ``WDM'' case ($\Delta z\sim 0.5$) also resulted in slower metal enrichment.

Important differences between ``WDM'' and BECDM due to the effect of the quantum potential are manifested on small scales. Smoking gun signatures of BECDM include: striated interference patterns seen in the cosmic web which result in enhanced small-scale power of dark matter fluctuations at low redshifts, formation of cylindrical and spherical solitonic cores inside haloes and filaments which are imprinted in the distribution of gas and stars \citep[see][]{moczPRL}. 

Our numerical method allows one to perform cosmological hydrodynamical simulations of BECDM with the same rigor as is done for CDM (although we are limited in cosmological box size due to the resolution requirements). The results of our simulations suggest new observational ways to test BECDM indicating that the key is in its small scale structure. Triaxiality of dark matter halos, rate of tidal disruption events, gravitational  lensing,  splash-back radius, abundance of supermassive and intermediate black holes might point out the nature of dark matter. We leave more quantitative study to future work. Our simulations are a firm step towards making robust constraints of this dark matter theory.

\section*{Acknowledgments}
We thank Jerry Ostriker, Mariangela Lisanti, David Spergel, Scott Tremaine, James Bullock, and Frenk van den Bosch for valuable discussions.
Support (P.M.) for this work was provided by NASA through Einstein Postdoctoral Fellowship grant number PF7-180164 awarded by the \textit{Chandra} X-ray Center, which is operated by the Smithsonian Astrophysical Observatory for NASA under contract NAS8-03060. 
A.F. is supported by the Royal Society University Research Fellowship.
M.B.K. acknowledges support from NSF grants AST-1517226, AST-1910346, and CAREER grant AST-1752913 and from NASA grants NNX17AG29G and HST-AR-14282, HST-AR-14554, HST-AR-15006, HST-GO-14191, and HST-GO-15658 from the Space Telescope Science Institute, which is operated by AURA, Inc., under NASA contract NAS5-26555. 
J.Z. acknowledges support by a Grant of Excellence from the Icelandic Research fund (grant number 173929).
F.M. is supported by the Program ``Rita Levi Montalcini'' of the Italian MIUR.
The authors acknowledge the Texas Advanced Computing Center (TACC) at The University of Texas at Austin for providing HPC resources that have contributed to the research results reported within this paper. URL: \url{http://www.tacc.utexas.edu}. XSEDE Allocation TG-AST170020.
Some of the computations in this paper were run on the Odyssey cluster supported by the FAS Division of Science, Research Computing Group at Harvard University.

\bibliography{mybib}{}

\label{lastpage}
\end{document}